\documentclass[12pt]{article}

\usepackage{amsmath}
\usepackage{graphicx}
\usepackage{enumerate}
\usepackage[authoryear,round]{natbib}
\usepackage{url} 

\usepackage{graphics}
\usepackage{amsmath}
\usepackage{amssymb}
\usepackage{xcolor}
\usepackage{mathtools}
\usepackage{bbm}
\usepackage{mathabx}
\usepackage{multirow}
\usepackage{siunitx}
\usepackage{tabularx}
\usepackage{array}
\usepackage{changepage}
\usepackage{tcolorbox}
\usepackage{enumitem}

\usepackage{xr}

\usepackage{algorithm2e}
\RestyleAlgo{ruled}
\DontPrintSemicolon

\SetCommentSty{mycommfont}
\usepackage{subcaption}
\usepackage{stmaryrd}
\usepackage[labelfont=bf, labelsep=colon, format=hang, textfont=singlespacing]{caption}

\usepackage{booktabs}

\newtheorem{theorem}{Theorem}

{\hfill{$\Box$}\\ \vspace*{0.3cm}} 

\newcounter{example}

\begin{document}


\newcommand{\blind}{1}

\def\spacingset#1{\renewcommand{\baselinestretch}%
	{#1}\small\normalsize} \spacingset{1}

\if1\blind
{
	\title{\bf Bounds for the regression parameters in dependently censored survival models}
	\author{Ilias Willems\thanks{Corresponding author (ilias.willems@kuleuven.be).}\\
		ORSTAT, KU Leuven, Belgium\\
		and \\
		Jad Beyhum\thanks{Jad Beyhum gratefully acknowledges financial support from the Research Fund KU Leuven through the grant STG/23/014.} \\
		Department of Economics, KU Leuven, Belgium\\
		and\\
		Ingrid Van Keilegom\thanks{Ingrid Van Keilegom gratefully acknowledges funding from the FWO and F.R.S. - FNRS (Excellence of Science programme, project ASTeRISK, grant no. 40007517), and financial support from the FWO (senior research projects fundamental research, grant no. G047524N).}\\
		ORSTAT, KU Leuven, Belgium}
	\maketitle
} \fi

\if0\blind
{
	\bigskip
	\bigskip
	\bigskip
	\begin{center}
		{\LARGE\bf Bounds for the regression parameters in dependently censored survival models}
	\end{center}
	\medskip
} \fi

\bigskip
\begin{abstract}
	We propose a semiparametric model to study the effect of covariates on the distribution of a censored event time while making minimal assumptions about the censoring mechanism. The result is a partially identified model, in the sense that we obtain bounds on the covariate effects, which are allowed to be time-dependent. Moreover, these bounds can be interpreted as classical confidence intervals and are obtained by aggregating information in the conditional Peterson bounds over the entire covariate space. As a special case, our approach can be used to study the popular Cox proportional hazards model while leaving the censoring distribution as well as its dependence with the time of interest completely unspecified. A simulation study illustrates good finite sample performance of the method, and several data applications in both economics and medicine demonstrate its practicability on real data. All developed methodology is implemented in R and made available in the package \texttt{depCensoring}.
\end{abstract}

\noindent%
\vfill

\newpage

\spacingset{1}


\section{Introduction} \label{sec: Introduction}

In the context of event time modeling, the \cite{Cox1972} proportional hazards (PH) model has enjoyed great popularity among practitioners. However, it has the downside of having to impose that the event time $T$ and censoring time $C$ are independent, which is an assumption that deserves careful scrutiny \citep{KaplanMeier1958}, as it may not hold in many applications. While several innovative works (\cite{HuangZhang2008}; \cite{Deresa2023}, ...) have succeeded in relaxing this assumption since the seminal publication by \cite{Cox1972}, a truly censoring-agnostic method for doing inference on the Cox PH model has eluded the literature thus far. With this paper, we aim to fill that gap. More precisely, we propose a framework that, as a special case, is able to perform inference in the Cox PH model without making any assumptions on both the censoring distribution as well as its dependence with respect to the event of interest. The framework, however, reaches much further as it includes many more models (e.g. the proportional odds model), and can handle time-dependent effects of covariates. 

Under the generality of the proposed approach -- specifically, by not imposing any assumptions on the censoring mechanism -- the parameters in the Cox PH model are not identified \citep{HuangZhang2008}. Therefore, we can only aim to \emph{partially identify} them. In practice, this means that we will estimate bounds on the covariate effects instead of obtaining point estimates. These bounds can be interpreted as regular confidence intervals and can still be informative provided they are sufficiently narrow.


From a technical point of view, the bounds are obtained starting from the conditional \cite{Peterson1976} bounds. These are generally known to be wide and uninformative, but by aggregating their information over the entire covariate space, we are able to convert them into meaningful bounds on the parameters. We achieve this by employing instrumental functions to cast the problem into one of testing moment restrictions and proving that the recent developments of \cite{Bei2024} are valid in this setting. The insights and structure used to construct the proofs are not specific to our setting and could be used by researchers who want to apply this test in a different setting. In particular, we replace a strong assumption made in \cite{Bei2024} by a more tractable one. This assumption from \cite{Bei2024} has already been the subject of detailed research \citep{Kaido2022}. As such, our theoretical results could be of independent interest.

The finite sample performance of the methodology is assessed by means of a simulation study. To demonstrate the practicability on real data, we provide two data applications in economics and medicine, where the independent censoring assumption may be dubious in the presented cases. The software is made available on CRAN in the \texttt{depCensoring} package in R, and details regarding the implementation can be found in Supplementary material \ref{supp: Algorithm and implementation details}.

\subsection{Literature review}\label{sec: Literature review}

Event time analysis under dependent censoring has received increasing amounts of attention over the last decade, and could be argued to have originated from the work by \cite{Tsiatis1975}, who shows that the distribution of $T$ is in general not identifiable in a fully nonparametric way. Therefore, additional assumption on the data generating process -- and, particularly, on the dependence between $T$ and $C$ -- often have to be made.

Many approaches are centered around copulas to model the dependence between $T$ and $C$, either specifying a fully known copula (\cite{ZhengKlein1995, RivestWells2001, BV2005, HuangZhang2008, SujicaVK2018}, ...), or allowing for a parametric copula class and estimating the dependence parameter (\cite{Czado2023, Lo2023, Deresa2023, Deresa2025, Ding2025}, ...). We refer to \cite{Crommen2024} for an in-depth review on copula-based methods for dependent censoring, as well as some other approaches, including transformation models (\cite{Deresa2020_a, Deresa2021, Crommen2023, Rutten2024+}, ...). Notably, all works discussed in their review paper impose restrictions on the dependence between the censoring time and the event time in order to obtain point identification. In contrast, we are fully agnostic on this dependence, but only reach partial identification.

Partial identification is mostly studied in the field of econometrics. By allowing models to only be partially identified, one is often able to omit stringent modeling assumptions which would otherwise be needed to attain point identification. The body of literature on partial identification can partly be categorized into two branches. One branch focuses on the special case of compact, convex identified sets and makes use of support functions, which allows for many computational advantages. The other branch considers more general identified sets, but as a trade-off needs to use a \emph{test inversion} method (cf. infra), which is computationally more burdensome. We refer to \cite{Molinari2020} for an extensive review on this literature.

Of most interest for this paper is the literature situated at the intersection of survival analysis and partial identification. In this line, several works exist, though in many, restrictive assumptions persist. \cite{HorowitzManski1998} use weighting or imputation to deal with the missing information that results from censoring, but this requires the specification of a weight function or imputation method, which in turn imposes assumptions on the censoring distribution. Other authors (\cite{Svidlowksi2019}; \cite{Kim2023}, ...) avoid this approach, but impose that covariates and/or outcomes are discrete, which does not allow practitioners to include typical continuous covariates such as age into their analysis. To our knowledge, the most general approach is provided by \cite{Sakaguchi2023}, who is able to partially identify several parameters of interest, including covariate effects, while leaving the censoring mechanism unspecified. Since his work is based on the assumption of time-independent effects and employs a different methodology as in this paper, our work is not comparable, but should rather be viewed as complementary. Finally, as an overarching remark, the works discussed in this paragraph have an econometric focus, whereas partially identified methods for statistics and biostatistics have received far less attention.

\subsection{Paper outline}
The remainder of this paper is structured as follows. In Section \ref{sec: The model} we introduce the model and explain the methodology used to estimate it. The section is self-contained and should suffice for researchers who want to apply our model in a practical setting. Section \ref{sec: Estimation procedure} contains the theoretical underpinnings, elaborates on several aspects of the method and provides a detailed discussion. Following that, we investigate finite sample performance in Section \ref{sec: Simulation study} using simulation studies. Section \ref{sec: Data applications} applies our methodology to two real data sets. Many additional analyses, remarks and discussions are deferred to the Supplementary material.

\section{Model and methodology}\label{sec: The model}
Let $T$ and $C$ be real-valued random variables. Suppose we have $n$ observations $\{W_i\}_{i = 1}^n$ which are independent and identically distributed according to $W = (Y, \Delta, \tilde{X})$, where $Y = \min(T, C)$, $\Delta = \mathbbm{1}(Y = T)$ and $\tilde{X}$ a $d$-dimensional vector of covariates, each element of which can either be discrete or continuous. Let $X = (1, \tilde{X}^\top)^\top$, which maps into the covariate space $\mathcal{X}$. For convenience, we will index the elements of $X$ starting from zero, so that $X_0 \equiv 1$ and $X_k = \tilde{X}_k$, for all $k \in \{1, \dots, d\}$. In this way, we can use $X_k$ and $\tilde{X}_k$ interchangeably. We will refer to random variables with uppercase letters and to realizations with lowercase letters. Fix a time point of interest $t$. We consider the model
\begin{equation} \label{eq: model}
	\mathbb{P}(T > t | X = x) = 1 - \Lambda_t(x^\top \beta_{\text{true}, t}),
\end{equation}
where $\Lambda_t: \mathbb{R} \to (0, 1)$ is a prespecified link function and $\beta_{\text{true}, t} = (\beta_{\text{true}, t,0}, \dots, \beta_{\text{true}, t,d}) \in \mathbb{R}^{d+1}$ is the true parameter vector. Under certain choices for $\Lambda_t$, this formulation includes some well-known models, such as the semiparametric Cox PH model (using $\Lambda_t(\cdot) = 1 - \exp(-\exp(\cdot)$) or the proportional odds model (cf. Supplementary material \ref{supp: Link function}), and has been studied before, for example by \cite{Delgado2022}. However, in contrast to these models, we stress again that we \emph{only} impose our model at a \emph{fixed} time point $t$. As a consequence, when choosing $\Lambda_t$ in order to specify model \eqref{eq: model} in the form of a Cox PH or proportional odds model, we emphasize that the proportional hazards or odds assumption are not actually imposed. Indeed, the model is only imposed at a single time point $t$, and hence the covariate effects are allowed to vary over time (more generally, the model can even be completely different at other time points). For practical settings where these proportionality assumptions would be valid, we extend the approach in Section \ref{sec: Time-independent effects of covariates} so that it can include the knowledge of covariate effects that are constant over time. For simplicity, we will suppress the subscript $t$ from the notation throughout the rest of the paper.

Specifying a choice for $\Lambda$ is not innocuous and should be motivated based on the practical setting. However, the mathematical structure provided by this minimal modeling assumption enables us to improve upon the generally uniformative \cite{Peterson1976} bounds, and provides the model with interpretable regression coefficients. Moreover, we obtain a specification test in Section \ref{sec: Discussion}. 

Naturally, the choice of $\Lambda$ determines the way in which the coefficients should be interpreted. For $\Lambda$ related to the Cox PH model as explained above, we have $\beta_\text{true} = (\log(H_0(t)), \beta_{\text{true}, 1}, \dots, \beta_{\text{true}, d})$, where $H_0(t) = \int_{-\infty}^t h_0(s) ds$ and the true baseline hazard $h_0(s)$ is allowed to vary in $s$ nonparametrically. This indeed results in a model that includes the Cox PH model:
\begin{equation*}
    S(t|x) = \exp(-\exp(\log(H_0(t)) + \tilde{x}^\top \tilde{\beta}_\text{true})) = \exp(-H_0(t) \exp(\tilde{x}^\top \tilde{\beta}_\text{true})),
\end{equation*}
where, analogously to the definition of $X = (1, \tilde{X})$, we define $\tilde{\beta}_\text{true} = (\beta_{\text{true}, 1}, \dots, \beta_{\text{true}, d})$.

To achieve point identification of $\beta_\text{true}$, additional assumptions on the data generating process will have to be made. For example, \cite{Delgado2022} impose the independence assumption. As already mentioned in the introduction, such identifying assumptions are stringent. We will therefore avoid them and allow model \eqref{eq: model} to be only partially identified instead of aiming for point identification.

\subsection{A preliminary note on partial identification}
We start by introducing some concepts regarding model identification. For comprehensive overview, we refer to \cite{Lewbel2019}. We will say that a conditional distribution $F_{T|X}$ implied by model \eqref{eq: model} is \emph{consistent} with the observed data distribution $F_{Y, \Delta, X}$ if there exists a conditional censoring distribution $F_{C|X}$ and copula $\mathcal{C}$ \citep{Sklar1959} such that $(\mathcal{C}(F_{T|X}, F_{C|X}), F_{X})$ together imply the observed distribution of $(Y, \Delta, X)$. The set of all distributions $F_{T|X}$ that are consistent with the observed data is referred to as the \emph{identified set of (conditional) distributions}. Each pair of elements in this set is called \emph{observationally equivalent}, since one cannot further distinguish these elements in terms of validity solely based on the observed data.

By model \eqref{eq: model} we can express the identified set of conditional distributions in terms of the parameter vector. That is, the identified set of interest becomes the set of candidate parameter vectors $\beta$ for $\beta_\text{true}$ for which model \eqref{eq: model} corresponds to a conditional distribution $F_{T|X}$ that is consistent with the data.

\subsection{Methodology}\label{sec: Methodology}
Let $\mathcal{B} \subset \mathbb{R}^{d + 1}$ denote the parameter space. Since our model is not identified, $\beta_\text{true}$ might not be the only parameter for which the model is consistent with the observed data generating process. That is, a different parameter vector $\beta^*$ might correspond to a distribution of $T|X$ which is also consistent with the observed data. Likewise, we call such values $\beta^*$ observationally equivalent to $\beta_\text{true}$. Consequently, there is no way to distinguish $\beta^*$ from $\beta_\text{true}$ in terms of validity and hence the best we can aim for in this context is to find the set $\mathcal{B}_I^*$ of all parameter vectors $\beta$ that are observationally equivalent to $\beta_\text{true}$. This set, which we refer to as the \emph{identified set of parameters}, or simply \emph{identified set}, will be our main object of interest.

To study and eventually estimate this identified set it is useful to first ask the question how we can mathematically formalize the concept of parameters being observationally equivalent to $\beta_\text{true}$. The answer is provided by \cite{Peterson1976}, who derives the bounds
\begin{equation} \label{eq: Peterson bounds}
	\mathbb{P}(Y \leq t, \Delta = 1 | X = x) \leq \Lambda(x^\top\beta) \leq \mathbb{P}(Y \leq t | X = x),
\end{equation}
based on which we define
\begin{equation}\label{eq: definition true identified set}
	\mathcal{B}^*_I = \{\beta \in \mathcal{B} \mid \forall x \in \mathcal{X}: \beta \text{ satisfies \eqref{eq: Peterson bounds}}\}.
\end{equation}
Equation \eqref{eq: definition true identified set} precisely characterizes the parameters that are observationally equivalent to $\beta_\text{true}$: all parameter vectors in $\mathcal{B}_I^*$ must lead to a model that is consistent with the bounds \eqref{eq: Peterson bounds}. Moreover, the bounds in \eqref{eq: Peterson bounds} are sharp. In other words, Peterson's bounds are the best obtainable bounds on $\mathbb{P}(T \leq t | X = x)$ that are based on the observed data alone.

With the formal characterization of the identified set \eqref{eq: definition true identified set} in hand, the idea behind estimating $\mathcal{B}^*_{I}$ is straightforward: we could consider a test for checking whether the model with a certain parameter vector $\beta$ satisfies \eqref{eq: Peterson bounds} and apply this test over the entire parameter space $\mathcal{B}$, collecting all values that are not rejected. The set of non-rejected values is then an estimator for $\mathcal{B}^*_{I}$. In the literature, such an approach is referred to as \emph{test inversion}.

Typically, however, researchers are only interested in one or a few elements of the vector $\beta_\text{true}$. Suppose that one of the elements of interest has index $k$, then this means that interest is only in the projection of $\mathcal{B}^*_I$ onto the $k$-th coordinate axis, denoted as $\mathcal{B}^*_{I, k}$. In the literature, studying $\mathcal{B}^*_{I, k}$ directly is often referred to as \emph{subvector inference} (as opposed to \emph{full vector inference}, where one studies $\mathcal{B}_I^*$). In this paper, we will also focus on the subvector problem, and, analogously to what is often done in the full vector case, use test inversion for the estimation. Note that multiple elements of $\beta_\text{true}$ might be of interest simultaneously when a categorical covariate enters the model through several dummy variables. To keep the exposition clear, however, we will restrict to the case where interest is only in the $k$-th element. All results can be generalized by replacing $\beta_k$ with a general subvector of dimension smaller than $d+1$.

Concretely, we define
\begin{align}
	\mathcal{B}^*_{I, k} & = \{r \in \mathcal{B}_k \mid \exists \beta \in \mathcal{B}_I^*: \beta_k = r\} \notag\\
	& = \{r \in \mathcal{B}_k \mid \exists \beta \in \mathcal{B}: \beta_k = r \text{ and } \forall x \in \mathcal{X}: \beta \text{ satisfies \eqref{eq: Peterson bounds}}\}, \label{eq: BIkstar}
\end{align}
where $\mathcal{B}_k$ denotes the projection of $\mathcal{B}$ onto the $k$-th coordinate. When testing the condition of the set defined in \eqref{eq: BIkstar} for the test inversion approach, one difficulty to note is that as soon as a continuous covariate enters the model, we require a continuum of restrictions to be tested: \eqref{eq: Peterson bounds} has to hold for each value of $x \in \mathcal{X}$. To our knowledge, subvector inference tests that can handle this case (i.e. testing conditional moment restrictions) do not exist in the literature yet. Therefore, following \cite{AndrewsShi2013}, we discretize this continuum of restrictions making use of instrumental functions. In this way, we obtain a condition that is similar to \eqref{eq: BIkstar}, only now $\beta$ has to satisfy finitely many restrictions. The cost of taking this approach is that the new condition will be slightly weaker than the original one, and hence the set $\mathcal{B}_{I, k}$ corresponding to this new condition will be a superset of $\mathcal{B}_{I, k}^*$. As a consequence, when regarding an eventual estimator for $\mathcal{B}_{I, k}$ as an estimator for $\mathcal{B}_{I, k}^*$, it will be slightly conservative.

In the following, we first elaborate on how the new condition is constructed and formally define $\mathcal{B}_{I, k}$. Next, a test for this modified condition is obtained. Lastly, we give an overview of the entire estimation procedure.

\subsection{Unconditional moment restrictions}\label{sec: Unconditional moment restrictions}
To solve the aforementioned challenge posed by continuous covariates entering the model, we first transform the possible continuum of restrictions in \eqref{eq: BIkstar} to a finite number of restrictions. To this end, we start by rewriting Peterson's bounds in Equation \eqref{eq: Peterson bounds} in the form of two conditional moment inequalities:
\begin{equation} \label{eq: conditional moment inequalities}
	\forall x \in \mathcal{X}:
	\begin{cases}
		\mathbb{E}[\mathbbm{1}(Y \leq t) - \Lambda(X^\top\beta) | X = x] \geq 0\\
		\mathbb{E}[\Lambda(X^\top \beta) - \mathbbm{1}(Y \leq t, \Delta = 1) | X = x] \geq 0.
	\end{cases}
\end{equation}
Next, these conditional moment restrictions are transformed to unconditional ones making use of a class $\mathcal{G} = \{g_j, j 
= 1, \dots, J\}$ of \emph{instrumental functions} \citep{AndrewsShi2013}, where $J$ denotes the number of instrumental functions considered:
\begin{equation} \label{eq: unconditional moment inequalities}
	\forall j \in \{1, \dots, J\}:
	\begin{cases}
		m_{j,1}(W, \beta) = \mathbb{E}[(\mathbbm{1}(Y \leq t) - \Lambda(X^\top\beta))g_j(X)] \geq 0\\
		m_{j,2}(W, \beta) = \mathbb{E}[(\Lambda(X^\top \beta) - \mathbbm{1}(Y \leq t, \Delta = 1))g_j(X)] \geq 0.
	\end{cases}
\end{equation}
Intuitively, the idea behind this transformation can be explained as follows. Expression \eqref{eq: conditional moment inequalities} defines two conditional moment inequalities, which can be viewed as two inequalities that should hold for each $x \in \mathcal{X}$. As already mentioned, this is problematic when $\mathcal{X}$ is an infinite set. The class $\mathcal{G}$ contains functions $g: \mathcal{X} \to \mathbb{R}_{\geq 0}$ that are only non-zero on $\mathcal{X}_g$, where $\mathcal{X}_g$ is a small region inside $\mathcal{X}$ such that $ \mathcal{X} \subseteq \bigcup_{g \in \mathcal{G}} \mathcal{X}_g$. The transformed unconditional moment restrictions thus only require that the inequalities are expected to hold on each of the regions $\mathcal{X}_g$ (instead of for all $x \in \mathcal{X}$), essentially discretizing the possible continuum of restrictions in \eqref{eq: conditional moment inequalities}. Also note that if $X$ contains only discrete elements, \eqref{eq: conditional moment inequalities} and \eqref{eq: unconditional moment inequalities} are equivalent when each instrumental function in $\mathcal{G}$ is only non-zero for one level of $X$. We define
\begin{align}
	\mathcal{B}_I & = \{\beta \in \mathcal{B} \mid \beta \text{ satisfies \eqref{eq: unconditional moment inequalities}}\}, \label{eq: BI}\\
	\mathcal{B}_{I, k} & = \{r \in \mathcal{B}_k \mid \exists \beta \in \mathcal{B}: \beta_k = r \text{ and } \beta \text{ satisfies \eqref{eq: unconditional moment inequalities}}\}. \label{eq: BIk}
\end{align}
In the process of transforming \eqref{eq: conditional moment inequalities} to \eqref{eq: unconditional moment inequalities} in the presence of continuous covariates, some information is lost. As a consequence, the condition in Equation \eqref{eq: BIk} is weaker than the one in \eqref{eq: BIkstar} and therefore $\mathcal{B}_{I, k}$ will be a superset of $\mathcal{B}_{I, k}^*$. The simulations show, however, that the information loss is limited, and data applications illustrate that inference based on the unconditional moment restrictions can still be informative. For a more detailed explanation of instrumental functions, we refer to Section 3.3 of \cite{AndrewsShi2013}.

\subsection{Testing procedure} \label{sec: Testing procedure}


To estimate $\mathcal{B}_{I, k}$ defined in \eqref{eq: BIk} by means of test inversion we require a test for the hypothesis
\begin{equation} \label{eq: hypothesis}
	\mathcal{H}_0(r): \exists \beta \in \mathcal{B}: \beta_k = r \text{ and } \beta \text{ satisfies \eqref{eq: unconditional moment inequalities}}.
\end{equation}
Such a test is described in \cite{Bei2024}, who uses the statistic $T_n(r) = \inf_{\beta \in \mathcal{B}(r)} S(\sqrt{n} \bar{m}(\beta), \hat{\sigma}(\beta))$, where $S(\sqrt{n}\bar{m}(\beta), \hat{\sigma}(\beta)) = \sum_{q = 1}^2 \sum_{j = 1}^J \max\{-\sqrt{n}\bar{m}_{j, q}(\beta)/\hat{\sigma}_{j, q}(\beta), 0\}^2$. In this notation, we define the vector-valued function $m = (m_{1, 1}, \dots, m_{J, 1}, m_{1, 2} \dots, m_{J, 2})$ and denote by $\bar{m}(\beta)$ and $\hat{\sigma}(\beta)$ the sample average and standard deviation of $m(W, \beta)$. The set $\mathcal{B}(r) = \{\beta \in \mathcal{B} \mid \beta_k = r\}$ contains all parameter vectors that have $r$ as their $k$-th entry.

Intuitively, the function $S$ can be viewed as a measure how much the sample moment restrictions are violated at a given $\beta$. The test statistic $T_n(r)$ considers the value of $\beta$ for which this violation is at a minimum and such that the $k$-th component of $\beta$ is equal to $r$. If this minimal violation is too large, $\mathcal{H}_0(r)$ can be rejected. To this end, a specialized bootstrap procedure to obtain the critical value $\gamma_{n, 1 - \alpha}(r)$ of $T_n(r)$ is developed. More detailed information on this test, and in particular on the bootstrap procedure, can be found in Supplementary material \ref{supp: Testing procedure of Bei}. We refer to \cite{Bei2024} for a full explanation.

\subsection{Overview}
In summary, after selecting a time point $t$ and a coefficient of interest $\beta_k$, the methodology of this paper consists of two parts. First, a test for the hypothesis \eqref{eq: hypothesis} is obtained. Next, this test is applied over the entire parameter subspace $\mathcal{B}_k$ and all non-rejected values are collected into the set $\hat{\mathcal{B}}_{I, k}$, which is then an estimator for $\mathcal{B}_{I, k}$. We refer to Supplementary material \ref{supp: Algorithm and implementation details} for details on how this can be done in practice via root finding algorithms. Notably, $\hat{\mathcal{B}}_{I, k}$ will, in most cases, be an interval so that it can be interpreted as a standard confidence interval (Section \ref{sec: Discussion}). 

\section{Estimation procedure}\label{sec: Estimation procedure}

\subsection{Modeling assumptions and theoretical results} \label{sec: Assumptions and theoretical results}

Throughout the rest of this paper, we will assume that interest lies in the $k$-th element of the parameter vector. We require the following set of assumptions:
\begin{enumerate}[noitemsep, label=(A\arabic*)]
	\item \label{Assumption iid} The observations $\{W_i, i = 1, \dots, n\}$ are independent and identically distributed.
	\item \label{Assumption Lambda} $\Lambda: \mathbb{R} \to (0,1)$ is a strictly increasing function that is twice continuously differentiable.
	\item \label{Assumption IF} $\mathcal{G}$ is a class of continuous functions $g_j: \mathbb{R}^{d+1} \to [0, M_\mathcal{G}]$, $j = 1, \dots, J$, for some $M_\mathcal{G} < \infty$.
	\item \label{Assumption support IF} Let $\mathcal{X}_{g, j}$ denote the support of the $j$-th instrumental function $g_j(\cdot)$. Then $\mathcal{X} \cap \mathcal{X}_{g, j} \neq \emptyset$ and $\mathcal{X} \subset \bigcup_{j = 1}^J \mathcal{X}_{g, j}$.
	\item \label{Assumption parameter space} The parameter space $\mathcal{B}$ is a non-empty, convex and compact subspace of $\mathbb{R}^{d + 1}$.
	\item \label{Assumption covariate space} The covariate space $\mathcal{X}$ is bounded.
	\item \label{Assumption variance of moments} $\exists \eta > 0: \forall j \in \{1, \dots, J\}, q \in \{1, 2\}: \inf_{\beta \in \mathcal{B}} \sigma_{j, q}(\beta) > \eta$, where $\sigma_{j, q}(\beta)$ denotes the standard deviation of $m_{j, q}(W, \beta)$.
\end{enumerate}

The first assumption is standard. Assumptions \ref{Assumption Lambda} and \ref{Assumption IF} hold for all link functions and classes of instrumental functions discussed in this paper. Assumption  \ref{Assumption support IF} imposes that the support of each instrumental function has non-empty intersection with the covariate space, and that each point in $\mathcal{X}$ is covered by at least one instrumental function. For a discussion on how to choose the family of instrumental functions in this way, we refer to Section \ref{sec: Instrumental functions}. Assumptions \ref{Assumption parameter space} and \ref{Assumption covariate space} are regularity conditions on the parameter and covariate space, respectively. In practice, one could select $\mathcal{B} = [-M_\mathcal{B}, M_\mathcal{B}]^{d + 1}$, for a sensible bound $M_\mathcal{B} > 0$. Assumption \ref{Assumption variance of moments} imposes that the variances of the moment functions are uniformly bounded away from zero. Example \ref{example: assumption variance of moments} in Supplementary material \ref{supp: Examples} illustrates that this is a mild and verifiable assumption.

Additionally, we require two assumptions of which we present the essence here and defer technical details to Supplementary material \ref{supp: Technical assumptions}.
\begin{enumerate}[noitemsep, label=(A8)]
	\item $\exists c > 0: \beta_{\text{true}, k} \in \tilde{\mathcal{L}}_0(c)$. \label{Assumption boundary BI}
\end{enumerate}
The definition of the set $\tilde{\mathcal{L}}_0(c) \subset \mathcal{B}_{I, k}$ can be found in Supplementary material \ref{supp: Technical assumptions}. Assumption \ref{Assumption boundary BI} deserves special attention, as its purpose is to account for the stringency of Assumption 2 in \cite{Bei2024}, which has already been the subject of detailed research \citep{Kaido2022}. In essence, $\tilde{\mathcal{L}}_0(c)$ serves to restrict the values of $r \in \mathcal{B}_{I, k}$ for which the hypothesis $\mathcal{H}_0(r)$ in \eqref{eq: hypothesis} can be tested with correct type-I error, and Assumption \ref{Assumption boundary BI} states that $\beta_{\text{true}, k}$ is an element of this set. Hence the estimated identified set will contain all values of $\tilde{\mathcal{L}}_0(c)$ with at least the specified confidence level $1 - \alpha$, in particular including $\beta_{\text{true}, k}$. Since $\tilde{\mathcal{L}}_0(c)$ may seem abstract at first glance, we provide an example (Example \ref{example: assumption boundary BI}) in Supplementary material \ref{supp: Examples} that illustrates which values are excluded. By nature of the unidentifiablity of $\beta_{\text{true}, k}$, Assumption \ref{Assumption boundary BI} is untestable. However, the aforementioned example illustrates that for well-behaved $\mathcal{B}_I$ the set of excluded values is very small.
\begin{enumerate}[label=(A9)]
	\item $\exists \eta_\mathcal{L} > 0: \forall r \in \tilde{\mathcal{L}}_0(c): \exists \beta \in \mathcal{B}(r): \mathbb{E}[m(W, \beta)] > \eta_\mathcal{L}$, for $c$ resulting from Assumption \ref{Assumption boundary BI} and where the inequality holds elementwise.
	\label{Assumption sufficient slackness}
\end{enumerate}
Lastly, Assumption \ref{Assumption sufficient slackness} facilitates certain technical difficulties in the theory of our method. It will in many cases be satisfied when Assumption \ref{Assumption boundary BI} holds, though it is not implied by it. We can now state the main theorem.

\begin{theorem} \label{theorem: main}
	Under Assumptions \ref{Assumption iid}--\ref{Assumption sufficient slackness} it holds that $\beta_{\text{true}, k}$ will be contained in $\hat{\mathcal{B}}_{I, k}$ with at least the specified confidence level $1-\alpha$.
\end{theorem}
The proof of this result, presented in Supplementary material \ref{supp: Lemmas and theorems}, consists in verifying that our context meets each of the underpinning assumptions of the test of \cite{Bei2024}, and can be of independent interest to researchers who want to apply this test in different contexts.

\subsection{Instrumental functions} \label{sec: Instrumental functions}
We elaborate on the family $\mathcal{G}$ of instrumental functions. First, we will consider a class of instrumental functions for a discrete covariate. Next, we discuss a class for a continuous covariate. An additional class of instrumental functions is discussed in Supplementary material \ref{supp: Instrumental functions}. By Assumption \ref{Assumption covariate space}, we consider covariates that take values in $[0, 1]$ after scaling. Instrumental functions for vectors of covariates can be obtained by multiplying the instrumental functions for each of its components:
\begin{equation} \label{eq: standard instrumental function}
	g_{j_0, j_1, \dots, j_d}: [0, 1]^{d+1} \to \mathbb{R}_+: (x_0, x_1, \dots, x_d) \mapsto g_{j_0, j_1, \dots, j_d}(x_0, x_1, \dots, x_d) = \prod_{k' = 0}^d g_{j_{k'}}(x_{k'}).
\end{equation}
Remark that $x_0 \equiv 1$ and hence we can simply set $g_{j_0}(\cdot) = 1$.

\textbf{Indicator functions.} The most natural class of instrumental functions for a discrete covariate $X_{k'}$ with $l$ levels -- say, $a_1, \dots, a_l$ -- are the indicator functions,
\begin{equation*}
	\mathcal{G}_\text{disc} = \{g_j: x_{k'} \mapsto g_j(x_{k'}) = \mathbbm{1}(x_{k'} = a_j), j = 1, \dots, l\}.
\end{equation*} When the discrete covariate is not ordinal, one usually encodes it with $l - 1$ dummy variables when including it in the analysis. Denote these as $X_{k', 1}, \dots, X_{k', l - 1}$. In this case, one could use the class
\begin{multline*}
	\mathcal{G}_\text{cat} = \bigg\{g_j: (x_{k', 1}, \dots, x_{k', l - 1}) \mapsto g_j(x_{k', 1}, \dots, x_{k', l - 1}) =\\ \mathbbm{1}(x_{k', j} = 1)\prod_{1 \leq j' < l,  j' \neq j}\mathbbm{1}(x_{k', j'} = 0),
	 j = 1, \dots, l\bigg\},
\end{multline*}
with $x_{k', l} \equiv 1$. Simply viewing each dummy variable as a separate binary covariate and considering $\mathcal{G}_\text{disc}$ for each will not work, since it can be seen that this will lead to instrumental functions of the form \eqref{eq: standard instrumental function} that will always be zero (which violates Assumption \ref{Assumption variance of moments}). Indeed, consider for example the case in which there is one discrete covariate $X_1$ that has four levels, which we encode using three binary variables $(X_{1, 1}, X_{1, 2}, X_{1, 3})$. Considering $\mathcal{G}_\text{disc}$ for each and combining them via Equation \eqref{eq: standard instrumental function}, we would obtain a class $\mathcal{G}$ that contains the instrumental function $g(x_1) = \mathbbm{1}(x_{1, 1} = 1)\mathbbm{1}(x_{1, 2} = 0)\mathbbm{1}(x_{1, 3} = 1)$. However, due to the dummy variable encoding, $(X_{1, 1} = 1, X_{1, 2} = 0, X_{1, 3} = 1)$ will never occur, so $g$ in this case is always zero.

\textbf{Spline functions.} A commonly used class of instrumental functions for continuous covariates is the class of cubic B-splines. Let $\{p_j\}_{j = -2, \dots, l + 2}$ denote an increasing sequence of knot points, where $p_0 = 0$ and $p_l = 1$, and let $B_3(\cdot; p_{j})$ denote the cubic B-spline defined by the knot points $p_{j - 2}, p_{j - 1}, p_{j}, p_{j + 1}, p_{j + 2}$. Then we can construct the class of cubic B-splines
\begin{equation*}
	\mathcal{G}_\text{spline} = \{g_j: x \mapsto g_j(x) = B_3(x; p_{j})\}.
\end{equation*}

\begin{figure}
	\centering
	\includegraphics[width = 0.9\linewidth]{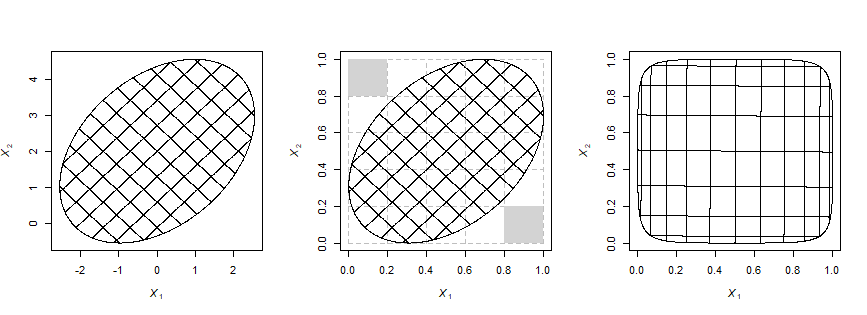}
	\caption{Left panel: the original covariate space. Middle panel: the transformed space using a component-wise min-max scaler. The gray, dashed grid depicts the supports of the considered instrumental functions. The shaded rectangles highlight the supports of two instrumental functions that fall entirely outside the transformed covariate space. Right panel: the transformed space using the PCA-based transformation.}
	\label{fig: transformation}
\end{figure}

\textbf{Instrumental functions on $\mathcal{X}$.} So far, we have defined all instrumental functions on $[0, 1]$, as by Assumption \ref{Assumption covariate space} $\mathcal{X}$ is bounded, and hence each component of $X$ can easily be rescaled to take values in the unit interval. We will denote this rescaling transformation as $N: \mathcal{X} \to [0, 1]^{d+1}$. In the simplest case, we can consider $N$ to be the component-wise min-max scaler (e.g. as proposed by \cite{AndrewsShi2013}), that is,
\begin{multline*}
	N: \mathcal{X} \to [0, 1]^{d+1}: (x_0, x_1, \dots, x_d) \mapsto (x_0, N_1(x_1), \dots, N_d(x_d)), \\ \text{with} \quad N_j(x_j) = \frac{x_j + M_j}{2M_j}, \forall j \in \{1, \dots, d\},
\end{multline*}
where $M_j \in \mathbb{R}$ are such that $\mathcal{X} \subset \prod_{j = 1}^{d+1} [-M_j, M_j]$. One problem, however, that could present itself is that $N$ is not \emph{sufficiently surjective}, in the sense that there exists an instrumental function of the form \eqref{eq: standard instrumental function} with support that falls entirely outside of $N(\mathcal{X})$. An example of this is given in the middle panel of Figure \ref{fig: transformation}. In what follows, we present a different rescaling transformation that is designed to overcome this issue.

Let $\mathcal{X}_\text{disc}$ and $\mathcal{X}_\text{cont}$ denote the subspaces of $\mathcal{X}$ of discrete and continuous covariates respectively and define their dimensions as $d_\text{disc}$ and $d_\text{cont}$. We will split the construction of $N$ into its effect on the continuous covariates on the one hand, and discrete covariates on the other hand. Denote the transformation $N$ with its domain restricted to $\mathcal{X}_\text{cont}$ ($\mathcal{X}_\text{disc}$) as $N_\text{cont}$  ($N_\text{disc}$). For $N_\text{disc}$, any injective transformation from $\mathcal{X}_\text{disc} \to [0, 1]^{d_\text{disc}}$ -- such as the min-max scaler -- can be selected. For $N_\text{cont}$, we start by noting that the problem in the middle panel of Figure \ref{fig: transformation} occurs due to $X_1$ and $X_2$ being correlated. The main idea in constructing $N_\text{cont}$ will therefore be to first decorrelate the continuous covariates using principal component analysis, and then proceed by transforming the decorrelated variables to $[0, 1]^{d_\text{cont}}$ in a sufficiently surjective way.

Precisely, $N_\text{cont}$ consists of several steps, which are outlined below. An illustration of the results of the transformation is given in the right panel of Figure \ref{fig: transformation}. Without loss of generality, suppose the first $d_\text{cont}$ elements of $\tilde{X}$ correspond to the continuous covariates. Furthermore, denote this subvector as $X^\text{cont}$.
\begin{enumerate}[noitemsep]
	\item Perform a principle component analysis (PCA) of the continuous covariates $\{x_i^\text{cont}\}_{i = 1}^n$. Each observation is transformed to its score $s_i$ on the principal components.
	\item Scale these scores elementwise, so that $\max_i s_{ij} - \min_i s_{ij} = 2$, for all $j \in \{1, \dots, d_\text{cont}\}$. Shift the scaled scores to $[-1, 1]^{d_\text{cont}}$, and denote the results as $s_i^*$.
	\item Typically, the points $s_i^*$ will be concentrated inside the inscribing sphere of $[-1, 1]^{d_\text{cont}}$, leading to a lower concentration of points near the corners of $[-1, 1]^{d_\text{cont}}$. To resolve this issue, we apply a function that has the effect of spreading out points in the unit sphere more evenly inside $[-1, 1]^{d_\text{cont}}$. In our implementation, we use $\sin(0.5\pi x)$ component-wise, but other options like $\text{arctan}(bx) / \text{arctan}(b)$ for $b > 0$ or the ones discussed in \cite{AndrewsShi2013} would also be possible.
	\item Finally, the points resulting from the previous step are scaled and shifted to $[0, 1]^{d_\text{cont}}$.
\end{enumerate}
Supplementary material \ref{supp: Instrumental functions} provides several remarks on handling dependent categorical covariates and on the dimensionality of $\mathcal{G}$.

\subsection{Time-independent effects of covariates} \label{sec: Time-independent effects of covariates}
In practice, it is often assumed that covariate effects do not change over time, as is for example the case in the traditional Cox proportional hazards model. It is therefore of great interest to study the proposed methodology under this additional information and, more specifically, to investigate if narrower bounds on the covariate effects can be obtained.

A straightforward modification of the methodology to a setting that includes the assumption of a time-independent covariate effect $\beta_k$ can be made by simply estimating the identified intervals of $\beta_k$ at multiple time points and later on combining all estimated intervals by means of intersection or majority vote. Specifically, suppose $A$ identified intervals $\hat{\mathcal{B}}_{I, k}^{1}, \dots, \hat{\mathcal{B}}_{I, k}^{A}$ are estimated at different time points $t_1, \dots, t_A$. If it is assumed that $\beta_k$ is independent of time, each of these identified intervals pertains to precisely the same quantity, and hence it is sensible to combine them via intersection:
\begin{equation} \label{eq: intersection rule}
	\hat{\mathcal{B}}_{I, k} = \bigcap_{a = 1}^A \hat{\mathcal{B}}_{I, k}^{a}.
\end{equation}
To obtain a correct level $\alpha$ of $\hat{\mathcal{B}}_{I, k}$, it is required that each separate identified interval is estimated with Bonferroni corrected confidence level $1 - \alpha/A$. When many time points are selected -- i.e., when $A$ is large -- the level of each identified interval will be very small, so that the identified intervals $\hat{\mathcal{B}}_{I, k}^{a}$ might become too wide. In this case, one might use the majority voting rule with threshold $1/2$ as described in \cite{Gasparin2024}:
\begin{equation} \label{eq: majority voting rule}
	\hat{\mathcal{B}}_{I, k} = \left\{\beta_k \in \mathcal{B}_k \mid \frac{1}{A} \sum_{a = 1}^A \mathbbm{1}(\beta_k \in \hat{\mathcal{B}}_{I, k}^{a}) > \frac{1}{2}\right\},
\end{equation}
where each identified interval $\mathcal{B}_{I, k}^{a}$ should be estimated only at a confidence level $1 - \alpha/2$.

Remark that there is a trade-off to be made when using these combination strategies. In the case of combination by Expression \eqref{eq: intersection rule}, we have on the one hand that $\hat{\mathcal{B}}_{I, k}$ can only decrease in size by adding more intervals $\hat{\mathcal{B}}_{I, k}^{a}$ to the intersection, making it seem desirable to consider as many additional time points as possible. On the other hand, in doing so, the estimation of each interval $\hat{\mathcal{B}}_{I, k}^{a}$ will have to be done at a larger confidence level, leading to wider intervals. Therefore, one could wonder about an appropriate value for $A$, as well as an appropriate set of time points $t_1, \dots, t_A$. In view of keeping the required computations manageable, we recommend to let $A \in \{3, 4, 5\}$. The selected time points $t_1, \dots, t_A$ can be spread over the support of $Y$, though we note that earlier time points are often more informative, since Peterson bounds widen as the considered time point $t_a$ increases, and hence the identified intervals $\hat{\mathcal{B}}_{I, k}^{a}$ tend to do so as well. An analysis of both combination approaches as well as the effect of the selected time points $t_1, \dots, t_A$ is presented in Supplementary material \ref{supp: Comparison of combination methods}.

Lastly, we note that the approach as described above only assumes that $\beta_k$ is independent of time. Supplementary material \ref{supp: Alternative approach to time-independent coefficients} describes a more direct approach that can take into account that all coefficients are time-independent. It works by augmenting the unconditional moment conditions in \eqref{eq: unconditional moment inequalities} to immediately include all considered time points $\{t_1, \dots, t_A\}$ and constructing a test for this augmented set of moment conditions. Such an approach, however, was found to be extremely computationally intensive, and it did not lead to worthwhile improvements.

\subsection{Discussion} \label{sec: Discussion}
Below we elaborate on several more aspects of the model. Going forward, we will refer to $\mathcal{B}_{I, k}$ and its estimator $\hat{\mathcal{B}}_{I, k}$ as the (estimated) \emph{identified interval} throughout the rest of this paper, in order to distinguish it from the identified set $\mathcal{B}_I$ in the full parameter space. While this is a slight abuse of terminology -- one can construct pathological examples where $\mathcal{B}_{I, k}$ is not an interval, cf. Example \ref{example: nonconvex identified interval} in Supplementary material \ref{supp: Examples} -- it will be valid in most practical cases. First, we discuss how the estimated identified interval should be interpreted. Next, we provide a specification test for the model. Lastly, we highlight several caveats. Supplementary material~\ref{supp: future research directions} discusses possible avenues for future research.

\textbf{Coverage.} In the partial identification literature there are two common coverage properties of the estimated identified interval (or, more generally, set). The first one states that $\hat{\mathcal{B}}_{I, k}$ covers each point in the true identified interval, $\mathcal{B}_{I, k}$, with specified confidence level at least $1 - \alpha$. The second one states that $\hat{\mathcal{B}}_{I, k}$ covers $\mathcal{B}_{I, k}$ entirely with specified confidence level at least $1 - \alpha$. It can be seen that by nature of the test inversion approach, our estimator satisfies the first coverage property. Since the true value $\beta_{\text{true}, k}$ is always an element of $\mathcal{B}_{I, k}$, our estimated identified interval covers $\beta_{\text{true}, k}$ with specified probability at least $1 - \alpha$ and in this sense can be interpreted as a regular confidence interval. For a more comprehensive discussion on identified sets, we refer to \cite{CanayShaikh2017}.

\textbf{Misspecification test.} Even though model \eqref{eq: model} with its underpinning assumptions \ref{Assumption iid}-\ref{Assumption sufficient slackness} is flexible, it could still be misspecified. Handily, as is the case in many partial identification models, one can immediately derive a specification test from $\hat{\mathcal{B}}_{I, k}$. Indeed, the model can be rejected when $\hat{\mathcal{B}}_{I, k} = \emptyset$. \cite{Bugni2015} study several specification tests for partially identified models. They refer to this type of test as the ``by-product'' test (test BP). In their paper, they provide alternative specification tests which outperform test BP in certain situations. When power is of the essence, we therefore recommend to use such an alternative. Nevertheless, test BP is still a useful front line tool.

\textbf{Pitfalls.} Lastly, we caution against two pitfalls of the method. A first pitfall is to include too many instrumental functions in the class $\mathcal{G}$. At first, this may seem counter-intuitive. Indeed, including more instrumental functions will lead to more information being transferred when transforming the conditional to the unconditional moments. Hence, $\mathcal{B}_{I, k}$ can only be made smaller in this way and it seems reasonable to assume that this effect will also be visible in the estimator $\hat{\mathcal{B}}_{I, k}$. While this is true up to some point, there is a balance to be struck: including more instrumental functions will result in more unconditional moments to test, increasing the variance of the test statistic, which in turn has the effect of making $\hat{\mathcal{B}}_{I, k}$ larger. It is therefore recommended to try out several choices for $\mathcal{G}$ when applying this method. Secondly, when many covariates are included in the analysis, we recommend that they be standardized so as to keep the size of their linear combinations manageable, since they are the argument to an exponential function.

\section{Simulation study} \label{sec: Simulation study}

To assess the performance of the proposed methodology, a simulation study is carried out. Throughout, we consider one continuous covariate $X_1 \sim \mathcal{N}(0, 1)$ and one binary covariate $X_2 \sim Ber(0.5)$, which are independent from each other. The parameter vector is specified as $\beta_\text{true}(t) = (\log(t), 1, -1)$. After generating pairs of covariates $(x_i)_{i = 1}^n = (1, x_{i1}, x_{i2})_{i = 1}^n$ according to their respective distributions, we simulate latent times from the conditional joint distribution $F_{T, C|X}(t, c|x) = \mathcal{C}_\theta(F_{T|X}(t|x), F_C(c))$. In this formulation, $F_{T|X}(\cdot|x)$ takes the form as in model \eqref{eq: model}, for which we will consider $\Lambda(\cdot) = 1 - \exp(-\exp(\cdot))$, referred to as the Cox link function. The censoring distribution $F_C$ is taken to be an exponential distribution with parameter $\lambda \in (0, 2)$, which is adapted in each design in order to control the percentage of censored observations. The copula $\mathcal{C}_\theta$ is taken to be a Frank copula where we let $\theta$ equal $6$, $0$ or $-6$, leading to Kendall's tau equal to $0.51$, $0$ or $-0.51$, respectively. Latent times are then simulated by first generating pairs $(u_{i1}, u_{i2})_{i = 1}^n$ of possibly dependent uniform variables $(U_1, U_2)$ according to the copula $\mathcal{C}_\theta$ and then applying the probability integral transform to obtain realizations $(t_i, c_i)_{i = 1}^n$ of $(T, C)$ as $t_i = F^{-1}_{T|X}(u_{i1}|x_i)$ and $c_i = F^{-1}_C(u_{i2})$. As usual, we construct $Y = \min(T, C)$ and $\Delta = \mathbbm{1}(Y = T)$. Inference will be done for $\beta_1$, at the selected time point of interest $t = 1$. Simulation analyses using $\Lambda(\cdot) = 1 - (1 + \exp(\cdot))^{-1}$, referred to as the AFT link function, are presented in Supplementary material \ref{supp: Simulations}, though in general, they lead to very similar conclusions.

The parameter space is defined as $\mathcal{B} = [-10, 10]^3$. The class of instrumental functions used is constructed by taking cubic B-splines for the continuous covariate and indicators for the categorical (binary) covariate, and combining them via Equation \eqref{eq: standard instrumental function}. Furthermore, the test by \cite{Bei2024} requires the specification of several hyperparameters. Where possible, we will use the values proposed in their paper. One notable hyperparameter that cannot be specified in this way relates to Assumption \ref{Assumption variance of moments} and defines the lower bound of the empirical variances. We set its value to $10^{-6}$. Preliminary simulations showed that smaller values could lead to overrejection. Other notable design choices are the number of bootstrap samples used to compute the critical values, $B = 600$, and the nominal level of the test, $\alpha = 0.05$. Finally, each simulation design is run with $500$ repetitions.

The results of the simulation can be found in Table \ref{tab: main Cox}. For each considered design, this table lists (\emph{Bounds}) the average lower and upper bound of the estimated identified intervals, (\emph{Var}) the variance of the width of the bounds, (\emph{Sig}) the percentage of repetitions in which zero was not contained in the bounds and (\emph{Cov}) the percentage of repetitions in which the true value, $\beta_{\text{true}, 1} = 1$, was contained in the computed bounds. Remark that all quantities are necessarily computed only based on the repetitions that did not conclude model misspecification (i.e. empty estimated identified interval). The model was incorrectly determined to be misspecified in no more than $5$ out of the $500$ repetitions, throughout all considered designs. The quantity \emph{Var} is interesting as it provides information on the stability of the method. \emph{Sig} quantifies the proportion of repetitions for which the derived bounds are informative, in the sense that the covariate effect is concluded to be different from zero (i.e. significant). Lastly, \emph{Cov} is equal to the proportion of repetitions in which the true value was contained within the estimated interval and, as such, provides information on the type-I error of the method. \emph{Cov} is equal to one in almost all designs. This is not indicative of the method being conservative, since the nominal level $\alpha$ holds in particular at the bounds of the identified interval, while the actual type-I error of the test on interior points of the interval will be much smaller.

From Table \ref{tab: main Cox} it can be seen that the bounds narrow and the method becomes more stable as the sample size increases. Interestingly, it can be seen that for the simulation designs considered here, an increase in the number of instrumental functions $(N_{IF})$ used -- specifically, using $8$ splines for $X_1$ instead of $6$ -- does not substantially narrow the bounds. This indicates that using a moderate amount of instrumental functions will already transfer most of the information from the conditional to the unconditional moments. It can also be seen that positive dependence between $T$ and $C$ leads to the narrowest bounds, while designs with negative dependence are substantially more difficult. This could be explained by remarking that it is easier to observe the full support of both $T$ and $C$ when they are positively dependent, while it will be difficult to observe the upper parts of their support in the negative dependence case. In this sense, data generated under positive dependence is more informative, which is reflected in the bounds. In the same line, bounds under heavy censoring $(65\%)$ are considerably wider than bounds under light censoring $(30\%)$. However, Table \ref{tab: main Cox} shows that even in cases with heavy censoring, bounds can still be informative. Supplementary material \ref{supp: True bounds} provides an analysis of the true identified interval of $\beta_{\text{true}, 1}$ for each design. We find that the true bounds are typically not substantially smaller than the estimated ones.

\begin{table}[ht]
	\centering
	\begin{tabular}{ccccccccccc}
		\toprule
		& & \multicolumn{4}{c}{$N_{IF} = 12$} & \phantom{\hspace{1pt}} & \multicolumn{4}{c}{$N_{IF} = 16$}\\
		\cmidrule{3-6} \cmidrule{8-11}
		Setting & $n$ & Bounds & Var & Sig & Cov & & Bounds & Var & Sig & Cov\\
		\midrule
		\multirow{3}{*}{\shortstack[c]{Indep.\\ $\sim 30\%$ cens.}} & 500 & [0.47, 1.56] & 0.03 & 1.00 & 1.00 && [0.50, 1.55] & 0.03 & 1.00 & 1.00\\
		& 1000 & [0.49, 1.39] & 0.01 & 1.00 & 1.00 && [0.52, 1.37] & 0.01 & 1.00 & 1.00\\
		& 2000 & [0.51, 1.29] & 0.01 & 1.00 & 1.00 && [0.53, 1.27] & 0.00 & 1.00 & 1.00\\
		\midrule
		\multirow{3}{*}{\shortstack[c]{Pos. dep.\\ $\sim 30\%$ cens.}} & 500 & [0.60, 1.69] & 0.03 & 1.00 & 1.00 && [0.66, 1.68] & 0.04 & 1.00 & 0.99\\
		& 1000 & [0.63, 1.50] & 0.01 & 1.00 & 1.00 && [0.68, 1.50] & 0.01 & 1.00 & 0.99\\
		& 2000 & [0.66, 1.40] & 0.01 & 1.00 & 1.00 && [0.70, 1.38] & 0.01 & 1.00 & 1.00\\
		\midrule
		\multirow{3}{*}{\shortstack[c]{Neg. dep.\\ $\sim 30\%$ cens.}} & 500 & [0.46, 1.64] & 0.04 & 1.00 & 1.00 && [0.49, 1.63] & 0.04 & 1.00 & 1.00\\
		& 1000 & [0.48, 1.46] & 0.01 & 1.00 & 1.00 && [0.50, 1.44] & 0.01 & 1.00 & 1.00\\
		& 2000 & [0.50, 1.35] & 0.01 & 1.00 & 1.00 && [0.51, 1.33] & 0.01 & 1.00 & 1.00\\
		\midrule
		\multirow{3}{*}{\shortstack[c]{Indep.\\ $\sim 65\%$ cens.}} & 500 & [-0.19,  3.69] &  2.61 &  0.04 &  1.00 && [-0.16,  2.99] &  0.57 &  0.09 &  1.00\\
		& 1000 & [-0.10,  2.46] &  0.12 &  0.11 &  1.00 && [-0.08,  2.23] &  0.07 &  0.14 &  1.00\\
		& 2000 & [-0.05,  2.14] &  0.07 &  0.20 &  1.00 && [-0.03,  1.90] &  0.03 &  0.28 &  1.00\\
		\midrule
		\multirow{3}{*}{\shortstack[c]{Pos. dep.\\ $\sim 65\%$ cens.}} & 500 & [0.19, 3.04] & 0.61 & 0.96 & 1.00 && [0.22, 2.75] & 0.20 & 0.97 & 1.00\\
		& 1000 & [0.23, 2.39] & 0.08 & 1.00 & 1.00 && [0.27, 2.25] & 0.06 & 1.00 & 1.00\\
		& 2000 & [0.26, 2.11] & 0.04 & 1.00 & 1.00 && [0.29, 2.00] & 0.02 & 1.00 & 1.00\\
		\midrule
		\multirow{3}{*}{\shortstack[c]{Neg. dep.\\ $\sim 65\%$ cens.}} & 500 & [-0.50,  9.76] &  0.76 &  0.00 &  1.00 && [-0.45,  9.36] &  2.21 &  0.00 &  1.00\\
		& 1000 & [-0.32,  7.52] &  5.18 &  0.00 &  1.00 && [-0.31,  5.45] &  4.69 &  0.00 &  1.00\\
		& 2000 & [-0.25,  5.00] &  3.46 &  0.00 &  1.00 && [-0.22,  3.35] &  0.41 &  0.00 &  1.00\\
		\bottomrule
	\end{tabular}
	\caption{Results of the main simulation using the Cox link function.}
	\label{tab: main Cox}
\end{table}

\subsection{Summary of additional simulations} \label{sec: Summary of additional simulations}
More simulations were carried out to study the performance of the proposed methodology under various settings. In this section, we will restrict to summarizing the conclusions we can draw from them. We note that some simulations were performed under modified settings with respect to the ones used above in order to ease computational burden. Detailed information on these simulations as well as the tables containing their results is deferred to Supplementary material \ref{supp: Simulations}.

\textbf{More instrumental functions.} In a first additional analysis, we investigate the effect of further increasing the number of instrumental functions used in transforming the conditional moment inequalities to unconditional ones. We find that using $10$ B-spline functions for the continuous covariate (hence $N_{IF} = 20$) can still help to decrease the width of the bounds slightly, while a further increase to $15$ B-spline functions can have adverse effects. This is in line with the remark made at the end of Section \ref{sec: Discussion}, cautioning against taking the class $\mathcal{G}$ too large. We also note that including more instrumental functions in the analysis will increase computational burden, as well as increase the risk of not satisfying Assumption \ref{Assumption variance of moments}. Higher numbers of instrumental functions should therefore be motivated by the availability of sufficient sample size and computing power. In light of these results, we suggest to use around $5-10$ instrumental functions per continuous covariate, and to use the indicator family of instrumental functions for the categorical covariates, irrespective of the link function used.

\textbf{Almost no censoring.} In a second analysis, we consider a setting in which there is only $2\%$ censoring. As a consequence, this setting is close to one without censoring, in which case the parameters in model \eqref{eq: model} are identified. In line with this observation, we see that the estimated identified intervals are very narrow in all cases. For small sample sizes, the coverage can be below its nominal value, though this problem disappears as the sample size increases.

\textbf{Time-independent effects of covariates.} Lastly, we investigate the modification of our approach to time-independent effects of covariates, proposed in Section \ref{sec: Time-independent effects of covariates}. We find that both the intersection and majority vote method can improve upon the estimation algorithm solely applied to a single point, though we emphasize that this improvement comes at the cost of having to assume that $\beta_{\text{true}, 1}$ is time-independent. The intersection method outperforms majority vote when the number of points in the considered grid of time points is small, whereas this relation reverses when the number of grid points increases, as indicated by the additional analysis based on the data application (Supplementary material \ref{supp: Comparison of combination methods}).

\section{Data applications} \label{sec: Data applications}

This section applies the developed methodology to two data sets. Throughout, the class of instrumental functions is constructed by using five spline functions for each continuous covariate, using indicator functions for the categorical covariates and combining them via Equation \eqref{eq: standard instrumental function}. Analogous to the simulation study, the lower and upper bounds for the regression parameters are set at $-10$ and $10$, respectively. All results correspond to a $95\%$ confidence level.

\subsection{Pancreas cancer data}
First, we consider a data set on pancreatic cancer retrieved from the Surveillance, Epidemiology and End Results (SEER) data base. In the construction of this data set, patients who were diagnosed with pancreatic cancer were followed up and their death time or possible censoring time was recorded. Moreover, for each patient in the study, demographic variables as well as variables pertaining to their disease were obtained. These data have previously been analyzed by \cite{Czado2023}, who focus on the subpopulation of black patients. For this data application, we will narrow the target group further by only considering black males. In this way, we obtain a data set of $\num{4490}$ observations.

We will model the time until a patient succumbs to their disease $(T)$ and view all other events as censoring. One possible event that is therefore viewed as censoring occurs when a patient receives a transplant. Since only those patients who are in the worst medical condition will be eligible for this treatment, the time until a patient undergoes transplantation and the time until they would have otherwise died is likely positively related. By extension, the validity of an independence assumption between $T$ and $C$ seems questionable.

We consider model \eqref{eq: model} using the Cox link function as discussed in Section \ref{sec: The model} and provide three separate analyses based on the stage of the cancer, which can either be local $(n = \num{476})$, regional $(n = \num{1404})$, or distant $(n = \num{2610})$. Each analysis further considers two continuous covariates, namely the size of the tumor $(X_1)$ and the age of the patient at diagnosis $(X_2)$. Both were standardized and had outliers -- defined as values outside the interval between the $0.025$-th and $0.975$-th quantile -- and missing values removed prior to the analysis. Interest will be in the survival at either $6$, $12$ and $18$ months after the initial diagnosis. We also consider the case in which covariate effects are assumed to be time-independent by combining identified intervals at each considered time point, estimated at a Bonferroni corrected level, by means of intersection.

\begin{figure}[ht]
	\centering
	\includegraphics[width = \linewidth]{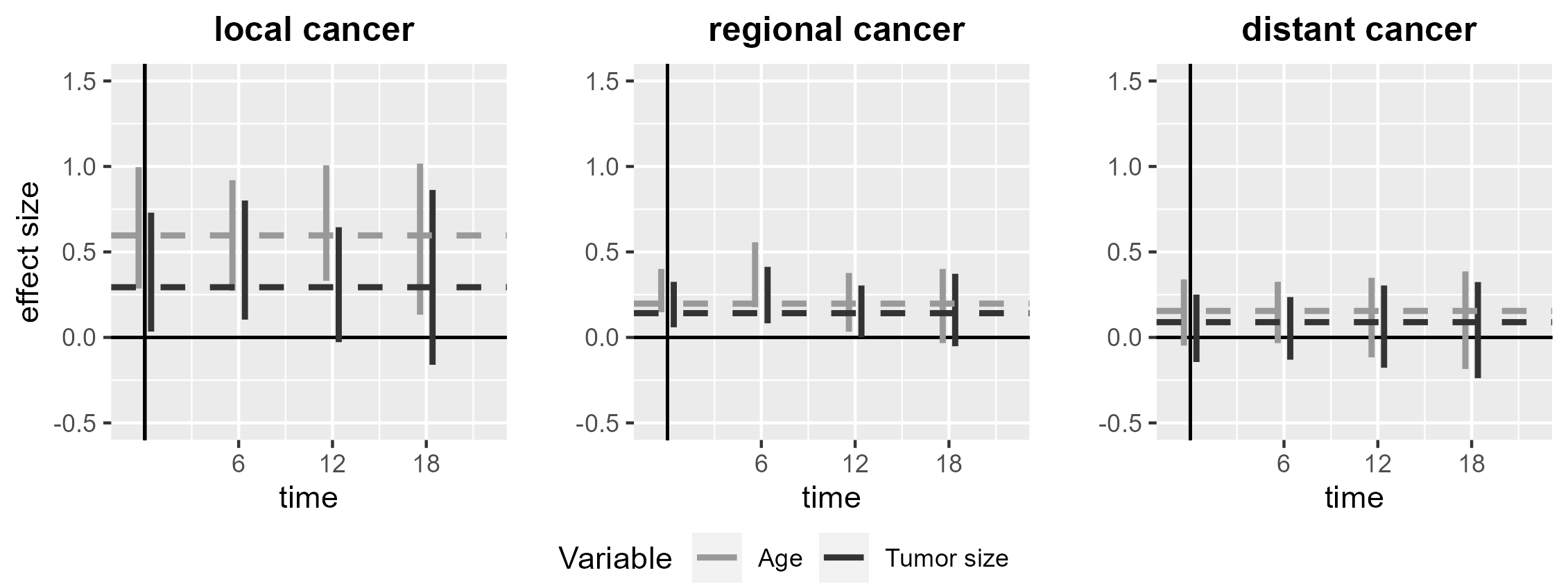}
	\caption{The results of the model using a Cox link function applied to the SEER Pancreas data set. The vertical bars indicate the estimated identified intervals at each time point of interest, while the dashed horizontal lines represent the point estimates of a Cox model assuming independence. The vertical bars at $t = 0$ pertain to the model assuming time-independent regression parameters.}
	\label{fig: Results Pancreas Cox}
\end{figure}

The results are shown in Figure \ref{fig: Results Pancreas Cox}. From the left panel, it can be seen that when the cancer is local, the identified intervals for the effects of $X_1$ and $X_2$ are entirely comprised of positive values when $t = 6$. As such, for this time point, even though we cannot be certain of the exact covariate effects on the probability of dying before $6$ months, we can be confident that the effect sizes are greater than zero. Furthermore, we can see that the effect of age remains significant also at the later time points, while the identified interval for the effect of the size of the tumor contains zero when $t = 12$ or $t = 18$. If one would be willing to assume that covariate effects do not change over time, the identified interval does become significant, albeit only barely so. The plots containing the results of the analyses for regional and distant cancer can be interpreted similarly. For reference, the results of a classical Cox model assuming independence are overlaid.

From a similar analysis using the AFT link function (delegated to Supplementary material \ref{supp: Pancreas data application}), similar conclusions may be drawn.

\subsection{NLSY data}
Next, we study the National Longitudinal Surveys Year 1979 (NLSY79) data, which is made available by the U.S. Bureau of Labour Statistics. Specifically, we are interested in unemployment spells, i.e. the time until people who enter the job market find a job. Censoring of these spells occurs at the end of the considered time period, or when subjects stop responding to the surveys. This latter cause may lead to a dependence between $T$ and $C$, as subjects who either found a job or have lost motivation to look for one might be more inclined to discontinue their participation in the study. This point is further corroborated by \cite{Frandsen2019}. Analogous to  the aforementioned paper, we obtain data sets of a workable size by narrowing our focus to the period $1989-1993$. We further divide this subset into different strata and run the method on each of them. Specifically, six strata are constructed based on the combinations of sex (male or female) and race (hispanic, black or other). Two variables are considered: a binary variable indicating whether an unemployed subject is highly educated (defined as having obtained a high school diploma before or during the unemployment spell), and a continuous covariate representing the age of the subject at the start of their unemployment spell. For the former, missing values were imputed based on the nearest-in-time observed values. The latter was standardized prior to the analysis. Interest is in the covariate effects on the probability of having an unemployment spell longer than $6$ months. The results of the analysis are shown in Table~\ref{tab: Results NLSY}.

For all strata pertaining to females, the identified intervals for the effect of education on the unemployment spell are strictly positive. We observe the same for black males. Hence for these subgroups in the population we conclude that having obtained a high school diploma significantly decreases the probability of being unemployed for longer than $6$ months. Notably, the model using the Cox link function was determined to be misspecified in the stratum ``Male--Other'' due to an empty identified interval.

\begin{table}
	\centering
	\begin{tabular}{rccccc}
		\toprule
		& \multicolumn{2}{c}{AFT link} & \phantom{a} & \multicolumn{2}{c}{Cox link}\\
		\cmidrule{2-3} \cmidrule{5-6}
		Stratum & Age & education && Age & Education\\
		\midrule
		Male--Hispanic & [-0.48, 0.09] & [-0.08, 0.88] && [-0.27, 0.04] & [-0.05, 0.42]\\
		Male--Black & [-0.32, 0.05] & [0.13, 0.74] && [-0.19, 0.01] & [0.07, 0.42]\\
		Male--Other & [-0.39, 0.24] & [-0.15, 0.58] && / & /\\
		Female--Hispanic & [-0.22, 0.34] & [0.34, 0.98] && [-0.14, 0.18] & [0.21, 0.59]\\
		Female--Black & [-0.19, 0.24] & [0.39, 0.91] && [-0.12, 0.15] & [0.24, 0.55]\\
		Female--Other & [-0.16, 0.36] & [0.01, 0.67] && [-0.09, 0.19] & [0.01, 0.37]\\
		\bottomrule
	\end{tabular}
	\caption{Results pertaining to the NLSY data application}
	\label{tab: Results NLSY}
\end{table}


\subsection*{Acknowledgements}
The computational resources and services used in this work were provided by the VSC (Flemish Supercomputer Center), funded by the Research Foundation Flanders (FWO) and the Flemish Government department EWI. Lastly, the authors thank Gilles Crommen and several attentive attendees at conferences where this research was presented for their useful advice and comments.

\subsection*{Disclosure statement}
The authors report that there are no competing interests to declare.


\bibliography{bibliographyArXiv}

\end{document}



\newcommand{\blind}{1}

\def\spacingset#1{\renewcommand{\baselinestretch}%
	{#1}\small\normalsize} \spacingset{1}

\if1\blind
{
	\title{\bf Supplement to ``Bounds for the regression parameters in dependently censored survival models''}
	\author{Ilias Willems\thanks{Corresponding author (ilias.willems@kuleuven.be).}\\
		ORSTAT, KU Leuven, Belgium\\
		and \\
		Jad Beyhum\thanks{Jad Beyhum gratefully acknowledges financial support from the Research Fund KU Leuven through the grant STG/23/014.} \\
		Department of Economics, KU Leuven, Belgium\\
		and\\
		Ingrid Van Keilegom\thanks{Ingrid Van Keilegom gratefully acknowledges funding from the FWO and F.R.S. - FNRS (Excellence of Science programme, project ASTeRISK, grant no. 40007517), and financial support from the FWO (senior research projects fundamental research, grant no. G047524N).}\\
		ORSTAT, KU Leuven, Belgium}
	\maketitle
} \fi

\if0\blind
{
	\bigskip
	\bigskip
	\bigskip
	\begin{center}
		{\LARGE\bf Supplement to ``Bounds for the regression parameters in dependently censored survival models''}
	\end{center}
	\medskip
} \fi


\newpage
\tableofcontents


\appendix

\setcounter{theorem}{1}
\setcounter{figure}{2}
\setcounter{table}{2}

\renewcommand\theequation{S\arabic{equation}}

\section{Technical assumptions} \label{supp: Technical assumptions}

\subsection{Details of Assumption \ref{Assumption boundary BI}}
To state Assumption \ref{Assumption boundary BI}, let $m_{j, 1}(W, \beta) = \left(\mathbbm{1}(Y \leq t) - \Lambda(X^\top \beta)\right)g_j(X)$ and $m_{j, 2}(W, \beta) = \left(\Lambda(X^\top \beta) -  \mathbbm{1}(Y \leq t, \Delta = 1)\right)g_j(X)$. For ease of notation, we will view these doubly indexed variables as elements of a $(2J)$-dimensional vector $m(W, \beta)$, of which the first $J$ elements correspond to the indices $(j, 1)$ and the next $J$ elements correspond to indices $(j, 2)$. Denote $\Sigma(\beta) \in \mathbb{R}^{2J \times 2J}$ as the covariance matrix of $m(W, \beta)$, and let $\bar{m}(\beta)$ and $\hat{\Sigma}(\beta)$ be the usual sample analogue estimators of $\mathbb{E}[m(W, \beta)]$ and $\Sigma(\beta)$, respectively. Furthermore, let $\sigma^2(\beta)$ and $\hat{\sigma}^2(\beta)$ be the diagonals of $\Sigma(\beta)$ and $\hat{\Sigma}(\beta)$. Define $(x)_- = \max(-x, 0)$ and $(x)_-^2 = ((x)_-)^2$, and let $S(\sqrt{n}m, \Sigma) = \sum_{q = 1}^2 \sum_{j = 1}^J \left(\sqrt{n}m_{j, q} / \sigma_{j, q}\right)^2_-$. The function $S$ plays an important role in constructing the test statistic of \cite{Bei2024}. Note that even though it only requires the elements on the diagonal of $\Sigma(\beta)$ and $\hat{\Sigma}(\beta)$ -- which are precisely $\sigma^2(\beta)$ and $\hat{\sigma}^2(\beta)$, respectively -- we will sometimes prefer to use $\Sigma(\beta)$ and $\hat{\Sigma}(\beta)$ in the notation in order to be consistent with existing literature, where other specifications for $S$ are current.

Denote $\partial^{(-k)} f(\beta) = \left(\frac{\partial}{\partial \beta_i} f(\beta)\right)_{i = 1, \dots, k - 1, k + 1, \dots, d}$, i.e. the vector of all partial derivatives of $f$ excluding the one with respect to $\beta_k$. Furthermore, define the orthogonal projection of $\beta \in \mathcal{B}(r)$ onto $\mathcal{B}_I(r) = \{\beta \in \mathcal{B}_I \mid \beta_k = r\}$ as $T^r\beta = \argmin_{\tilde{\beta} \in \mathcal{B}_I(r)} d(\beta, \tilde{\beta})$, with $d(\cdot, \cdot)$ denoting the usual Euclidean distance. For $c > 0$, define
\begin{equation} \label{eq: parameter set correct inference}
	\tilde{\mathcal{L}}_0(c) = \{r \in \mathcal{B}_{I, k} \hspace{0.1cm} | \hspace{0.1cm} \forall \beta \in \mathcal{B}(r)\setminus\mathcal{B}_I(r): \norm{\partial^{(-k)}S(\mathbb{E}[m(W, T^r\beta)], \Sigma(T^r\beta))} \geq c\}.
\end{equation}
Definition \eqref{eq: parameter set correct inference}, together with Assumption \ref{Assumption boundary BI} can be simplified to:
\begin{enumerate}[noitemsep, label=(A8')]
	\item $\exists c > 0: \forall \beta \in \mathcal{B}(\beta_{\text{true}, k})\setminus\mathcal{B}_I(\beta_{\text{true}, k}): \norm{\partial^{(-k)}S(\mathbb{E}[m(W, T^r\beta)], \Sigma(T^r\beta))} > c$. \label{Assumption boundary BI bis}
\end{enumerate}

\subsection{Details of Assumption \ref{Assumption sufficient slackness}}
Assumption \ref{Assumption sufficient slackness} is imposed in order to avoid certain technical difficulties in the theory of our method. The condition requires that for each element $r$ of $\tilde{\mathcal{L}}_0(c)$ there exists a corresponding value $\beta \in \mathcal{B}(r)$ for which each moment restriction is \emph{sufficiently slack} \citep{AndrewsSoares2010}. Since assumption \ref{Assumption boundary BI} already excludes regions close to the boundary of $\mathcal{B}_{I, k}$ from $\tilde{\mathcal{L}}_0(c)$ (which is typically where moment restrictions are not slack), Assumption \ref{Assumption sufficient slackness} will in many cases be satisfied when Assumption \ref{Assumption boundary BI} is satisfied, though it is technically not implied by it.

\section{Testing procedure of Bei (2024)} \label{supp: Testing procedure of Bei}
The hypothesis to be tested in the estimation procedure of $\mathcal{B}_{I, k}$ (cf. Section \ref{sec: Methodology}) is given in Equation \eqref{eq: hypothesis}, and repeated below for ease of reading:
\begin{equation*}
	\mathcal{H}_0(r): \exists \beta \in \mathcal{B}: \beta_k = r \text{ and } \beta \text{ satisfies \eqref{eq: unconditional moment inequalities}}.
\end{equation*}
In what follows, we give a brief overview of the testing procedure of \cite{Bei2024}. We refer to their paper for a more detailed description.

We start by making the test statistic and critical value introduced in Section \ref{sec: Testing procedure} more concrete. To do so, we recall the notation introduced in Section \ref{supp: Technical assumptions}, and further define $D(\beta) = \sigma^2(\beta)\mathbb{I}_{2J}$, where $\mathbb{I}_{2J}$ represents a $(2J \times 2J)$-dimensional identity matrix, and let $\hat{D}_n(\beta)$ be its empirical counterpart. 

The test statistic $T_n(r)$ is defined as
\begin{equation} \label{eq: test statistic}
	T_n(r) = \inf_{\beta \in \mathcal{B}(r)} S(\sqrt{n}\bar{m}(\beta), \hat{\Sigma}(\beta)),
\end{equation}
The function $S(\sqrt{n}\bar{m}(\beta), \hat{\Sigma}(\beta))$ can be viewed as an accumulative measure of how much each of the sample analogues to the unconditional moment restrictions violate the requirement of being positive for a given parameter vector $\beta$. $T_n(r)$ subsequently defines this measure for a single element of the parameter vector by considering the best possible one that has $r$ as its $k$-th entry.

Large values of $T_n(r)$ indicate that there likely does not exist a vector $\beta$ with $\beta_k = r$ for which the unconditional moment restrictions are satisfied. The critical value $\gamma_{n, 1 - \alpha}(r)$ of $T_n(r)$ makes precise the threshold value from which point we reject that $r \in \mathcal{B}_{I, k}$, and is obtained through a bootstrapping procedure that is modified in order to obtain precise and fast results.

Formally, the critical value $\gamma_{n, 1 - \alpha}(r)$ is defined as the $(1 - \alpha)$-th quantile of
\begin{equation} \label{eq: critical value distribution}
	J_n^{LL}(r) = \inf_{\beta_b \in \hat{\mathcal{B}}_{I, k}(r)} S(v_n(\beta_b) + \phi_n(\beta_b) + \hat{G}_n(\beta_b)\xi(\beta_b), \hat{\Omega}_n(\beta_b)).
\end{equation}
We explain each of the elements of \eqref{eq: critical value distribution} below.
\begin{itemize}
	\item The set $\hat{\mathcal{B}}_{I, k}(r)$ contains all values of $\beta$ for which the infimum in \eqref{eq: test statistic} is attained.
	\item The stochastic process $v_n(\cdot): \mathcal{B} \to \mathbb{R}^{2J}$ is defined as
	\begin{equation*}
		v_n(\beta_b) = n^{-1/2} \sum_{i = 1}^n \hat{D}_n^{-1/2}(\beta_b) (m(W_i, \beta_b) - \bar{m}_n(\beta_b))\zeta_i,
	\end{equation*}
    where $\zeta_i|\{W_i\}_{i = 1}^n \sim \mathcal{N}(0, 1)$ i.i.d.
	\item The function $\phi_n(\cdot): \mathcal{B} \to \mathbb{R}^{2J}$ denotes the generalized moment selection (GMS) function. Its aim is to determine for each of the $2J$ unconditional moment restrictions whether they are binding (less than or equal to zero) or slack (larger than zero). By setting the elements of $\phi_n(\beta_b)$ corresponding to slack moment inequalities equal to a positive number (proportional to \emph{how slack} the moment restrictions are), it can be seen from Equation \eqref{eq: test statistic} that these moment conditions will have a smaller contribution to $J_n^{LL}(r)$. This is a desirable property that can lead to an increase in power \citep{Canay2023}. For more information on the choice of GMS function we refer to Section 3.2 in \cite{Bei2024}. An in-depth study of them can be found in \cite{AndrewsSoares2010}.
	\item The matrix $\hat{G}_n(\beta_b)$ consistently estimates $G(\beta_b) = \nabla_\beta \left(D(\beta)^{-1/2}\mathbb{E}[m(W, \theta)]\right)\vert_{\beta = \beta_b}$. Moreover, $\hat{\Omega}_n(\beta_b)$ denotes the empirical correlation matrix and $\xi(\beta_b)$ is defined as
	\begin{equation*}
		\xi(\beta_b) = \argmin_{\xi \in \Xi_n^r(\beta_b)} \left(S(v_n(\beta_b) + \phi_n(\beta_b) + \hat{G}_n(\beta_b) \xi,  \hat{\Omega}_n(\beta_b)) + \frac{\lambda_n}{n} ||\xi||^2\right),
	\end{equation*}
	for $\Xi_n^r(\beta_b) = \sqrt{n}(\mathcal{B}(r)^{-\epsilon_n} - \beta_b) \cup \{0_{d+1}\}$, and where $\mathcal{B}(r)^{-\epsilon_n}$ denotes the subset of points in $\mathcal{B}(r)$ whose distance to the boundary of $\mathcal{B}(r)$ is at least $\epsilon_n (= \sqrt{\log(\log(n))/n})$ and $0_{d+1}$ represents a $(d+1)$-dimensional vector of zeros. The term $\hat{G}_n(\beta_b)\xi(\beta_b)$ stems from the fact that $v_n(\beta_b) + \phi_n(\beta_b) + \hat{G}_n(\beta_b)\xi(\beta_b)$ is a linearization of $D(\beta)^{-1/2} \sqrt{n} \bar{m}_n(\beta)$ around the minimizer of the test statistic $T_n(r)$ under $\mathcal{H}_0(r)$. Intuitively, $\xi(\beta_b)$ is the vector that points from this hypothetical minimizer to $\beta_b$. The penalty term in the expression for $\xi(\beta_b)$ keeps the bootstrap version of $\beta_b$ close to its corresponding quantity under $\mathcal{H}_0(r)$.
\end{itemize}
The linearization of the test statistic when bootstrapping its distribution allows for significant computational advantages over other existing methods (notably \cite{Kaido2019} and \cite{Bugni2017}) and is the main reason why we selected the method of \cite{Bei2024} over its competitors.

\section{Link function} \label{supp: Link function}
\textbf{Proportional odds model.} The choice $\Lambda(\cdot) = 1 - S_0(\exp(\cdot))$ will lead to a semiparametric accelerated failure time (AFT) model, where $S_0$ is a specified baseline survival function. One particular AFT model of interest is the proportional odds model, which can be represented in log-linear form as:
\begin{equation*}
	\log(T) = \mu - \tilde{x}^\top \tilde{\beta} + \sigma_\Psi \Psi,
\end{equation*}
where $\Psi$ follows a standard logistic distribution. A simple computation shows that this implies that $S_{T|X}(t|x) = (1 + \exp(x^\top \beta))^{-1}$, where $\beta = \sigma_\Psi^{-1}(\log(t) - \mu, \beta_1, \dots, \beta_d)$. Therefore, to obtain the proportional odds model, one should select $S_0(\cdot) = (1 + (\cdot))^{-1}$, in which case the covariate effects can be estimated up to a scale parameter $\sigma_\Psi^{-1}$. Analogous to the Cox link function, we emphasize again that because the model is only imposed at a fixed time $t$, it only takes the form of a proportional odds model but does not actually impose the proportional odds assumption. If appropriate, the proportional odds assumption can be imposed through the modification discussed in Section \ref{sec: Time-independent effects of covariates}. Throughout the rest of this paper, we will let $\sigma_\Psi = 1$ when referring to the model using this link function. Note that even in practical settings where $\sigma_\Psi$ is not known this is not restrictive, since interest will often only be in inferring whether or not $\beta_k = 0$.

\textbf{Remark on model flexibility.} In general, the flexibility introduced by the baseline cumulative hazard $H_0$ and centrality parameter $\mu$ above will lead to wider identified sets for the parameters $(\beta_1, \dots, \beta_d)$. If one is willing to make additional assumptions on $H_0$ or $\mu$, the obtained identified sets will be smaller. For example, taking $\mu = 0$, the intercept parameter is completely fixed and hence the dimension of the problem is reduced. This facilitates the estimation of the other parameters, leading to smaller identified sets. Likewise, specifying a specific form for $H_0$, the value of $\log(H_0(t))$ is know and hence also in this case the dimension of the problem can be reduced. The drawback, of course, is also a reduction in model flexibility.

\section{Instrumental functions} \label{supp: Instrumental functions}

\textbf{Box functions.} Another commonly used class of instrumental functions for continuous covariates are the box functions, defined as
\begin{equation*}
	\mathcal{G}_\text{box} = \{g_j: x \mapsto g_j(x) = \mathbbm{1}(x \in [a_j, b_j]), j = 1, \dots, l\},
\end{equation*}
where $a_1 = 0$, $b_l = 1$ and $\forall j \in \{2, \dots, l\}: b_{j - 1} = a_{j}$. Typically, the set of knot points $\{a_{j}\}_{j = 1, \dots, l}$ is chosen equidistantly. Note that box functions are not continuous and hence violate Assumption \ref{Assumption IF}. However, smoothed continuous modifications of these functions would be appropriate. For example, one could consider additional knot points $\{c_{1, j}\}_{j = 1, \dots, l}$ and $\{c_{2, j}\}_{j = 1, \dots, l}$ such that $\forall j \in \{1, \dots, l\}: a_{j} < c_{1, j} < c_{2, j} < b_{j}$, and redefine
\begin{align*}
	& \mathcal{G}_\text{box} = \{g_j: g_j(x) = g_{j}^\text{mod}(x)\}, && g_{j}^\text{mod}(x) = \begin{cases}
		\frac{x - a_j}{c_{1, j} - a_j} & \text{ if } x \in [a_j, c_{1, j}],\\
		1 & \text{ if } x \in (c_{1, j}, c_{2, j}],\\
		\frac{b_j - x}{b_j - c_{2, j}} & \text{ if } x \in (c_{2, j}, b_j],\\
		0 & \text{elsewhere.}
	\end{cases}
\end{align*}

In our applications, we prefer to use the cubic B-spline class of instrumental functions since, in contrast to the box-type instrumental functions, their supports are wider, and can overlap. These properties are desirable, since they will facilitate the validity of Assumption \ref{Assumption variance of moments}. At the same time, due to the shape of the spline function, the influence of points far from the center of the spline is reduced, hence preventing oversmoothing.

\textbf{A remark on dependent covariates}. In Section \ref{sec: Instrumental functions} an approach is discussed to make the methodology more robust against correlated continuous covariates. However, it provides no guarantee, nor does it help the absence of instrumental functions whose support lie outside of $N(\mathcal{X})$ when categorical covariates are taken up in the model. Should such problematic instrumental functions remain present in the method, we propose to remove them from the considered class $\mathcal{G}$. We emphasize that this is merely a practical work-around of the problem, and that the development of more optimal methods in the sense of statistical efficiency might be a fruitful subject for future research. Nevertheless, simulations indicate that this crude solution is already sufficient in many cases (Supplementary material \ref{supp: Dependent covariates}).

\textbf{A remark on the dimensionality.} From Equation \eqref{eq: standard instrumental function} it can be seen that the number of instrumental functions used in the analysis will in general be $\mathcal{O}(l^d)$, where $l$ represents the size of the largest univariate class of instrumental functions used. This can become problematic when $d$ is large. As suggested in Section 9 of \cite{AndrewsShi2013}, when $d > 3$, one can opt to replace \eqref{eq: standard instrumental function} by
\begin{equation} \label{eq: many instrumental functions}
	g^{k', k''}: [0, 1]^{d+1} \to \mathbb{R}_+: (x_0, x_1, \dots, x_d) \mapsto g^{k', k''}(x_0, x_1, \dots, x_d) = g_{j_{k'}}(x_{k'})g_{j_{k''}}(x_{k''}),
\end{equation}
for each pair of distinct $k', k'' > 0$, giving rise to supports of the form
\begin{equation*}
	\mathcal{X}_g = \mathcal{X}_0 \times \dots, \mathcal{X}_{k' - 1} \times \mathcal{X}_{g, k'} \times \mathcal{X}_{k' + 1} \times \dots \times \mathcal{X}_{k'' - 1} \times \mathcal{X}_{g, k''} \times \mathcal{X}_{k'' + 1} \times \dots \times \mathcal{X}_{d},
\end{equation*}
and hence forming a class of instrumental functions on all possible combinations of $2$ out of the $d$ dimensions. It can be shown that the total number of instrumental functions of this form is $\mathcal{O}(l^2 d^2)$.

\section{Examples} \label{supp: Examples}

\begin{example}[Variance of moments.]\label{example: assumption variance of moments}
	Let $j \in \{1, \dots, J\}$ and $q \in \{1, 2\}$. Observe that $g_j(X) > 0 \implies m_{j, q}(W, \beta) \neq 0$, since $\Lambda(\cdot)$ maps into $(0, 1)$ and $\mathbbm{1}(\cdot)$ into $\{0, 1\}$. Therefore, conditionally on $g_j(X) > 0$, we have that $m_{j, q}(W, \beta): \mathcal{W} \times \mathcal{B} \to (-1, 0) \cup (0, 1)$. Additionally, taking $\Lambda(\cdot)$ as in Section \ref{sec: The model} or \ref{supp: Link function}, we can sharpen this statement to $m_{j, q}(W, \beta): \mathcal{W} \times \mathcal{B} \to [l_1, u_1] \cup [l_2, u_2]$ for some $-1 < l_1 < u_1 < 0$ and $0 < l_2 < u_2 < 1$, since $\mathcal{X}$ and $\mathcal{B}$ are bounded by Assumption \ref{Assumption covariate space} and Assumption \ref{Assumption parameter space}, respectively. As a result, $g_j(X) > 0 \implies \inf_{\beta \in \mathcal{B}} |m_{j, q}(W, \beta)| = l_3$ for some $l_3 > 0$. Clearly on the other hand, $g_j(X) = 0 \implies m_{j, q}(W, \beta) = 0$.
	
	Assumption \ref{Assumption support IF} guarantees that $\mathbb{P}(g_j(X) > 0) > 0$. Hence, for all $\beta \in \mathcal{B}$ it holds that $|m_{j, q}(W, \beta)| > l_3$ with positive probability. If also $\mathbb{P}(g_j(X) = 0) > 0$, it holds that $m_{j, q}(W, \beta) = 0$ with positive probability. It follows that $\inf_{\beta \in \mathcal{B}} \sigma_{j}(\beta) > 0$. Lastly, note that the condition $\mathbb{P}(g_j(X) = 0) > 0$ simply requires $\mathcal{X} \setminus \mathcal{X}_{g, j} \neq \emptyset$, which is easily satisfied.
	
	In summary, this illustrates that Assumption \ref{Assumption variance of moments} will be satisfied when using either the Cox or AFT link function and provided that there is no instrumental function in $\mathcal{G}$ with a support that covers the entire covariate space.
\end{example}

\begin{example}[Illustration of Assumption \ref{Assumption boundary BI}.]\label{example: assumption boundary BI}
	In this example, we consider the case in which there is one covariate and hence two model parameters, $(\beta_0, \beta_1)$. We will assume that interest is in obtaining the identified interval for $\beta_0$. Define $T\beta_1 = (T^r(r, \beta_1))_2$. It follows from this setting that
	\begin{multline*}
		\tilde{\mathcal{L}}_0(c) = \bigg\{r \in \mathcal{B}_0 \mid\\
		\forall \beta_1 \in \mathcal{B}_1 \text{ for which } (r, \beta_1) \notin \mathcal{B}_I: \bigg|\frac{\partial}{\partial \tilde{\beta}_1} S(\mathbb{E}[m(W, (r, \tilde{\beta}_1))], \Sigma((r, \tilde{\beta}_1)))\Big|_{\tilde{\beta_1} = T\beta_1}\bigg| > c\bigg\}.
	\end{multline*}
	
	Figure \ref{fig: illustration assumption boundary BI} gives two examples of sets $\tilde{\mathcal{L}}_0(c)$ for a certain choice of $c > 0$. In the left panel, the identified set is a circle.\footnote{This example of a problematic case was already discussed in \cite{Bei2024}} At the left and right edges of the boundary of the circle, the derivative of $S(\mathbb{E}[m(W, \beta)], \Sigma(\beta))$ will have a vertical component smaller than $c$. Hence in these regions the condition in the definition of $\tilde{\mathcal{L}}_0(c)$ is not satisfied. The identified set in the middle panel has a less regular shape. The right panel zooms in on one of the problematic regions. Again it is clear that the derivative of $S(\mathbb{E}[m(W, \beta)], \Sigma(\beta))$ evaluated at points projected on this part of the boundary will have a vertical component that is too small.
	
	It is important to note that $c$ can be chosen arbitrarily close to $0$. In other words, it will be possible to let $\tilde{\mathcal{L}}_0(c)$ be close to $\mathcal{B}_{I, k}$, implying that Assumption \ref{Assumption boundary BI} is not stringent.
	\begin{figure}
		\centering
		\includegraphics[width = \linewidth]{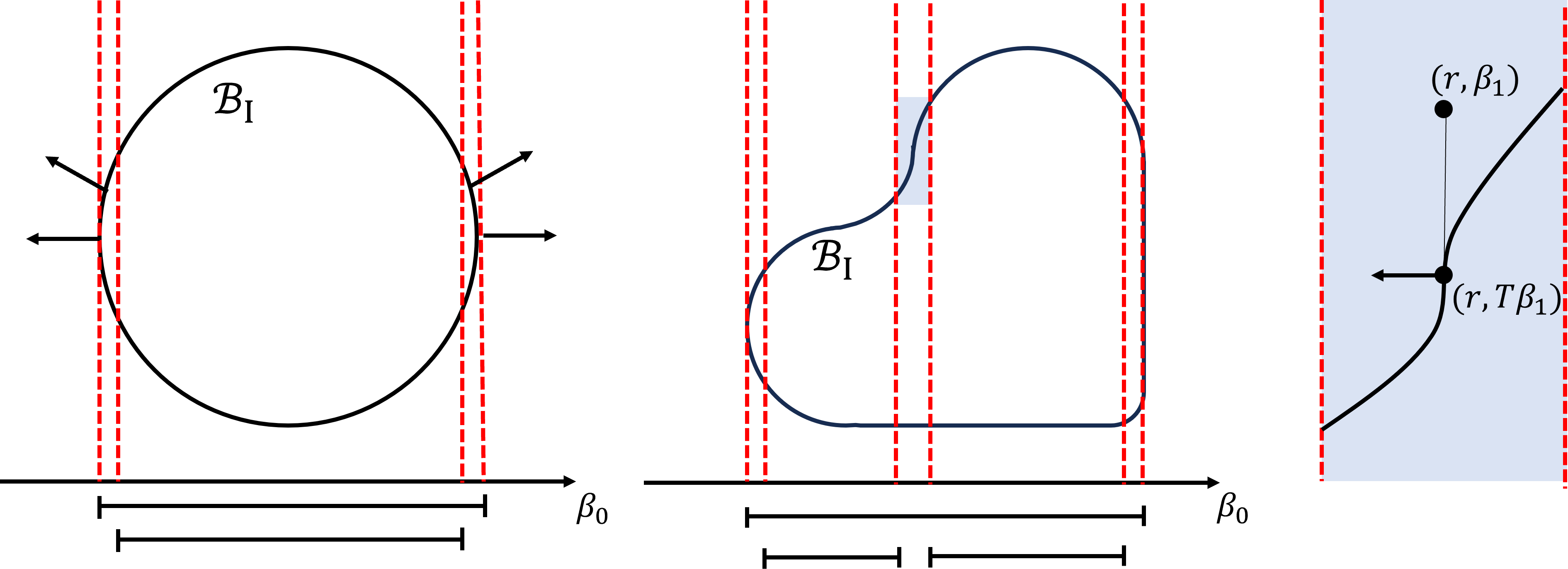}
		\caption{Left panel: example of a circular identified set. In this case, only in a region near the left and right-most parts of the boundary will there be no guarantee of the test having a valid level. Middle panel: more irregularly shaped identified region for which also a part in the middle will be problematic. Right panel: zoom-in of the problematic part of the boundary in the middle panel. In the left and middle panel, the first bar underneath the horizontal axis represents $\mathcal{B}_{I, 0}$. The second bar represents $\tilde{\mathcal{L}}_0(c)$. Note that the problematic parts correspond to regions in which the boundary is (nearly) vertical (or, more generally, nearly parallel to $\mathcal{B}_I(r)$).}
		\label{fig: illustration assumption boundary BI}
	\end{figure}
\end{example}

\begin{example}[Nonconvex identified set.]\label{example: nonconvex identified interval}
	Consider the setting in which we have one single continuous covariate $X_1$ taking values in $[-1, 1]$ with density function $f_{X_1}(x_1) = 1 - |x_1|$. Let $\mathcal{G}$ be the class of instrumental functions, defined as
	\begin{equation*}
		\mathcal{G} = \left\{g_1(x) = \mathbbm{1}(-0.4 \leq x \leq 0.2), \quad g_2(x) = \mathbbm{1}(x \leq 0.5), \quad g_3(x) = \mathbbm{1}(0.25 \geq x)\right\}.
	\end{equation*}
	Suppose the data generating process for $Y$ and $\Delta$ is such that
	\begin{align*}
		\mathbb{E}[\mathbbm{1}(Y \leq t)g_1(X)] = 0.40, & \qquad \mathbb{E}[\mathbbm{1}(Y \leq t, \Delta = 1)g_1(X)] = 0,\\
		\mathbb{E}[\mathbbm{1}(Y \leq t)g_2(X)] = 0.90, & \qquad \mathbb{E}[\mathbbm{1}(Y \leq t, \Delta = 1)g_2(X)] = 0.60,\\
		\mathbb{E}[\mathbbm{1}(Y \leq t)g_3(X)] = 0.30, & \qquad \mathbb{E}[\mathbbm{1}(Y \leq t, \Delta = 1)g_3(X)] = 0.23.\\
	\end{align*}
	We can compute the true identified set by considering a fine grid for the coefficient vector $(\beta_0, \beta_1)$ and approximating the expectations numerically. The result is plotted as the black region in Figure \ref{fig: non pathconnected identified set}, from which it can be seen that the identified set is not path connected and as a result, its projection onto the vertical axis is not an interval.
	
	It can be argued that this example is rather pathological in the sense that the true values of the unconditional moments were chosen specifically for this problem to occur. Perturbing any of these values only slightly can cause the identified set to become connected again. Moreover, the class of instrumental functions used is unnatural. The authors could not find an example of a disconnected identified set using the more natural instrumental functions of Section \ref{sec: Instrumental functions}, supporting the hypothesis that situations as illustrated in this example occur only rarely.
	
	\begin{figure}
		\centering
		\includegraphics[width = \linewidth]{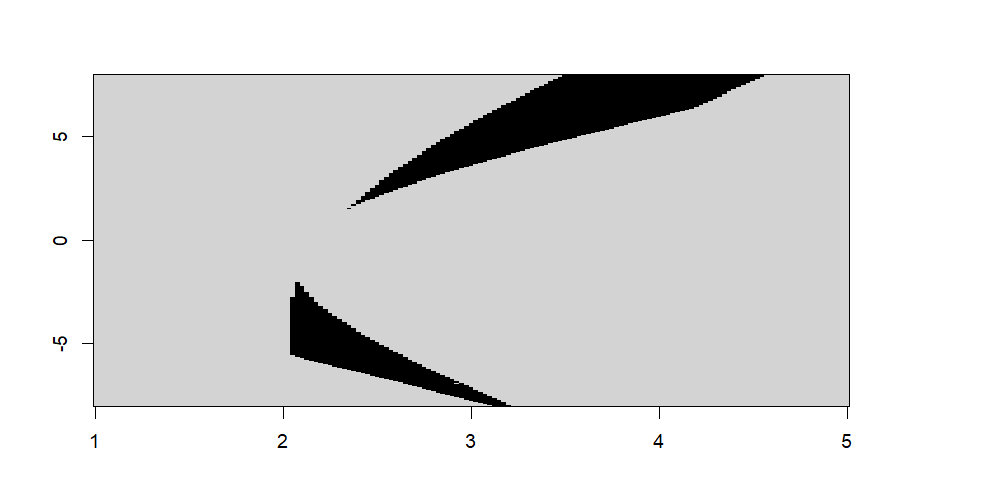}
		\caption{Plot of the identified set (in black) corresponding to the setting described in Example \ref{example: nonconvex identified interval}. The horizontal axis represents $\beta_0$, and the vertical axis $\beta_1$.}
		\label{fig: non pathconnected identified set}
	\end{figure}
\end{example}

\section{Lemmas and theorems} \label{supp: Lemmas and theorems}
This section contains all theorems and lemmas underpinning Theorem \ref{theorem: main}, alongside their proof. Specifically, each of the theoretical results below pertain to a certain assumption imposed in \cite{Bei2024}. A notable exception is Bei's assumption $2$, which is not shown to hold but is rather avoided by means of Assumption \ref{Assumption boundary BI} and theorem \ref{theorem: Polynomial minorant}. To conclude, Section \ref{sec: proof of main theorem} gives a short proof of theorem \ref{theorem: main}. Lemma \ref{lemma: Convexity of combined intervals} pertains to a result on the modification to time-independent effects of covariates.

\subsection{Notation}
We start by giving an overview of all notation introduced thus far, and introducing several other useful definition. Recall that we sometimes view doubly indexed variables (like $m_{j, q}(W, \beta)$) as elements of a $(2J)$-dimensional vector ($m(W, \beta)$), of which the first $J$ elements correspond to the indices $(j, 1)$ and the next $J$ elements correspond to indices $(j, 2)$. We will use double indices $(j, q)$ and single indices $j$ interchangeably, where for the single index, $j = 1, ..., J$ corresponds to double index $(j, 1)$ and $j = J + 1, ..., 2J$ corresponds to double index $(j - J, 2)$. Like in Section \ref{sec: Assumptions and theoretical results}, we let $k$ be the index of the coefficient of interest.

Let $\mathcal{W}$ denote the set of possible values of $W$ and $\mathcal{A}$ be the Borel $\sigma$-algebra on $\mathcal{W}$. Further define $\mathbb{P}$ as the true, unknown probability measure on $(\mathcal{W}, \mathcal{A})$, giving rise to the probability space $(\mathcal{W}, \mathcal{A}, \mathbb{P})$. Unless stated otherwise, $\mathbb{E}$ denotes the expectation with respect to $\mathbb{P}$. Lastly, let $c$ be as in Assumption \ref{Assumption boundary BI}.

Denote $\partial^{(-k)} f(\beta) = \left(\frac{\partial}{\partial \beta_i} f(\beta)\right)_{i = 1, \dots, k - 1, k + 1, \dots, d}$, i.e. the vector of all partial derivatives of $f$ excluding the one with respect to $\beta_k$. We list all relevant notation to state and prove the results below.
\begin{align*}
	& m_{j, 1}(W, \beta) = \left(\mathbbm{1}(Y \leq t) - \Lambda(X^\top \beta)\right)g_j(X),\\
	& m_{j, 2}(W, \beta) = \left(\Lambda(X^\top \beta) -  \mathbbm{1}(Y \leq t, \Delta = 1)\right)g_j(X),\\
	& \Sigma(\beta) = \mathbb{E}[(m(W, \beta) - \mathbb{E}[m(W, \beta)])(m(W, \beta) - \mathbb{E}[m(W, \beta)])^\top],\\
	& \sigma_{j, q}^2(\beta) = \mathbb{E}\left[\left(m_{j, q}(W, \beta) - \mathbb{E}[m_{j, q}(W, \beta)]\right)^2\right], \text{ the diagonal of $\Sigma(\beta)$},\\
	& \bar{m}(\beta) = \frac{1}{n}\sum_{i = 1}^n m(W_i, \beta),\\
	& \hat{\Sigma}(\beta) = \frac{1}{n}\sum_{i = 1}^n\left(m(W_i, \beta) - \bar{m}(\beta)\right)\left(m(W_i, \beta) - \bar{m}(\beta)\right)^\top,\\
	& \hat{\sigma}_{j, q}(\beta) = \frac{1}{n}\sum_{i = 1}^n m_{j, q}(W_i, \beta)^2 - \left(\bar{m}_{j, q}(\beta)\right)^2, \text{ the diagonal of $\hat{\Sigma}(\beta)$},\\
	& S(\sqrt{n}m, \Sigma) = \sum_{q = 1}^2 \sum_{j = 1}^J \left(\sqrt{n}m_{j, q} / \sigma_{j, q}\right)^2_-,\\
	& \mathcal{B}(r) = \{\beta \in \mathcal{B} \mid \beta_k = r\}, \quad \mathcal{B}_I(r) = \{\beta \in \mathcal{B}_I \mid \beta_k = r\},\\
	& T^r\beta = \argmin_{\tilde{\beta} \in \mathcal{B}_I(r)} d(\beta, \tilde{\beta}), \text{ with $d(\cdot, \cdot)$ denoting the usual Euclidean distance,}\\
\end{align*}
Moreover, we will define the empirical process $v_n(\beta)$ to have $(j, q)$-th component
\begin{equation*}
	v_{n, j, q}(\beta) = n^{-1/2} \sigma_{j, q}^{-1}(\beta) \sum_{i = 1}^n \left(m_{j, q}(W_i, \beta) - \mathbb{E}[m_{j, q}(W, \beta)]\right).
\end{equation*}
Following the convention introduced before, the process $v_n(\beta)$ can be thought of as a vector with $2J$ components. Moreover, we define the variance semimetric $\rho_{j, q}(\beta_1, \beta_2)$ as
\begin{equation*}
	\rho_{j, q}(\beta_1, \beta_2) = Var\left[\frac{m_{j, q}(W, \beta_1)}{\sigma_{j, q}(\beta_1)} - \frac{m_{j, q}(W, \beta_2)}{\sigma_{j, q}(\beta_2)}\right]^{1/2}
\end{equation*}
and the variance semimetric $\rho(\beta_1, \beta_2)$ as
\begin{equation*}
	\rho(\beta_1, \beta_2) = \norm{\{\rho_{j, q}(\beta_1, \beta_2)\}_{j = 1, \dots, J; q = 1, 2}}.
\end{equation*}

\subsection{Lemmas}

\begin{lemma} \label{lemma: F is Borel}
	The set $\left\{\frac{m_{j, q}(\cdot, \beta)}{\sigma_{j, q}(\beta)}\bigg|\beta \in \mathcal{B}\right\}$ is composed of Borel measurable functions for all $j \in \{1, \dots, J\}$ and $q \in \{1, 2\}$.
\end{lemma}
\begin{proof}
	By Assumption \ref{Assumption Lambda} the link function $\Lambda(\cdot)$ is Borel measurable. The statement follows immediately, as compositions of measurable functions remain measurable.
\end{proof}

\begin{lemma}\label{lemma: existence 2+c moment}
	For some constant $\tilde{c} > 0$ and all $(j, q) \in \{1, \dots, J\} \times \{1, 2\}$ it holds that
	\begin{equation*}
		\mathbb{E}\left[\sup_{\beta \in \mathcal{B}}\left|\frac{m_{j, q}(W, \beta)}{\sigma_{j, q}(\beta)}\right|^{2 + \tilde{c}}\right] < \infty.
	\end{equation*}
\end{lemma}
\begin{proof}
	The supremum $\sup_{\beta \in \mathcal{B}}\left|\frac{m_{j, q}(W, \beta)}{\sigma_{j, q}(\beta)}\right|^{2+\tilde{c}}$ is a bounded function on $\mathcal{W}$ for all $\tilde{c} > 0$, and hence has finite expectation for all $\tilde{c} > 0$.
\end{proof}

\begin{lemma}\label{lemma: unform convergence correlation matrix}
	The matrix $\Omega$ satisfies
	\begin{equation*}
		\lim_{\delta \to 0} \sup_{\norm{(\beta_1, \beta_1^*) - (\beta_2, \beta_2^*)} < \delta} \norm{\Omega(\beta_1, \beta_1^*) - \Omega(\beta_2, \beta_2^*)} = 0.
	\end{equation*}
\end{lemma}
\begin{proof}
	Let $\epsilon > 0$. We start by noting the following:
	\begin{align*}
		\norm{\Omega(\beta_1, \beta_1^*) - \Omega(\beta_2, \beta_2^*)} \leq \epsilon & \iff \sum_{i, j = 1}^{2J} (\Omega(\beta_1, \beta_1^*) - \Omega(\beta_2, \beta_2^*))^2_{i, j} < \epsilon^2\\
		& \impliedby \forall i, j = 1, \dots, 2J: (\Omega(\beta_1, \beta_1^*) - \Omega(\beta_2, \beta_2^*))^2_{i, j} \leq \frac{\epsilon^2}{2J}.
	\end{align*}
	Hence it suffices to show that each element of $\Omega(\beta, \beta^*)$ is a uniformly continuous function. Recall its definition:
	\begin{align*}
		\Omega(\beta, \beta^*)_{i, j} & = \frac{\mathbb{E}[(m(W, \beta) - \mathbb{E}[m(W, \beta)])(m(W, \beta^*) - \mathbb{E}[m(W, \beta^*)])^\top]_{i, j}} {\sigma_i(\beta)\sigma_j(\beta^*)}\\
		& = \frac{\mathbb{E}[(m_i(W, \beta) - \mathbb{E}[m_i(W, \beta)]) (m_j(W, \beta^*) - \mathbb{E}[m_j(W, \beta^*)])]} {\sigma_i(\beta)\sigma_j(\beta^*)}.
	\end{align*}
	By assumption \ref{Assumption variance of moments}, $\sigma_i(\beta)\sigma_j(\beta^*)$ is uniformly bounded away from zero. Both the numerator and denominator are continuous functions of $(\beta, \beta^*)$ and as the parameter space $\mathcal{B}$ is compact by 
    Assumption \ref{Assumption parameter space}, they are uniformly continuous and bounded. Together, this implies uniform continuity of $\Omega(\beta, \beta^*)_{i, j}$.
\end{proof}

\begin{lemma}\label{lemma: Lower semicontinuity}
	$\forall n \geq 1: S(\sqrt{n} \bar{m}_n(\beta), \hat{\Sigma}_n(\beta))$ is a lower semicontinuous function of $\beta \in \mathcal{B}$.
\end{lemma}
\begin{proof}
	Both $\bar{m}_n(\beta)$ and $\hat{\Sigma}_n(\beta))$ are continuous functions of $\beta$. The statement follows by remarking that $S$ is continuous in both of its arguments.
\end{proof}

\begin{lemma}\label{lemma: G matrix}
	For each $j \in \{1, \dots, 2J\}$, $G_j(\beta) = \nabla_{\beta^*} \left(\frac{\mathbb{E}[m_j(W, \beta^*)]}{\sigma_j(\beta^*)}\right)\bigg|_{\beta^* = \beta}$ exists and its estimator $\hat{G}_{n, j}(\beta)$ is such that 
	\begin{equation*}
		\sup_{\beta \in \mathcal{B}} \norm{\hat{G}_{n, j}(\beta) - G_j(\beta)} = O_P\left(\frac{\kappa_n}{\sqrt{n}}\right),
	\end{equation*}
	where $\kappa_n = \sqrt{\log(n)}$ is a GMS tuning parameter (cf. Section \ref{supp: Testing procedure of Bei}). Moreover, $G_j(\cdot)$ is a bounded, Lipshitz continuous function for all $j \in \{1, \dots, 2J\}$. 
\end{lemma}
\begin{proof}
	A straightforward computation shows that the elements in the Hessian matrix (with respect to $\beta$) of $\frac{\mathbb{E}[m_j(W, \beta)]}{\sigma_j(\beta)}$ are continuous as compositions of continuous functions (Assumption \ref{Assumption Lambda}). Therefore, $G_j(\beta)$ is continuously differentiable on $\mathcal{B}$ and since $\mathcal{B}$ is compact by Assumption \ref{Assumption parameter space}, the derivative is bounded. A fortiori, $G_j(\cdot)$ is Lipshitz continuous and bounded.
	
	We conclude the proof by showing $\sup_{\beta \in \mathcal{B}} \norm{\hat{G}_{n, j}(\beta) - G_j(\beta)} = O_P\left(\frac{\kappa_n}{\sqrt{n}}\right)$, for which we first define the estimator $\hat{G}_{n, j}(\beta)$. To this end, we compute:
	\begin{equation*}
		\frac{\partial}{\partial \beta_k} \frac{\mathbb{E}[m_j(W, \beta)]}{\sigma_j(\beta)} = \frac{\sigma_j(\beta) \frac{\partial}{\partial \beta_k} \mathbb{E}[m_j(W, \beta)] - \mathbb{E}[m_j(W, \beta)] \frac{\partial}{\partial \beta_k} \sigma_j(\beta)}{\sigma_j^2(\beta)},
	\end{equation*}
	leading to the plug-in estimator
	\begin{equation*}
		\hat{G}_{n, j, k}(\beta) = \frac{\hat{\sigma}_j(\beta) \frac{\partial}{\partial \beta_k} \bar{m}_j(\beta) - \bar{m}_j(\beta)\frac{\partial}{\partial \beta_k} \hat{\sigma}_j(\beta)}{\hat{\sigma}_j^2(\beta)},
	\end{equation*}
	where we define
	\begin{align*}
		\bar{m}_j(\beta) & = \frac{1}{n} \sum_{i = 1}^n m_j(W_i, \beta),\\
		\frac{\partial}{\partial \beta_k} \bar{m}_j(\beta) & = \frac{1}{n} \sum_{i = 1}^n \frac{\partial}{\partial \beta_k}m_j(W_i, \beta),\\
		\hat{\sigma}^2_j(\beta) & = \frac{1}{n} \sum_{i = 1}^n (m_j(W_i, \beta) - \bar{m}_j(\beta))^2,\\
		\hat{\sigma}_j(\beta) & = \sqrt{\hat{\sigma}_j^2(\beta)},\\
		\frac{\partial}{\partial \beta_k} \hat{\sigma}_j(\beta) & = \frac{1}{\hat{\sigma}_j(\beta)} \sum_{i = 1}^n \bigg(m_j(W_i,\beta) - \bar{m}_j(\beta)\bigg)\bigg(\frac{\partial}{\partial \beta_k} m_j(W_i, \beta) - \frac{\partial}{\partial \beta_k} \bar{m}_j(\beta)\bigg).
	\end{align*}
	In order to derive the uniform rate of convergence of $\hat{G}_{n, j, k}$ to $G_j$, it will be useful to have some preliminary results at hand. Specifically, we start by observing that the classes of functions
	\begin{equation*}
		\left\{m_j(\cdot, \beta) \mid \beta \in \mathcal{B}\right\}, \quad \left\{m^2_j(\cdot, \beta) \mid \beta \in \mathcal{B}\right\}, \quad \left\{\frac{\partial}{\partial \beta_k}m_j(\cdot, \beta) \mid \beta \in \mathcal{B}\right\},
	\end{equation*}
	are all Donsker. Indeed this can be seen by noting that the functions in all three classes are Lipshitz continuous in $\beta$ (which follows from Assumption \ref{Assumption Lambda}), $\mathcal{B}$ is bounded and an application of theorem 2.7.11 in \cite{vdVaartWellner1996}. From these results we obtain uniform convergence rates for each component in the plug-in estimator $\hat{G}_{n, j, k}$ above:
	\begin{align*}
		\sup_{\beta \in \mathcal{B}} \norm{\bar{m}_j(\beta) - \mathbb{E}[m_j(W, \beta)]} = O_P(n^{-1/2}),\\
		\sup_{\beta \in \mathcal{B}} \norm{\hat{\sigma}_j^2(\beta) -\sigma^2_j(\beta)]} = O_P(n^{-1/2}),\\
		\sup_{\beta \in \mathcal{B}} \norm{\hat{\sigma}_j(\beta) -\sigma_j(\beta)]} = O_P(n^{-1/2}),\\
		\sup_{\beta \in \mathcal{B}} \norm{\frac{\partial}{\partial \beta_k}\bar{m}_j(\beta) - \frac{\partial}{\partial \beta_k}\mathbb{E}[m_j(W, \beta)]} = O_P(n^{-1/2}),\\
		\sup_{\beta \in \mathcal{B}} \norm{\frac{\partial}{\partial \beta_k}\hat{\sigma}_j(\beta) - \frac{\partial}{\partial \beta_k}\sigma_j(\beta)} = O_P(n^{-1/2}).
	\end{align*}
	To obtain the final result, we use the following notations, to be viewed independently from previous definitions for these variables:
	\begin{align*}
		& A_n = \hat{\sigma}_j(\beta) \frac{\partial}{\partial \beta_k} \bar{m_j}(\beta), && A = \sigma_j(\beta) \frac{\partial}{\partial \beta_k} \mathbb{E}[m_j(W, \beta)],\\
		& B_n = \bar{m}_j(\beta)\frac{\partial}{\partial \beta_k} \hat{\sigma}_j(\beta), && B = \mathbb{E}[m_j(W, \beta)] \frac{\partial}{\partial \beta_k} \sigma_j(\beta),\\
		& C_n = \hat{\sigma}_j^2(\beta), && C = \sigma_j^2(\beta).\\
	\end{align*}
	For example, $A$ denoted the total number of time points in the grid in Section \ref{sec: Time-independent effects of covariates}, but is unrelated to $A$ as defined above. We suppressed the dependence on $\beta$ for each of these variables for ease of notation. We can compute
	\begin{align*}
		\hat{G}_{n, j, k}(\beta) - G_{j, k}(\beta) & = \frac{A_n - B_n}{C_n} - \frac{A - B}{C}\\
		& = \frac{A_n - A}{C_n} + (A - B)\left(\frac{1}{C_n} - \frac{1}{C}\right) - \frac{B_n - B}{C_n}\\
		& = \frac{1}{C_n}\left((A_n - A) - (C_n - C)\frac{(A - B)}{C} - (B_n - B)\right)\\
	\end{align*}
	so that
	\begin{multline*}
		\sup_{\beta \in \mathcal{B}} \norm{\hat{G}_{n, j, k}(\beta) - G_j(\beta)}\\ \leq \frac{1}{\inf_{\beta \in \mathcal{B}} C_n} \left(\sup_{\beta \in \mathcal{B}} \norm{A_n - A} + \sup_{\beta \in \mathcal{B}} \norm{C_n - C}\sup_{\beta \in \mathcal{B}} \norm{\frac{A - B}{C}} + \sup_{\beta \in \mathcal{B}} \norm{B_n - B}\right).
	\end{multline*}
	From the uniform convergence rates of each component in $\hat{G}_{n, j, k}$, noting that $\inf_{\beta \in \mathcal{B}} C_n$ is bounded away from zero in probability as $n \to \infty$ and noting that $(A - B)/C$ is bounded as a function of $\beta$, it follows that $\sup_{\beta \in \mathcal{B}} \norm{\hat{G}_{n, j, k}(\beta) - G_j(\beta)} = O_P(n^{-1/2})$. The result to be proven follows.
\end{proof}

\begin{lemma} \label{lemma: Convexity of combined intervals}
	Let $k \in \{0, \dots, d\}$. When the true coefficient vector $\beta_\text{true}$ is constant over time, combining a finite number of identified intervals for $\beta_k$ by intersection or majority vote will again result in an interval.
\end{lemma}
\begin{proof}
	Denote the identified intervals as $\mathcal{I}_1, \dots, \mathcal{I}_A$. We start by remarking that $\forall a \in \{1, \dots, A\}: \beta_{\text{true}, k} \in \mathcal{I}_a$. Therefore, $\bigcap_{l = 1}^A \mathcal{I}_a \neq \emptyset$.
	
	Consider the majority vote combination rule in Equation \eqref{eq: majority voting rule} with cut-off $\tau \in (0, 1)$. That is,
	\begin{equation*}
		\mathcal{I} = \left\{\beta_k \in \mathcal{B}_k \mid \frac{1}{A} \sum_{a = 1}^A \mathbbm{1}(\beta_k \in \mathcal{I}_{a}) > \tau\right\}.
	\end{equation*}
	By the previous remark, we have that for any $\tau \in (0, 1)$, $\beta_{\text{true}, k} \in \mathcal{I}$. Suppose further that there exists $\beta^*_k \in \mathcal{I}$. Moreover, assume without loss of generality that $\beta^*_k \leq \beta_{\text{true}, k}$. We will prove that $[\beta^*_k, \beta_{\text{true}, k}] \subset \mathcal{I}$.
	
	Since $\beta^*_k \in \mathcal{I}$, there exist indices $a_1, \dots, a_L$, with $L \geq A\tau$, such that $\forall l \in \{1, \dots, L\}: \beta^*_k \in \mathcal{I}_{a_l}$. Because also $\beta_{\text{true}, k} \in \mathcal{I}_{a_l}$ and $\mathcal{I}_{a_l}$ is an interval, it holds that $[\beta^*_k, \beta_{\text{true}, k}] \subset \mathcal{I}_{a_l}$. As a consequence, each element in $[\beta^*_k, \beta_{\text{true}, k}]$ is contained in at least $A\tau$ identified intervals, and hence $[\beta^*_k, \beta_{\text{true}, k}] \subset \mathcal{I}$. The proof concludes by writing $\mathcal{I} = \bigcup_{\beta_k \in \mathcal{I}} [\beta_k, \beta_{\text{true}, k}] \cup [\beta_{\text{true}, k}, \beta_k]$, defining $[a,b] = \emptyset$ when $a > b$, and noting that a union of intersecting intervals is again an interval.
	
	The result for the intersection of intervals follows immediately by noting that it is the special case where $\mathcal{I}$ is constructed using $\tau > \frac{A - 1}{A}$.
\end{proof}

\subsection{Theorems}

\begin{theorem} \label{theorem: Donsker}
	The class of functions $\mathcal{F} = \left\{\frac{m_{j, q}(\cdot, \beta)}{\sigma_{j, q}(\beta)}\bigg|\beta \in \mathcal{B}\right\}$ is Donsker (\cite{vdVaartWellner2023}, p.130) for any $j \in \{1, \dots, J\}$ and $q \in \{1, 2\}$.
\end{theorem}
\begin{proof}
	In the following, let $j \in \{1, \dots, J\}$ and $q = 2$. It is straightforward to modify the arguments for the case that $q = 1$. Below we prove the uniform entropy bound
	\begin{equation}\label{eq: uniform entropy bound}
		\int_0^\infty \sup_Q \sqrt{\log N(\epsilon\norm{F}_{Q, 2}, \mathcal{F}, L_2(Q))} d\epsilon < \infty,
	\end{equation}
	where the supremum is taken over all finitely discrete probability measures $Q$ such that $\int F^2 dQ > 0$ and $F$ is a square integrable envelope function of $\mathcal{F}$.
	Theorem \ref{theorem: Donsker} then follows by an application of Theorem $2.5.2$ in \cite{vdVaartWellner2023}.
	
	\textbf{Step 1: the envelope function $F$.}\\
	We define the envelope function $F$ of $\mathcal{F}$ as follows.
	\begin{align*}
		\sup_{\beta \in \mathcal{B}} \frac{m_{j, 2}(W, \beta)}{\sigma_{j, 2}(\beta)} & \leq \frac{\sup_{\beta \in \mathcal{B}}(\Lambda(X^\top\beta) - \mathbbm{1}(Y \leq t, \Delta = 1))g_j(X)}{\inf_{\beta \in \mathcal{B}}\sigma_{j, 2}(\beta)}\\
		& \overset{def}{=} F(W)
	\end{align*}
	
	Let $P$ be any probability measure on $(\mathcal{W}, \mathcal{A})$. The following shows $F$ is square integrable w.r.t. $P$:
	\begin{equation}\label{eq: envelope square integrable}
		\int_\mathcal{W} F^2 dP \leq \left(\frac{1}{\eta}\right)^2 \int_\mathcal{W} (1 - \mathbbm{1}(y \leq t, \delta = 1))g^2_j(x) dP(w) \leq \frac{M_\mathcal{G}^2}{\eta^2} < \infty.
	\end{equation}
	The first inequality follows from Assumption \ref{Assumption variance of moments} and \ref{Assumption Lambda}. The second inequality invokes Assumption \ref{Assumption IF} and uses that $(1 - \mathbbm{1}(y \leq t, \delta = 1)) \leq 1$. Next, we show lower and upper bounds on $\norm{F}_{Q, 2}$. Recall that $Q$ represents a finite, discrete probability measure. Let $\mathcal{X}_Q = \{w_i \in \mathcal{W}, i \in \{1, \dots, m_Q\}\}$ be the union of singletons that are assigned a non-zero measure by $Q$. Since $Q$ must be such that $\norm{F}_{Q,2}^2 > 0$, we must have at least one $w_i \in \mathcal{X}_Q$ with $g_j(x_i) \neq 0$. We compute
	\begin{align*}
		\norm{F}_{Q, 2} & = \left(\int_\mathcal{W} F^2 dQ\right)^{1/2}\\
		& = \left(\sum_{i = 1}^{m_Q} F^2(w_i) q(w_i)\right)^{1/2}\\
		& = \left(\sum_{i = 1}^{m_Q} \frac{\sup_{\beta \in \mathcal{B}}\left(\Lambda(x_i^\top\beta) - \mathbbm{1}(y_i \leq t, \delta_i = 1)\right)^2g_j^2(x_i)}{\eta^2} q(w_i)\right)^{1/2},
	\end{align*}
	from which we obtain the bounds
	\begin{equation}\label{eq: bounds on Qnorm of envelope}
		\frac{L}{\eta} \left(\sum_{i = 1}^{m_Q} g_j^2(x_i) q(w_i)\right)^{1/2} \leq \norm{F}_{Q, 2} \leq \frac{U}{\eta} \left(\sum_{i = 1}^{m_Q} g_j^2(x_i) q(w_i)\right)^{1/2}.
	\end{equation}
	For for ease of notation, we defined $L = \min_{i \in \{1, \dots, m_Q\}} \sqrt{\sup_{\beta \in \mathcal{B}}\left(\Lambda(x_i^\top\beta) - \mathbbm{1}(y_i \leq t, \delta_i = 1)\right)^2}$ and $U = \max_{i \in \{1, \dots, m_Q\}} \sqrt{\sup_{\beta \in \mathcal{B}} \left(\Lambda(x_i^\top\beta) - \mathbbm{1}(y_i \leq t, \delta_i = 1)\right)^2}$.
	
	\textbf{Step 2: Uniform entropy bound on $\mathcal{F}$}\\
	Define $B(f, \epsilon\norm{F}_{Q, 2}) = \left\{g \in \mathcal{F}: \left(\int_\mathcal{W} (f - g)^2 dQ\right)^{1/2} < \epsilon\norm{F}_{Q, 2}\right\}$. We will look for a characterization of an appropriate subset of $B(f, \epsilon\norm{F}_{Q, 2})$, based on which we obtain an upper bound for the uniform entropy number $N(\epsilon\norm{F}_{Q, 2}, \mathcal{F}, L_2(Q))$. With this upper bound, we can show that the uniform entropy bound in Equation \eqref{eq: uniform entropy bound} holds.
	
	First, we make the following calculations:
	\begin{align*}
		& \sup_{w \in \mathcal{W}, \beta \in \mathcal{B}} \frac{\partial}{\partial\beta_k} \frac{\Lambda(x^\top\beta) - \mathbbm{1}(y \leq t, \delta = 1)}{\sigma_{j, 2}(\beta)}\\
		& \quad = \sup_{w \in \mathcal{W}, \beta \in \mathcal{B}} \frac{\Lambda'(x^\top \beta) x_k \sigma_{j, 2}(\beta) - (\Lambda(x^\top\beta) - \mathbbm{1}(y \leq t, \delta = 1)) \frac{\partial}{\partial\beta_k} \sigma_{j, 2}(\beta)}{\sigma^2_{j, 2}(\beta)}\\
		& \quad \leq \frac{K_{\Lambda'} K_{\mathcal{X}_k}}{\eta} + \frac{K_{\frac{\partial}{\partial\beta_k} \sigma_{j, 2}}}{\eta^2}\\
		& \quad \overset{def}{=} K^* < \infty,
	\end{align*}
	where we defined
	\begin{equation*}
		K_{\Lambda'} = \sup_{x} \Lambda'(x), \quad K_{\mathcal{X}_k} = \sup_{x_k \in \mathcal{X}_k} |x_k|, \quad K_{\frac{\partial}{\partial\beta_k} \sigma_{j, 2}} = \sup_{\beta^* \in \mathcal{B}} \left|\frac{\partial}{\partial\beta_k} \sigma_{j, 2}(\beta)\bigg|_{\beta = \beta^*}\right|.
	\end{equation*}
	The constant $K^*$ being finite follows from Assumptions \ref{Assumption Lambda}, \ref{Assumption covariate space}, \ref{Assumption variance of moments} and a straight-forward computation showing  that $K_{\frac{\partial}{\partial\beta_k} \sigma_{j, 2}} < \infty$. The above allows to derive a characterization of a subset of $B(f, \epsilon\norm{F}_{Q, 2})$. To this end, we first compute an upper bound on the $L_2(Q)$-distance between two functions in $\mathcal{F}$:
	\begin{align*}
		\left[\int_\mathcal{W} \left(\frac{m_{j, 2}(w, \beta_1)}{\sigma_{j, 2}(\beta_1)} - \frac{m_{j, 2}(w, \beta_2)}{\sigma_{j, 2}(\beta_2)}\right)^2 dQ\right]^{1/2} & \leq \left[\sum_{i = 1}^{m_Q} (K^*)^2 (\beta_1 - \beta_2)^2 g_j^2(x_i) q(w_i)\right]^{1/2}\\
		& = K^* |\beta_1 - \beta_2|  \left[\sum_{i = 1}^{m_Q} g_j^2(x_i) q(w_i)\right]^{1/2}\\
		& \leq K^* |\beta_1 - \beta_2| \frac{\eta}{L} \norm{F}_{Q, 2},
	\end{align*}
	where in the first inequality we used that a differentiable function is Lipshitz continuous with Lipshitz constant $K^*$ and the third follows from the bounds derived in \eqref{eq: bounds on Qnorm of envelope}. With the derived bound on the $L_2(Q)$-distance in hand, we obtain the following:
	\begin{align*}
		B(f, \epsilon\norm{F}_{Q, 2}) & = \left\{g \in \mathcal{F}: \left(\int_\mathcal{W} (f - g)^2 dQ\right)^{1/2} < \epsilon\norm{F}_{Q, 2}\right\}\\
		& = \left\{\beta_g \in \mathcal{B}: \left(\int_\mathcal{W} \left(\frac{m_{j, 2}(w, \beta_f)}{\sigma_{j, 2}(\beta_f)} - \frac{m_{j, 2}(w, \beta_g)}{\sigma_{j, 2}(\beta_g)}\right)^2 dQ\right)^{1/2} < \epsilon\norm{F}_{Q, 2}\right\}\\
		& \supseteq \left\{\beta_g \in \mathcal{B}: K^*\frac{\eta}{L}\norm{F}_{Q, 2}|\beta_f - \beta_g| < \epsilon\norm{F}_{Q, 2}\right\}\\
		& = \left\{\beta_g \in \mathcal{B}:|\beta_f - \beta_g| < \frac{L}{\eta K^*}\epsilon \right\}\\
		& \overset{def}{=} \underline{B}(f, \epsilon).
	\end{align*}
	It will take $\mathcal{O}(\epsilon^{-(d + 1)})$ sets $\underline{B}(f, \epsilon)$ to cover the entire parameter space $\mathcal{B}$. Therefore, $\mathcal{F}$ is a polynomial class as defined in \cite{vdVaartWellner2023}, page 135.
	
	\textbf{Step 3: $\mathbb{P}$-measurability of $\mathcal{F}$, $\mathcal{F}_\delta$ and $\mathcal{F}_{\infty}^2$:}\\
	Definition 2.3.3 in \cite{vdVaartWellner2023} states:
	
	\quad
	
	\begin{minipage}{0.95\linewidth}
		\textbf{Measurable class:} \quad A class $\mathcal{F}$ of measurable functions $f: \mathcal{W} \to \mathbb{R}$ on a probability space $(\mathcal{W}, \mathcal{A}, \mathbb{P})$ is called $\mathbb{P}$-measurable if the function
		\begin{equation*}
			\mathcal{H}: \mathcal{W}^n \to [0, \infty): (W_1, \dots, W_n) \mapsto \norm{\sum_{i = 1}^n e_i f(W_i)}_\mathcal{F}
		\end{equation*}
		is measurable on the completion of $(\mathcal{W}^n, \mathcal{A}^n, \mathbb{P}^n)$ for every $n$ and every $(e_1, \dots, e_n) \in \mathbb{R}^n$ (Here we define $\norm{\sum_{i = 1}^n e_i f(W_i)}_\mathcal{F} = \sup_{f \in \mathcal{F}} \left|\sum_{i = 1}^n e_i f(W_i)\right|$. Also note that $\mathcal{F}$ is not trivially $\mathbb{P}$-measurable as the supremum is taken over a possibly uncountable set).
	\end{minipage}
	
	\quad
	
	\noindent In the following, we will show that $\mathcal{F}$ is $\mathbb{P}$-measurable. Similar arguments can be used to obtain $\mathbb{P}$-measurability of $\mathcal{F}_\delta = \{f - g \mid f, g \in \mathcal{F}, \norm{f - g}_{P, 2} < \delta\}$ and $\mathcal{F}^2_\infty = \{(f - g)^2 \mid f, g \in \mathcal{F}\}$. Denote $h_\beta(W_1, \dots, W_n) = \sum_{i = 1}^n e_i \frac{(\Lambda(x_i^\top\beta) - \mathbbm{1}(y_i \leq t, \delta_i = 1))g_j(x_i)}{\sigma_{j, 2}(\beta)}$ for some vector $(e_1, \dots, e_n)$.
	
	By Lemma \ref{lemma: F is Borel}, $\mathcal{F}$ is a class of Borel measurable functions. Moreover, from its definition, it is clear that $m_{j, 2}$ is possibly discontinuous on $\{w \in \mathcal{W}: y = t\}$ but continuous elsewhere. By extension, $h_\beta$ can only be discontinuous on $D = \{w^{(n)} \in \mathcal{W}^n \mid \exists i \in \{1, \dots, n\}: y^{(n)}_i = t\}$, which is a null set. Let $h_{\beta, D}$ denote the restriction of $h_\beta$ to $D$ and $h_{\beta, C}$ analogously, for $C = \mathcal{W} \setminus D$. Then $h_{\beta|C}$ is a continuous function, so $H_C = \sup_{\beta \in \mathcal{B}} h_{\beta|C}$ is lower semicontinuous and therefore measurable. Further define $H_D = \sup_{\beta \in \mathcal{B}} h_{\beta|D}$ and note that
	\begin{equation*}
		\mathcal{H}(w^{(n)}) = \begin{cases}
			H_C(w^{(n)}) & \text{if } w^{(n)} \in C,\\
			H_D(w^{(n)}) & \text{if } w^{(n)} \in D.
		\end{cases}
	\end{equation*}
	
	We can now show that $\mathcal{H}$ is measurable in the completion of $(\mathcal{W}^n, \mathcal{A}^n, \mathbb{P}^n)$. To this end, let $Z \subset \mathbb{R}$ be a Borel set. Then
	\begin{equation}
		\mathcal{H}^{-1}(Z) = H^{-1}_C(Z) \cup H^{-1}_D(Z).
	\end{equation}
	As $H_C$ is a measurable function, $H^{-1}_C(Z) \in \mathcal{A}^n$. Moreover, $H^{-1}_D(Z) \subseteq H^{-1}_D(\mathbb{R}) \subseteq D$. Hence $H^{-1}_D(Z)$ is a null set and therefore measurable with respect to the completion of $(\mathcal{W}^n, \mathcal{A}^n, \mathbb{P}^n)$. It follows that also $\mathcal{H}^{-1}(Z)$ is measurable with respect to the completion of $(\mathcal{W}^n, \mathcal{A}^n, \mathbb{P}^n)$.
	
	\textbf{Step 4: Finishing up}\\
	Combining steps 1-3 and Theorem 2.5.2 in Van der Vaart and Wellner (2023) we obtain that $\mathcal{F}$ is ($\mathbb{P}$-)Donsker. As also mentioned in the beginning of this proof, the arguments used here can be adapted to the case $q = 2$. 
\end{proof}

\begin{corollary}\label{corrolary: Asymptotic equicontinuity}
	The empirical process $v_n(\beta)$ is asymptotically $\rho$-equicontinuous. That is, $$\forall \epsilon > 0: \lim_{\delta \overset{>}{\to} 0} \limsup_{n \to \infty} \mathbb{P}^*\left(\sup_{\rho(\beta_1, \beta_2) < \delta} \norm{v_n(\beta_1) - v_n(\beta_2)} > \epsilon\right) = 0.$$
\end{corollary}
\begin{proof}
	By Theorem \ref{theorem: Donsker} it follows from Theorem 1.5.7 and Problem 2.1.5 in \cite[p.139]{vdVaartWellner2023} that the statement holds elementwise, in the sense that
	\begin{equation*}
		\forall (j, q) \in \{1, \dots, J\} \times \{1, 2\}: \forall \epsilon > 0: \lim_{\delta \overset{>}{\to} 0} \limsup_{n \to \infty} \mathbb{P}^*\left(\sup_{\rho_{j, q}(\beta_1, \beta_2) < \delta} \norm{v_{n, j, q}(\beta_1) - v_{n, j, q}(\beta_2)} > \epsilon\right) = 0.
	\end{equation*}
	We conclude by making the following observations.
	\begin{align*}
		\mathbb{P}^*\left(\sup_{\rho(\beta_1, \beta_2) < \delta} \norm{v_n(\beta_1) - v_n(\beta_2)}^2 > \epsilon^2\right) & = \mathbb{P}^*\left(\sup_{\rho(\beta_1, \beta_2) < \delta} \sum_{q = 1}^2 \sum_{j = 1}^J |v_{n, j, q}(\beta_1) - v_{n, j, q}(\beta_2)|^2 > \epsilon^2\right)\\
		& \leq \mathbb{P}^*\left(\sum_{q = 1}^2 \sum_{j = 1}^J \sup_{\rho(\beta_1, \beta_2) < \delta} |v_{n, j, q}(\beta_1) - v_{n, j, q}(\beta_2)|^2 > \epsilon^2\right)\\
		& \leq \sum_{q = 1}^2 \sum_{j = 1}^J \mathbb{P}^*\left(\sup_{\rho(\beta_1, \beta_2) < \delta} |v_{n, j, q}(\beta_1) - v_{n, j, q}(\beta_2)|^2 > \frac{\epsilon^2}{2J}\right)\\
		& \leq \sum_{q = 1}^2 \sum_{j = 1}^J \mathbb{P}^*\left(\sup_{\rho_{j, q}(\beta_1, \beta_2) < \delta} |v_{n, j, q}(\beta_1) - v_{n, j, q}(\beta_2)|^2 > \frac{\epsilon^2}{2J}\right).
	\end{align*}
\end{proof}

\begin{theorem} \label{theorem: Polynomial minorant}
	There exist $\tilde{c}, d_\text{max} > 0$ such that for all $r \in \tilde{\mathcal{L}}_0(c)$ and for all $\beta \in \mathcal{B}(r)$ it holds that
	\begin{equation*}
		S(\mathbb{E}[m(W, \beta)], \Sigma(\beta)) \geq \tilde{c} \min\{d_\text{max}, d(\beta, \mathcal{B}_I(r))\}^2
	\end{equation*}
\end{theorem}
\begin{proof}
	Let $\beta \in \mathcal{B}(r)$. Note that the statement is trivial when $\beta \in \mathcal{B}_I(r)$ since then both sides of the inequality are zero, so we can restrict to considering $\beta \in \mathcal{B}(r) \setminus \mathcal{B}_I(r)$. Write
	\begin{equation*}
		S(\mathbb{E}[m(W, \beta)], \Sigma(\beta)) = \sum_{j = 1}^{2J} \left[\frac{m_{j}(W, \beta)}{\sigma_{j}(\beta)}\right]^2_-,
	\end{equation*}
	using the single index notation. Recall that $T^r\beta$ is the orthogonal projection of $\beta$ onto $\mathcal{B}_I(r)$. That is,
	\begin{equation*}
		T^r\beta \in \argmin_{\beta^* \in \mathcal{B}_I(r)} d(\beta, \beta^*).
	\end{equation*}
	Expanding $S(\mathbb{E}[m(W, \beta)], \Sigma(\beta))$ around $T^r \beta$, we obtain
	\begin{align}
		S(\mathbb{E}[m(W, \beta)], \Sigma(\beta)) & = S(\mathbb{E}[m(W, T^r\beta)], \Sigma(T^r\beta)) + \nabla_{\beta^*} S(\mathbb{E}[m(W, \beta^*)], \Sigma(\beta^*)) \big|_{\beta^* = \bar{\beta}}^\top (\beta - T^r\beta)\notag\\
		& = \nabla_{\beta^*} S(\mathbb{E}[m(W, \beta^*)], \Sigma(\beta^*)) \big|_{\beta^* = T^r\beta}^\top (\beta - T^r\beta) + \mathcal{O}(\norm{\beta - T^r\beta}^2) \label{eq: computation S (1)}.
	\end{align}
	The first equality follows from the mean value theorem. The second equality follows from Lipshitz continuity of $\nabla_{\beta} S(\mathbb{E}[m(W, \beta)], \Sigma(\beta))$ and the fact that $T^r\beta \in \mathcal{B}_I(r)$. As both $\beta$ and $T^r\beta$ are elements of $\mathcal{B}(r)$, we can write $\beta - T^r\beta = (\epsilon_0, \dots, \epsilon_{k - 1}, 0, \epsilon_{k + 1}, \dots, \epsilon_d)^\top$, for $\epsilon_0, \dots, \epsilon_d \in \mathbb{R}$. Hence in the vector product above, the $k$-th element of $\nabla_{\beta^*} S(\mathbb{E}[m(W, \beta^*)], \Sigma(\beta^*)) \big|_{\beta^* = T^r\beta}$ will have no contribution. Therefore, we define
	\begin{equation*}
		\partial^{(-k)}S(\mathbb{E}[m(W, \beta)], \Sigma(\beta)) = \left[\frac{\partial}{\partial \beta_i} S(\mathbb{E}[m(W, \beta^*)], \Sigma(\beta^*))\big|_{\beta^* = \beta}\right]_{i = 0, \dots, k - 1, k + 1, \dots, d},
	\end{equation*}
	or, equivalently,
	\begin{equation*}
		\partial^{(-k)}S(\mathbb{E}[m(W, \beta)], \Sigma(\beta)) = T^r \nabla_{\beta^*} S(\mathbb{E}[m(W, \beta^*)], \Sigma(\beta^*)) \big|_{\beta^* = \beta}.
	\end{equation*}
	
	Let $\mathcal{C}$ denote the boundary of $\mathcal{B}_I(r)$. As a working assumption, we can impose that $\mathcal{C}$ is smooth, in the sense that it has a well-defined surface normal everywhere. Indeed, it is always possible to ``smooth out'' any sharp regions in the boundary by adding artificial inequality restrictions (see Figure \ref{fig: illustration working smoothness assumption}). The resulting identified set including these artificial restrictions can be made to lie arbitrarily close to the original identified set $\mathcal{B}_I$.
	\begin{figure}
		\centering
		\includegraphics[width = 0.4\linewidth]{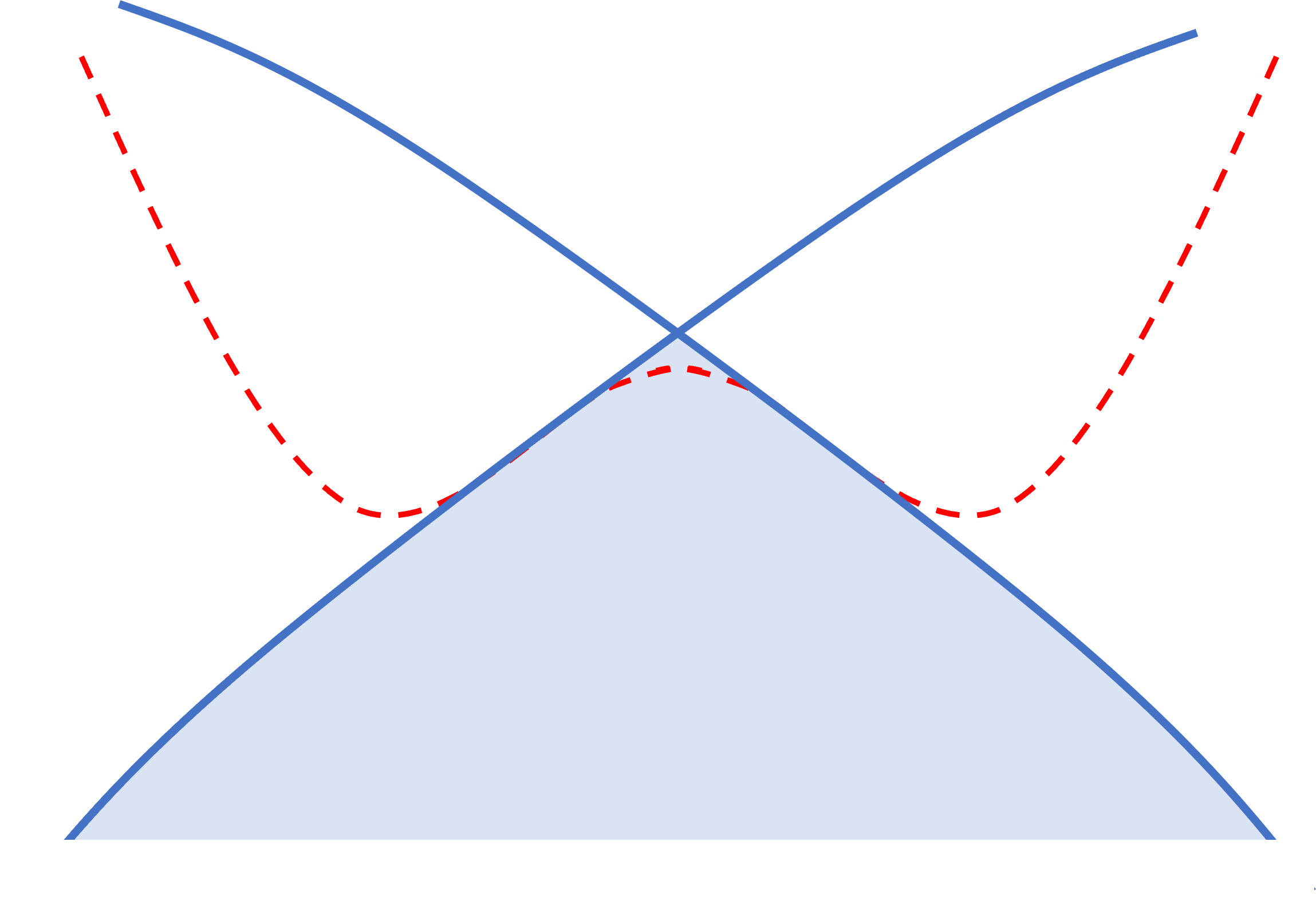}
		\caption{Illustration of the working smoothness assumption. The shaded area represents a part of the original identified set with a non-smooth boundary. By adding the restriction that the identified set must lie below the dashed line, the sharp edge in the boundary can be ``smoothed out''.}
		\label{fig: illustration working smoothness assumption}
	\end{figure}
	
	Since $T^r\beta$ is an orthogonal projection of $\beta$ onto $\mathcal{B}_I(r)$ and the boundary of $\mathcal{B}_I(r)$ is smooth by our working assumption, the vector $\beta - T^r\beta$ will be parallel to the surface normal of $\mathcal{C}$ at the point $T^r\beta$, which is precisely $\partial^{(-k)}S(\mathbb{E}[m(W, T^r\beta)], \Sigma(T^r\beta))$. We can continue our computation in \eqref{eq: computation S (1)} as follows:
	\begin{align*}
		& S(\mathbb{E}[m(W, \beta)], \Sigma(\beta))\\
		& \quad = \partial^{(-k)}S(\mathbb{E}[m(W, T^r\beta)], \Sigma(T^r\beta)) (\epsilon_0, \dots, \epsilon_{k - 1}, \epsilon_{k + 1}, \dots, \epsilon_d)^\top + \mathcal{O}(||\beta - T^r\beta||^2)\\
		& \quad = ||\partial^{(-k)}S(\mathbb{E}[m(W, T^r\beta)], \Sigma(T^r\beta))|| \times ||(\epsilon_0, \dots, \epsilon_{k - 1}, \epsilon_{k + 1}, \dots, \epsilon_d)|| + \mathcal{O}(||\beta - T^r\beta||^2)\\
		& \quad = ||\partial^{(-k)}S(\mathbb{E}[m(W, T^r\beta)], \Sigma(T^r\beta))|| \times d(\beta, \mathcal{B}_I(r)) + \mathcal{O}(d(\beta, \mathcal{B}_I(r))^2)\\
		& \quad \geq c \times d(\beta, \mathcal{B}_I(r)) + \mathcal{O}(d(\beta, \mathcal{B}_I(r))^2)\\
		& \quad \geq  \tilde{c} \min(\tilde{d}_\text{max}, d(\beta, \mathcal{B}_I(r))).
	\end{align*}
	The first step rewrites and the second step uses that the vectors in the vector product are parallel. The third step observes that $\norm{\beta - T^r\beta}$ is equal to the distance of $\beta$ to $\mathcal{B}_I(r)$ by definition of $T^r\beta$. The fourth step invokes Assumption \ref{Assumption boundary BI} and in the last step we define $\tilde{d}_\text{max}$ such that for $d(\beta, \mathcal{B}_I(r)) < \tilde{d}_\text{max}$ the term of order $d(\beta, \mathcal{B}_I(r))^2$ is smaller than $c \times d(\beta, \mathcal{B}_I(r))$. Lastly, we defined $\tilde{c} = 2c$. Finally, let $d_\text{max} = \min(1, \tilde{d }_\text{max})$. We obtain
	\begin{equation*}
		S(\mathbb{E}[m(W, \beta)], \Sigma(\beta)) \geq \tilde{c} \min\{d_\text{max}, d(\beta, \mathcal{B}_I(r))\}^2
	\end{equation*}
\end{proof}

\begin{theorem} \label{theorem: Continuity of limiting distribution}
	For any sequence $\{r_n\}_n \subset \tilde{\mathcal{L}}_0(c)$ with limit $r_n \to r$, let $\gamma^S(1 - \alpha)$ be the $(1 - \alpha)$-th quantile of
	\begin{equation*}
		T^* = \inf_{\beta \in \mathcal{B}(r)} S(\nu_\Omega(\beta) + l(\beta), \mathbb{I}),
	\end{equation*}
	where $\nu_\Omega: \mathcal{B} \to \mathbb{R}^{2J}$ is a tight, zero-mean Gaussian process with correlation kernel $\Omega$, $\mathbb{I}$ is a $(2J \times 2J)$ identity matrix and $l(\beta) = \lim_{n \to \infty}l_n(\beta)$, with
	\begin{equation*}
		l_n(\beta): \mathcal{B} \to \mathbb{R}^{2J}: l_{n, j}(\beta) = \sqrt{n}\sigma_j^{-1/2}(\beta)\mathbb{E}[m_j(W, \beta)].
	\end{equation*}
	The following statements hold:
	\begin{enumerate}
		\item[{\ref{theorem: Continuity of limiting distribution}.1}] If $\gamma^S(1 - \alpha) > 0$, the distribution of $T^*$ is continuous at $\gamma^S(1 - \alpha)$.
		\item[\ref{theorem: Continuity of limiting distribution}.2] If $\gamma^S(1 - \alpha) = 0$, $\liminf_{n \to \infty} \mathbb{P}(T_n(r_n) = 0) \geq 1 - \alpha$, where $T_n(r)$ is as in \eqref{eq: test statistic}.
	\end{enumerate}
\end{theorem}
\begin{proof}
	We start by noting that the $j$-th component $l_j$ of the function $l$ takes the following form:
	\begin{equation*}
		l_j(\beta) = \begin{cases}
			+\infty & \text{if } \mathbb{E}[m(W, \beta)] > 0,\\
			0 & \text{if } \mathbb{E}[m(W, \beta)] = 0,\\
			-\infty & \text{if } \mathbb{E}[m(W, \beta)] < 0.\\
		\end{cases}
	\end{equation*}
	This leads to the following expression for $T^*$:
	\begin{equation*}
		T^* \in \begin{cases}
			\{0\} & \text{if } \exists \beta \in \mathcal{B}(r): \forall j \in \{1, \dots, 2J\}: \mathbb{E}[m_j(W, \beta)] > 0,\\
			\{+\infty\} & \text{if } \forall \beta \in \mathcal{B}(r): \exists j \in \{1, \dots, 2J\}: \mathbb{E}[m_j(W, \beta)] < 0,\\
			[0, +\infty) & \text{otherwise}.
		\end{cases}
	\end{equation*}
	As a working assumption, let $\tilde{\mathcal{L}}_0(c)$ be closed (if not, we can approximate $\tilde{\mathcal{L}}_0(c)$ arbitrarily well by a closed subset), so that $r \in \tilde{\mathcal{L}}_0(c)$. By Assumption \ref{Assumption sufficient slackness}, it then follows that $T^* \equiv 0$, so that $\gamma^S(1 - \alpha) = 0$. As a consequence, statement 4.1 is vacuously true. To show statement 4.2, we first define $\beta_n \in \mathcal{B}(r_n), n \in \mathbb{N}$ as the parameter vector for which Assumption \ref{Assumption sufficient slackness} is satisfied. Next, we make the following computations:
	\begin{align*}
		& \liminf_{n \to \infty} \mathbb{P}(T_n(r_n) = 0)\\
		& \quad = \liminf_{n \to \infty} \mathbb{P}\left(\inf_{\beta \in \mathcal{B}(r_n)} S(\hat{D}^{-1/2}(\beta)\sqrt{n}\bar{m}_n(\beta), \mathbb{I}) = 0\right)\\
		& \quad \geq \liminf_{n \to \infty} \mathbb{P}\left( S(\hat{D}^{-1/2}(\beta_n)\sqrt{n}\bar{m}_n(\beta_n), \mathbb{I}) = 0\right)\\
		& \quad = \liminf_{n \to \infty} \mathbb{P}\left(\forall j \in \{1, \dots, 2J\}: (\bar{m}_n(\beta_n))_j \geq 0\right)\\
		& \quad = \liminf_{n \to \infty} \mathbb{P}\left(\forall j \in \{1, \dots, 2J\}: \mathbb{E}[m_j(W, \beta_n)] - (\bar{m}_n(\beta_n))_j \leq \mathbb{E}[m_j(W, \beta_n)]\right)\\
		& \quad \geq \liminf_{n \to \infty} \mathbb{P}\left(\forall j \in \{1, \dots, 2J\}: \mathbb{E}[m_j(W, \beta_n)] - (\bar{m}_n(\beta_n))_j \leq \eta_\mathcal{L}]\right)\\
		& \quad = 1
	\end{align*}
	The first line is straight-forward. In the second line follows from the fact that $\beta_n \in \mathcal{B}(r_n)$. The third line uses that $S(\cdot, \mathbb{I})$ is equal to zero if and only if each of its $2J$ arguments are positive. The fourth line is again straight-forward and the fifth invokes Assumption \ref{Assumption sufficient slackness}. Finally, the sixth line follows from the classes $\{m_j(\cdot, \beta) \mid \beta \in \mathcal{B}\}$ being Glivenko-Cantelli, which can be proved using similar techniques as in Theorem \ref{theorem: Donsker}.
\end{proof}

\subsection{Proof of theorem \ref{theorem: main}} \label{sec: proof of main theorem}
Below we restate our main theorem for readability and provide a proof.

\textbf{Theorem \ref{theorem: main} (repeated)} With $\tilde{\mathcal{L}}_0(c)$ as in \eqref{eq: parameter set correct inference} and under Assumptions \ref{Assumption iid}--\ref{Assumption sufficient slackness} it holds that each element in $\tilde{\mathcal{L}}_0(c)$, in particular $\beta_{\text{true}, k}$, will be contained in $\hat{\mathcal{B}}_{I, k}$ with at least the specified confidence level $1-\alpha$.

\begin{proof}
    The proof can be constructed using the same ideas as in Theorem 1 of \cite{Bei2024}, replacing the set $\mathcal{L}_0$ in their proof with the set $\tilde{\mathcal{L}}_0(c)$ of our paper. For completeness, we list each of the assumptions used in their paper alongside the corresponding assumptions/theorems/lemmas of this paper. Importantly, Assumption 2 in \cite{Bei2024} is not shown to hold. Instead, it is replaced by our Assumption \ref{Assumption boundary BI} and Theorem \ref{theorem: Polynomial minorant}.

    \begin{center}
	\begin{tabular}{|r|l|}
		\hline
		Assumption & Result(s)\\
		\hline
		$1$ & Assumption \ref{Assumption parameter space}\\
		$2^*$ & Theorem \ref{theorem: Polynomial minorant} (using Assumption \ref{Assumption boundary BI})$^*$\\
		$3$ & Lemma \ref{lemma: G matrix}\\
		$4$ & [holds by construction; cf. Section 4.1 in \cite{Bei2024}]\\
		$5$ & [holds by construction; cf. Section 3.2 in \cite{Bei2024}]\\
		$6$ & Assumptions \ref{Assumption iid}, \ref{Assumption variance of moments} + Lemmas \ref{lemma: F is Borel}, \ref{lemma: existence 2+c moment}, \ref{lemma: unform convergence correlation matrix} + Theorem \ref{theorem: Donsker} and Corollary \ref{corrolary: Asymptotic equicontinuity}\\
		$7$ & Theorem \ref{theorem: Continuity of limiting distribution}\\
		$8$ & [holds by construction] + Lemma \ref{lemma: Lower semicontinuity}\\
		$9$ & [holds by construction]\\
		\hline
        \multicolumn{2}{l}{$^*$ The assumption of Bei is not implied by our result. Instead, it is replaced by it.}
	\end{tabular}
\end{center}
\end{proof}

\section{Estimation algorithm and implementation details}\label{supp: Algorithm and implementation details}
This section describes the algorithm used to compute
\begin{equation}\label{eq: BIk_hat}
	\hat{\mathcal{B}}_{I, k} = \left\{r \in \mathcal{B}_k \mid T_n(r) \leq \gamma_{n, 1 - \alpha}(r)\right\}.
\end{equation}

Formulation \eqref{eq: BIk_hat} requires the computation of the test statistic and corresponding critical value for each $r \in \mathcal{B}_k$, which is not directly feasible when $\mathcal{B}_k$ is an infinite set. Typically, one proceeds by checking the condition in \eqref{eq: BIk_hat} for a finite grid in $\mathcal{B}_k$, but this is computationally intensive when a high accuracy is required. Alternatively, one could rewrite the condition as $T_n(r) - \gamma_{n, 1 - \alpha}(r) \leq 0$ and transform the problem to finding the roots of the violation curve $V(r) = T_n(r) - \gamma_{n, 1 - \alpha}(r)$. If $\mathcal{B}_{I, k}$ is an interval, we can equivalently write
\begin{equation*}
	\hat{\mathcal{B}}_{I, k} = \left[\min_{r \in \mathcal{B}_k: V(r) = 0} r, \max_{r \in \mathcal{B}_k: V(r) = 0} r\right].
\end{equation*}
This raises the question: will $\mathcal{B}_{I, k}$ always be an interval? When all covariates are binary and $\mathcal{G}_\text{disc}$ and/or $\mathcal{G}_\text{cat}$ are used, it can be shown that the region of parameters satisfying each moment restriction will be halfspaces, implying that $\mathcal{B}_I$ is convex and hence $\mathcal{B}_{I, k}$ is indeed an interval. When there are continuous covariates in the model, example \ref{example: nonconvex identified interval} in Section \ref{supp: Examples} of the Supplementary material shows that $\mathcal{B}_I$ might not be path connected, in which case $\mathcal{B}_{I, k}$ could fail to be an interval. The case when there are continuous covariates in the model will therefore require additional attention. We will first present the algorithm in the case that $\mathcal{B}_{I, k}$ is an interval and then discuss how to proceed when continuous covariates are present.

To start, we define some more notation. At each point in the algorithm, let $E$ be the set of all evaluations of the violation curve that are already computed. Specifically, store each evaluation as the tuple $(r, T_n(r), \gamma_{n, 1-\alpha}(r))$. As we need to search for both the lower and upper bound of the identified interval, it will be convenient to define a variable $dir$ that equals $1$ when searching for the upper bound and $-1$ when searching for the lower bound. Recall that this was equivalent to finding the two roots of the violation curve $V(r)$. Therefore, let $s(E, dir)$ denote any root finding algorithm. More specifically, $s(E, dir)$ returns a candidate value $r_\text{next}$ for the upper or lower root of $V(r)$, depending on $dir$ and the current knowledge about the violation curve stored in $E$. We will discuss different options for $s$ in Section \ref{supp: Root finding algorithms}. Lastly, let $\text{conv}(E, dir)$ be a function that determines convergence of $s$. Algorithm \ref{alg: Main algorithm} presents pseudo-code for the entire estimation routine.

\begin{algorithm}
	\footnotesize
	\caption{Implementation of methodology}\label{alg: Main algorithm}
	\KwIn{$N_\text{init.evals} > 0$}
	\;
	\tcp{Step 0: Initialization}
	$E \gets \emptyset$\;
	$n_\text{evals} \gets 0$\;
	stop $\gets$ \textbf{false}\;
	$dir \gets 1$\;
	\;
	\tcp{Step 1: search for initial feasible point, needed in the root finding algorithm in the second step.}
	\While{$(n_\text{evals} < N_\text{init.evals}) \textbf{ and } (\textbf{not} \text{ stop})$}{
		Select $r_\text{init} \in \mathcal{B}_k$ as the $(n_\text{evals})$-th grid point.\;
		
		Evaluate $T_n(r_\text{init})$ and $\gamma_{n, 1 - \alpha}(r_\text{init})$. Store $(r_\text{init}, T_n(r_\text{init}), \gamma_{n, 1-\alpha}(r_\text{init}))$ in $E$.\;
		\;
		\uIf{$T_n(r_\text{init}) - \gamma_{n, 1 - \alpha}(r_\text{init}) \leq 0$}{
			$\text{stop} \gets \textbf{true}$
		}
		\Else{$n_\text{evals} \gets n_\text{evals} + 1$}
	}
	\;
	\tcp{If no initial point was found, conclude model misspecification and stop algorithm.}
	\If{$n_{evals} = N_\text{init.evals}$}{\textbf{stop}("Model is misspecified"))}
	\;
	\tcp{Step 2: root finding algorithm}
	$converged \gets \textbf{false}$\;
	\While{\textbf{not} converged}{
		$r_\text{next} \gets s(E, dir)$\;
		Evaluate $T_n(r_\text{next})$ and $\gamma_{n, 1 - \alpha}(r_\text{next})$. Store $(r_\text{next}, T_n(r_\text{next}), \gamma_{n, 1-\alpha}(r_\text{next}))$ in $E$.\;
		\;
		$converged \gets \text{conv}(E, dir)$\;
		\If{$converged$}{
			\tcp{If upper bound found, start search for lower bound. Otherwise, leave the while loop.}
			\If{$dir = 1$}{
				$converged \gets \textbf{false}$\;
				$dir \gets -1$
			}
		}
	}
	\;
	\tcp{Step 3: compute the bounds}
	$E_\text{feas} \gets [\text{subset of } E \text{ for which } T_n(r) \leq \gamma_{n, 1-\alpha}(r)]$\;
	$LB \gets \min(r : (r, T_n(r), \gamma_{n, 1 - \alpha}) \in E_\text{feas})$\;
	$UB \gets \max(r : (r, T_n(r), \gamma_{n, 1 - \alpha}) \in E_\text{feas})$\;
	\;
	\KwOut{$[LB, UB]$}
\end{algorithm}

When there are continuous covariates in the model, we need to take into account that $\mathcal{B}_{I, k}$ takes the general form $\mathcal{B}_{I, k} = \bigcup_{i \in \mathcal{I}} [r_{l, i}, r_{u, i}]$ for some index set $\mathcal{I}$ and $r_{l, i} \leq r_{u, i}$. In this case, we have to slightly adapt algorithm \ref{alg: Main algorithm}. Specifically, when searching for an initial feasible point (see step 1 in Algorithm \ref{alg: Main algorithm}), we should no longer stop after a single point $r_\text{init}$ is found but rather thoroughly scan $\mathcal{B}_k$ to possibly find multiple starting values. Steps 2 and 3 can then be run (in parallel) for each starting value that was found in step 1. An additional fourth step computes and returns the union of each interval obtained in step 3. We reiterate that cases like Example \ref{example: nonconvex identified interval} are rather pathological.

Lastly, we remark the influence of the input parameter $N_\text{init.evals}$. It determines the granularity of the grid on which to search for an initial feasible point, and should be taken large enough since otherwise, the model might be incorrectly determined to be misspecified. In our implementation, we set $N_\text{init.evals} = 100$.

\subsection{Root finding algorithms} \label{supp: Root finding algorithms}
The root finding algorithm $s(E, dir)$ could take many forms. As a simple option, one can select it to be the binary search algorithm. Given points $r_l$, $r_u$ such that $V(r_l) < 0$ and $V(r_u) > 0$, binary search works by evaluating $r_m = (r_l + r_u)/2$. If $V(r_m) = 0$, the root has been found. Otherwise, if $V(r_m) < 0$, redefine $r_l = r_m$ and if $V(r_m) > 0$, redefine $r_u = r_m$. Iterating this procedure, $r_m$ will converge to a root of $V(\cdot)$. Moreover, as the maximum distance from $r_m$ to the root of $V(\cdot)$ is halved at each iteration, the algorithm will have logarithmic time complexity.

The most important criterion when selecting $s$ for this methodology should be its efficiency in terms of function evaluations. That is, one should look for a root finding algorithm that can approximate the true root to the desired accuracy using the least amount of evaluations of $V(\cdot)$ as possible. The reason is that evaluating $T_n(r)$ and $\gamma_{n, 1 - \alpha}(r)$ even for a single value of $r$ is computationally costly, so we want to avoid it as much as possible. In this regard, binary search as described above can be a simple yet potent option. A more complex algorithm that has been used in this context is the Estimation-Approximation-Maximization (EAM) algorithm \citep{Kaido2019}. It works by approximating the violation curve using a Kriging model, based on which a more accurate candidate root $r_m$ can be proposed, speeding up convergence. A middle ground between binary search and the EAM algorithm could be interpolation search, which uses a simple piecewise linear model to approximate $V(\cdot)$ and proceeds like binary search, letting $r_m$ be the root of the approximation.

In preliminary simulations we compared binary search against the EAM algorithm. Surprisingly, the binary search algorithm proved superior, at least given our implementation of the EAM algorithm. For this reason, our estimation algorithm is implemented using binary search when applying it throughout the simulations and data applications.

\subsection{Implementation in R}
A user-friendly framework implementing the proposed methodology of this paper is made available under the package \texttt{depCensoring} in R \citep{R:depCensoring}. In this section, we will elaborate on its implementation. The full documentation, including example code, can be found on CRAN.

The implementation can be divided into several core components. Firstly, the MatLab code provided by \cite{Bei2024} to perform their test is translated into the R programming language. Since their code is only written for separable moment functions of the form $m(W, \beta) = m_1(W) + m_2(\beta)$, we extend it to the general case. We also note that the computation of both the test statistic and its critical value include optimization problems. To this end, several algorithms are available in R under the function \texttt{optim}, and the \texttt{nloptr} package greatly extends this assortment. We find that our implementation using \texttt{optim} can fail to find the global minimum, and therefore we recommend to use the NEWUOA \citep{Powell2006} method provided in \texttt{nloptr}, which, despite being only a local optimization scheme, appears to have no problem finding the global minimum while enjoying great computational advantages over global optimization methods.

Besides the implementation of the test, we also require a root finding algorithm. We provide implementations of standard algorithms like grid search, binary search and interpolation search, as well as the more advanced EAM algorithm described in \cite{Kaido2019}.

Lastly, we extend the framework to include our methodology under the extra assumption of time-independent effects of covariates, as described in Section \ref{sec: Time-independent effects of covariates}.

The implementation is designed to be highly customizable. Moreover, the modularity of the code should allow users to adapt it to other settings with relative ease. Given the substantial computations required in running our method, parallelization is implemented both in steps 1 and 2 of Algorithm \ref{alg: Main algorithm}. Moreover, if many computing cores are available, convergence of the algorithm can be sped up by considering a large number of initial points $N_\text{init.evals}$ and not stopping after a single feasible point has been found, since the initial point search can be done completely (\emph{embarrassingly}) parallel. A visualization of the estimation procedure is provided as well, which can help new users to understand the inner workings of the algorithm. Lastly, we precede the procedure with an extensive input validation, ensuring that our implementation is operated as intended, and hence avoiding accidental misuse.

\section{Future research directions} \label{supp: future research directions}

This section supplements Section \ref{sec: Discussion} with several additional topics, focusing on aspects that are not addressed in our work and which could be fruitful topics for future research.

\textbf{Survival probabilities.} As a by-product of the proposed approach, it is possible to obtain identified sets for the conditional probabilities of survival. For a given value of the covariate, $X = x$, all possible values for $x^\top \beta$ can be computed based on the identified interval of each element in the parameter vector. Evaluating $\Lambda(\cdot)$ at each possible value of $x^\top \beta$ and collecting the results, we obtain the identified set for the probability of survival at time $t$, conditionally on $X = x$. Precisely,
\begin{equation}\label{eq: identified set probabilities}
	\hat{\mathcal{S}}_{I, x} = \left\{1 - \Lambda(x^\top \beta) \mid \beta \in \hat{\mathcal{B}}_{I, 0} \times \hat{\mathcal{B}}_{I, 1} \times \dots \times \hat{\mathcal{B}}_{I, d}\right\}.
\end{equation}
Further thought reveals that this approach is rather naive as is. Firstly, (i) if each identified set $\mathcal{B}_{I, k}$ is estimated at a level $\alpha$, the level of $\hat{\mathcal{S}}_{I, x}$ could be larger than $\alpha$ due to the induced multiple testing issue. On the other hand, (ii) by essentially taking the Cartesian product of all estimated identified intervals as an estimator for $\mathcal{B}_I$, it is possible that the estimator $\hat{\mathcal{S}}_{I, x}$ is (very) conservative. Indeed, in an extreme case, suppose $\mathcal{B}_I$ is the line segment $\{\beta \in \mathcal{B}: 0 \leq \beta_0 \leq 1, \beta_0 = \beta_1 = \dots = \beta_d\}$, then $\mathcal{B}_{I, k} = [0, 1]$ for all $k \in \{0, \dots, d\}$, yet $\mathcal{B}_{I, 0} \times \dots \times \mathcal{B}_{I, d}$ is a gross overestimation of $\mathcal{B}_I$. Therefore, so will be $\hat{\mathcal{B}}_{I, 0} \times \dots \times \hat{\mathcal{B}}_{I, d}$ as an estimator for $\hat{\mathcal{B}}_I$. Lastly, (iii) even if $\mathcal{B}_I$ could be estimated at a correct level, $\hat{\mathcal{S}}_{I, x}$ might still be conservative: with probability $\alpha$ it will hold that $\beta \in \mathcal{B}_I$ is not an element of $\hat{\mathcal{B}}_I$. However, it is still possible that $\Lambda(x^\top \beta) \in \hat{\mathcal{S}}_{I, x}$, implying that the level of $\hat{\mathcal{S}}_{I, x}$ is at most $\alpha$.

To make the previous idea into a valid approach, some adaptations are required. The first point is easily addressed by applying Bonferroni correction when estimating the identified intervals $\mathcal{B}_{I, k}$. That said, it is widely known that Bonferroni correction can be crude, so much can be gained by estimating $\mathcal{B}_I$ directly instead of focusing on each component separately. To this end, many methods exist in the literature. One possible approach would be to use the conditional moment inequality test of \cite{AndrewsShi2013}. While this now addresses the first and second point of the previous paragraph, it does not address the last one and hence might still lead to conservative estimates.

\textbf{Cure models.} It is noteworthy that despite potentially being conservative, the estimator in \eqref{eq: identified set probabilities} could still be informative if the bounds on each of the coefficients are small enough. One particularly interesting case would occur when the chosen time point of interest $t$ is such that it can be safely assumed that no events of interest can take place beyond this time point. That is, if a subject does not experience the event by time $t$, it can be assumed that they never will. These subjects are called \emph{cured}, alluding to the typical application of this context when studying the time until relapse of a certain disease (e.g. cancer). Much like the setting with dependent censoring, many models in survival analysis that can deal with this type of data (so-called \emph{cure models}) rely on stringent assumptions. The estimator \eqref{eq: identified set probabilities} would allow one to find bounds on the probability of being cured in an assumption-lean manner.

\textbf{Further extensions.} Based on the previous discussion, it is natural to ask how the methodology can be extended in other directions. For example, in some practical settings it might be known that any dependence between $T$ and $C$ should be positive, or it might occur that there is additional left truncation and/or missingness in the data on top of the random right censoring. While these extensions seem worthwhile for future research, we note that the extra information or generalizations in the given examples will change the form of the original bounds in \eqref{eq: Peterson bounds}. This can be problematic, since all of the methodology described in this paper relies on the fact that the bounds in \eqref{eq: Peterson bounds} can be transformed to (un)conditional moment inequalities. Without this instrumental property, a different approach would have to be developed.

\section{Simulations} \label{supp: Simulations}

This section contains the results of the additional simulations discussed in Section \ref{sec: Summary of additional simulations}, presented here in Tables \ref{tab: moreIF, Cox}, \ref{tab: almost no censoring, Cox} and \ref{tab: results multiple testing Cox}, as well as detailed discussions thereof. It also contains analogous analyses pertaining to the model using the AFT link as described in Section \ref{supp: Link function}. All tables are presented at the end of this document.

\subsection{Main simulation} \label{supp: Main simulation}

First, we consider the same simulation as described in Section \ref{sec: Simulation study}, this time under the AFT link. The results are tabulated in Table \ref{tab: main AFT}, which can be interpreted in the same way as Table \ref{tab: main Cox}. From Table \ref{tab: main AFT}, many of the same conclusions can be drawn as already presented in Section \ref{sec: Simulation study}. A noteworthy difference with the simulations using the Cox link is that the model using the AFT link function in general seems to be more stable.

The traditional proportional odds model assuming independence was estimated alongside the identified intervals. As expected, the estimates from these models are contained in the identified intervals in nearly all repetitions, over all designs. Therefore, for brevity, the results are not taken up in the table below. For the same reason, we did not include this information in Table \ref{tab: main Cox} either.

\subsection{True bounds} \label{supp: True bounds}
For some simulation designs in the main simulation, the bounds on the coefficients can be wide. This section aims to answer the question whether the width of the estimated bounds can be attributed to the corresponding true identified set being wide, or whether it is due to sampling variability. To this end, we compute an approximation of the true identified set for the coefficient of interest, for each of the considered designs in the main simulation.

\begin{figure}
    \centering
    \includegraphics[width=\linewidth]{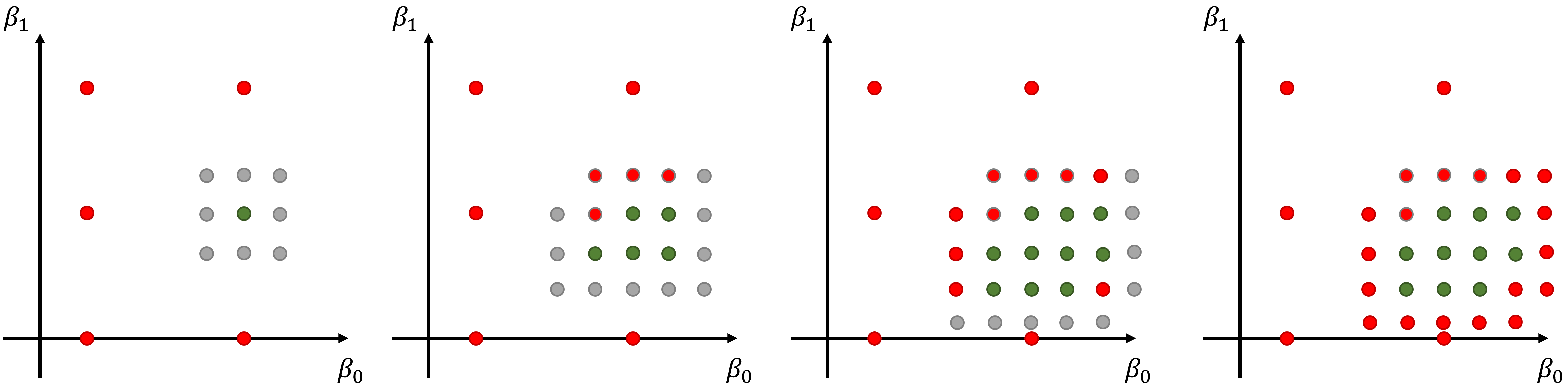}
    \caption{Illustration of the adaptive grid search algorithm. From left to right, each panel illustrates a consecutive iteration of the algorithm. Red points indicate parameter vectors that have been determined to lie outside of $\mathcal{B}_I$, green points correspond to parameter vectors which have been determined to lie inside of $\mathcal{B}_I$ and gray points indicate parameter vectors to be considered in the next iteration. Note that in the final panel (iteration 4) no gray points remain and hence the algorithm stops.}
    \label{fig: adaptive grid search}
\end{figure}

Our method is as follows. First, we generate a large data set $(n = \num{50000})$ according to the design under investigation. Next, we consider a grid of coefficients in $\mathcal{B}$ and, for each point in the grid, compute the unconditional moment restrictions in Equation \ref{eq: unconditional moment inequalities} by Monte Carlo integration. After an initial point in $\mathcal{B}_I$ has been found (called a feasible point), the grid is iteratively refined by expanding it around feasible points, until the elementwise errors are smaller than $0.05$. Figure \ref{fig: adaptive grid search} illustrates this procedure. For each considered design, we repeat the process for $5$ generated data sets and report the average results. Given the small variance in the obtained bounds and the significant computational burden, we argue that this modest number of repetitions suffices.

The results are shown in Table \ref{tab: true bounds}. Comparing the bounds in this table with the results of the main simulations in Tables \ref{tab: main Cox} and \ref{tab: main AFT} it can be seen that the true bounds are contained within the estimated ones. Moreover, proportionally to the true bounds, the estimated bounds are not substantially wider than the true ones, apart from some exceptions in the case of heavy censoring. This leads us to conclude that the width of the estimated bounds can mostly be attributed to the width of the true bounds.

We note that due to computational complexities, the identified set of the design with AFT link, 65\% censoring and negative dependence between $T$ and $C$ could only be computed up to an error smaller than $0.10$. We also remark that identified intervals for $\beta_2$ are wider than those for $\beta_1$. This is due to $\beta_2$ pertaining to a discrete covariate, which generally contains less information in comparison to a continuous covariate.

\begin{table}[ht]
    \centering
    \begin{tabular}{ccccc}
        \toprule
        Link & $\%$ cens. & dependence & $\mathcal{B}_{I, 1}$ & $\mathcal{B}_{I, 2 }$\\
        \cmidrule{1-5}
        \multirow{6}{*}{AFT} & \multirow{3}{*}{$30\%$} & Indep. & [0.89, 1.24]  & [-1.15, -0.72]\\
        & & Pos. dep. & [0.93, 1.24]  & [-1.11, -0.74] \\
        & & Neg. dep. & [0.90, 1.24]      & [-1.13, -0.72]\\
        \cmidrule{2-5}
        & \multirow{3}{*}{$65\%$} & Indep. & [0.27, 1.30]  & [-2.29, 0.72] \\
        & & Pos. dep. & [0.70, 1.30]   & [-1.38, -0.61]\\
        & & Neg. dep. & [0.00, 2.02]     & [-4.18, 2.50]  \\
        \cmidrule{1-5}
        \multirow{6}{*}{Cox} & \multirow{3}{*}{$30\%$} & Indep. & [0.64, 1.24]  & [-1.18, -0.49]\\
        & & Pos. dep. & [0.86, 1.26]   & [-1.12, -0.72] \\
        & & Neg. dep. & [0.63, 1.24]     & [-1.26, -0.45]\\
        \cmidrule{2-5}
        & \multirow{3}{*}{$65\%$} & Indep. & [0.11, 1.28]  & [-2.04, 0.72] \\
        & & Pos. dep. & [0.46, 1.39]  & [-1.58, -0.21]\\
        & & Neg. dep. & [-0.02, 1.32] & [-2.64, 1.62] \\
        \bottomrule
    \end{tabular}
    \caption{Approximations of the true identified intervals of $\beta_1$ and $\beta_2$ in the designs considered in the main simulations.}
    \label{tab: true bounds}
\end{table}
 
\subsection{More instrumental functions} \label{supp: More instrumental functions}

Based on the simulations presented in Tables \ref{tab: main Cox} and \ref{tab: main AFT} (referred to as the \emph{main simulations} throughout), one could wonder whether a further increase in instrumental functions can improve the informativeness and stability of the bounds. Tables \ref{tab: moreIF, Cox} and \ref{tab: moreIF, AFT} show the results of the designs considered in the main simulations with the Cox and AFT link, respectively, but using a larger number of instrumental functions. For completeness, we also performed the simulations in a setting with almost no censoring. The results of these will be discussed in Section \ref{supp: Almost no censoring}.

Comparing the results of Tables  \ref{tab: moreIF, Cox} and \ref{tab: moreIF, AFT} to the ones from the main simulation, specifically when $N_{IF} = 20$, the average bounds can be seen to narrow, but often only slightly. On the other hand, the average bounds obtained using $N_{IF} = 30$ can sometimes widen compared to those obtained using a smaller $N_{IF}$. This illustrates the remark made in Section \ref{sec: Discussion}, cautioning against taking the class $\mathcal{G}$ too large. This point is further substantiated by comparing the variance of the bounds between the results using $N_{IF} = 20$ and $N_{IF} = 30$, which can be seen to increase in some designs. Moreover, the significance and coverage rates can drop slightly. We conclude that an increase to $20$ instrumental functions can have a positive effect (at the cost of a heavier computational burden), while a further increase to $30$ instrumental functions does not seem worthwhile and can even be harmful in some cases. A notable difference is that the AFT link seems to suffer more from this issue than the Cox link does.

\subsection{Almost no censoring} \label{supp: Almost no censoring}

We again consider the same designs as in the main simulations, now in a case with almost no censoring ($\sim 2\%$ censoring). In this setting, the identified intervals are very small, making it difficult for step 1 of Algorithm \ref{alg: Main algorithm} to find an initial feasible point and as a consequence, misspecification of the model was (incorrectly) concluded in $2-12\%$ of repetitions, varying across designs. We therefore recommend to increase $N_\text{init.eval}$ when applying our model to data with a very low censoring percentage.

The results are shown in Tables \ref{tab: almost no censoring, Cox} and \ref{tab: almost no censoring, AFT}. From these tables it can be seen that the bounds are indeed very narrow in all cases. This is to be expected, since little uncertainty about $T$ remains in this setting. It should be noted that the coverage values can be substantially below their nominal value, especially when a larger class of instrumental functions is used (cf. Tables \ref{tab: moreIF, Cox} and \ref{tab: moreIF, AFT}). Further investigation reveals that in the cases where the true value was not included in the estimated identified interval, it was only excluded barely so, often lying within a distance of $0.05$. Moreover, the coverage problem disappears when sample size increases.

That said, it is clear that the necessary caution should be exercised when the sample size is small. In this setting, we hence recommend to supplement our method with a sensitivity analysis based on (point identified) copula models \citep{HuangZhang2008}. If the sensitivity analysis concludes that the results remain similar under varying degrees of dependence, point identified models assuming known dependence (independence) will provide fast and approximately correct results, and might hence be preferred over the methodology presented here.

\subsection{Time-independent effects of covariates} \label{supp: Time-independent effects of covariates simulation}

Section \ref{sec: Time-independent effects of covariates} discusses an approach to obtain more informative bounds if one is willing to assume that covariate effects are time-independent. In this case, the methodology described in our paper could be applied over different time points, and the obtained identified intervals can subsequently be intersected or combined via a majority voting rule. As already noted in the aforementioned section, the identified intervals at each of the considered time points should then be estimated with an appropriately corrected type-I error. 

The designs of the simulations considered in this section are mostly the same as in the main simulation. To reduce computational cost, we only consider a sample size of $n = 1000$ and run $200$ repetitions of each design. Based on the recommendation in Section \ref{sec: Summary of additional simulations}, we use eight instrumental functions for the continuous covariate and two for the binary covariate, resulting in a total of $N_{IF} = 16$ instrumental functions after combining them via Equation \eqref{eq: standard instrumental function}.

The results of the simulations are shown in Tables \ref{tab: results multiple testing Cox} and \ref{tab: results multiple testing AFT}. For brevity, the settings of the main simulation have been encoded in pairs $a-b$, where $a = 1$ and $a = 2$ refer to light ($30\%$) and heavy $(65\%)$ censoring, respectively, and $b = 1$, $b = 2$ and $b = 3$ refer to the cases in which $T$ and $C$ are independent, positively dependent and negatively dependent, respectively. We consider two time points of interest. To follow the same line as the previous simulations, we first let $t = 1$. Next, we also investigate $t = 5$. This latter time point is interesting to investigate since Peterson's bounds widen as time increases, and hence obtaining informative bounds on the covariate effects becomes more difficult. Therefore, combining identified intervals over different time points might be particularly useful when $t$ lies in or near the right tail of the distribution of $Y$. In our case, $t = 5$ lies between the $80^\text{th}$ and $99^\text{th}$ quantile of $Y$, varying across the simulation designs, and in some designs even exceeds the largest observation.

When $t = 1$, we consider additional time points at $0.333$ and $0.667$ when using the intersection or majority voting method. For $t = 5$, we additionally use time points $1$ and $3$. It should be noted that both the intersection method as well as the majority vote method will always again result in an identified \emph{interval} (in view of the remark made in Section \ref{supp: Algorithm and implementation details} regarding Example \ref{example: assumption boundary BI}; Lemma \ref{lemma: Convexity of combined intervals}). Therefore, reporting $\hat{\mathcal{B}}_{I, 1}$ using its (average) upper and lower bound analogous to the previous simulations is sensible.

From Tables \ref{tab: results multiple testing Cox} and \ref{tab: results multiple testing AFT} it can be seen that both the intersection method as well as the majority vote method can improve upon the estimation algorithm solely applied to a single point. We do emphasize that this improvement comes at the cost of having to assume that $\beta_1$ is time-independent. Notably, in the designs with heavy censoring, the intersected bounds can improve the informativeness of the bounds substantially. Also the majority vote method appears to be able to achieve this, though at least in this simulation to a far lesser extent.

Finally, we remark that the bounds listed under \emph{Intersection} and \emph{Majority vote} for a given setting and either $t = 1$ or $t = 5$ are by construction assumed to hold for all time points, and are hence directly comparable. Such a comparison yields that the bounds constructed using time points $0.333$, $0.667$ and $1$ are always more informative than the ones obtained when using $1$, $3$ and $5$. This confirms the remark made in Section \ref{sec: Time-independent effects of covariates} that it can be useful to include early time points when applying our methodology under the assumption of time-independent coefficients.

Comparing Tables \ref{tab: results multiple testing Cox} and \ref{tab: results multiple testing AFT}, one can notice that the Single point and Majority vote method perform substantially better for the model using the AFT link. This is explained by the fact that $t = 5$ lies further in the right tail of $Y$ under the Cox link than the AFT link: for the model using the Cox link, $t = 5$ always exceeds the $90^\text{th}$ quantile.

\subsection{Dependent covariates} \label{supp: Dependent covariates}

We investigate the effect of possible dependence between covariates on the results of the method. Similar to the main simulations, we consider the setting of two covariates, $X_1 \sim \mathcal{N}(0, 1)$ and $X_2 \sim Ber(0.5)$. However, we now introduce dependence between $X_1$ and $X_2$ through a Gaussian copula with correlation parameter equal to $0.8$. Moreover, to ease the computational burden, we restrict to running $200$ repetitions of each design. All other aspects are left unchanged with respect to the main simulation.

In the context of dependent covariates, Section \ref{sec: Instrumental functions} discusses the possible problem of instrumental functions having empty intersection with the observed covariate space, alongside a method to resolve the issue. However, the solution is not applicable to this simulation as it assumes that the dependent covariates are all continuous. Therefore, we opt to remove instrumental functions whose support does not intersect the observed covariate space from the analysis. Preliminary simulation showed that this is indeed a necessary step in order to prevent a violation of Assumption \ref{Assumption variance of moments}.

From Tables \ref{tab: Dep cov. Cox. 1}, \ref{tab: Dep cov. Cox. 2}, \ref{tab: Dep cov. AFT. 1} and \ref{tab: Dep cov. AFT. 2} it can be seen that the results of the simulations under dependent covariates are similar to the ones in their counterparts that consider independent covariates (Tables \ref{tab: main Cox}, \ref{tab: moreIF, Cox}, \ref{tab: main AFT} and \ref{tab: moreIF, AFT}), though generally the bounds tend to be slightly wider in the case considered here. The widening of the bounds can be attributed to several causes, such as dependent covariates having less variance, or the removal of several instrumental functions from the analysis. Regardless, we view these results as satisfactory and as a validation of the proposal to remove certain instrumental functions from the analysis.



\begin{table}[ht]
	\centering
	\begin{tabular}{ccccccccccc}
		\toprule
		& & \multicolumn{4}{c}{$N_{IF} = 20$} & \phantom{c} & \multicolumn{4}{c}{$N_{IF} = 30$}\\
		\cmidrule{3-6} \cmidrule{8-11}
		Setting & $n$ & Bounds & Var & Sig & Cov & & Bounds & Var & Sig & Cov\\
		\midrule
		\multirow{3}{*}{\shortstack[c]{Indep.\\ $\sim 30\%$ cens.}} & 500 & [0.54, 1.56] & 0.03 & 1.00 & 1.00 && [0.59, 1.60] & 0.04 & 1.00 & 0.99\\
		& 1000 & [0.54, 1.37] & 0.01 & 1.00 & 1.00 && [0.59, 1.40] & 0.02 & 1.00 & 1.00\\
		& 2000 & [0.55, 1.28] & 0.01 & 1.00 & 1.00 && [0.57, 1.29] & 0.01 & 1.00 & 1.00\\
		\midrule
		\multirow{3}{*}{\shortstack[c]{Pos. dep.\\ $\sim 30\%$ cens.}} & 500 & [0.70, 1.68] & 0.05 & 1.00 & 0.96 && [0.78, 1.74] & 0.06 & 1.00 & 0.89\\
		& 1000 & [0.71, 1.50] & 0.02 & 1.00 & 0.99 && [0.76, 1.53] & 0.03 & 1.00 & 0.94\\
		& 2000 & [0.72, 1.39] & 0.01 & 1.00 & 1.00 && [0.76, 1.40] & 0.01 & 1.00 & 0.98\\
		\midrule
		\multirow{3}{*}{\shortstack[c]{Neg. dep.\\ $\sim 30\%$ cens.}} & 500 & [0.51, 1.64] & 0.03 & 1.00 & 1.00 && [0.55, 1.70] & 0.04 & 1.00 & 1.00\\
		& 1000 & [0.52, 1.43] & 0.01 & 1.00 & 1.00 && [0.53, 1.47] & 0.02 & 1.00 & 1.00\\
		& 2000 & [0.52, 1.33] & 0.01 & 1.00 & 1.00 && [0.55, 1.35] & 0.01 & 1.00 & 1.00\\
		\midrule
		\multirow{3}{*}{\shortstack[c]{Indep.\\ $\sim 65\%$ cens.}} & 500 & [-0.15,  2.81] &  0.34 &  0.11 &  1.00 && [-0.16,  2.85] &  0.30 &  0.10 &  1.00\\
		& 1000 & [-0.07,  2.15] &  0.07 &  0.18 &  1.00 && [-0.06,  2.16] &  0.06 &  0.25 &  1.00\\
		& 2000 & [-0.01,  1.84] &  0.02 &  0.37 &  1.00 && [-0.01,  1.84] &  0.03 &  0.39 &  1.00\\
		\midrule
		\multirow{3}{*}{\shortstack[c]{Pos. dep.\\ $\sim 65\%$ cens.}} & 500 & [0.27, 2.71] & 0.19 & 0.98 & 1.00 && [0.34, 2.79] & 0.20 & 0.98 & 1.00\\
		& 1000 & [0.31, 2.24] & 0.07 & 1.00 & 1.00 && [0.36, 2.25] & 0.07 & 1.00 & 1.00\\
		& 2000 & [0.32, 1.98] & 0.02 & 1.00 & 1.00 && [0.37, 1.97] & 0.03 & 1.00 & 1.00\\
		\midrule
		\multirow{3}{*}{\shortstack[c]{Neg. dep.\\ $\sim 65\%$ cens.}} & 500 & [-0.44,  8.39] &  4.35 &  0.00 &  1.00 && [-0.46,  7.86] &  4.90 &  0.00 &  1.00\\
		& 1000 & [-0.29,  4.68] &  2.96 &  0.00 &  1.00 && [-0.30,  4.08] &  1.18 &  0.01 &  1.00\\
		& 2000 & [-0.21,  2.98] &  0.24 &  0.00 &  1.00 && [-0.21,  2.83] &  0.19 &  0.01 &  1.00\\
		\midrule
		\multirow{3}{*}{\shortstack[c]{Indep.\\ $\sim 2\%$ cens.}} & 500 & [0.92, 1.37] & 0.02 & 1.00 & 0.74 && [0.99, 1.43] & 0.03 & 1.00 & 0.53\\
		& 1000 & [0.92, 1.24] & 0.01 & 1.00 & 0.81 && [0.94, 1.25] & 0.01 & 1.00 & 0.72\\
		& 2000 & [0.93, 1.16] & 0.00 & 1.00 & 0.87 && [0.95, 1.18] & 0.00 & 1.00 & 0.77\\
		\midrule
		\multirow{3}{*}{\shortstack[c]{Pos. dep.\\ $\sim 2\%$ cens.}} & 500 & [0.89, 1.39] & 0.02 & 1.00 & 0.79 && [0.94, 1.43] & 0.03 & 1.00 & 0.69\\
		& 1000 & [0.90, 1.25] & 0.01 & 1.00 & 0.89 && [0.93, 1.28] & 0.01 & 1.00 & 0.75\\
		& 2000 & [0.91, 1.18] & 0.00 & 1.00 & 0.93 && [0.93, 1.19] & 0.00 & 1.00 & 0.84\\
		\midrule
		\multirow{3}{*}{\shortstack[c]{Neg. dep.\\ $\sim 2\%$ cens.}} & 500 & [0.90, 1.37] & 0.02 & 1.00 & 0.81 && [0.98, 1.42] & 0.02 & 1.00 & 0.59\\
		& 1000 & [0.91, 1.22] & 0.01 & 1.00 & 0.83 && [0.95, 1.26] & 0.01 & 1.00 & 0.71\\
		& 2000 & [0.93, 1.17] & 0.00 & 1.00 & 0.88 && [0.95, 1.18] & 0.00 & 1.00 & 0.81\\
		\bottomrule
	\end{tabular}
	\caption{Results for the model using the Cox link under an increased number of instrumental functions.}
	\label{tab: moreIF, Cox}
\end{table}

\begin{table}[ht]
	\centering
	\begin{tabular}{ccccccccccc}
		\toprule
		& & \multicolumn{4}{c}{$N_{IF} = 12$} & \phantom{c} & \multicolumn{4}{c}{$N_{IF} = 16$}\\
		\cmidrule{3-6} \cmidrule{8-11}
		Setting & $n$ & Bounds & Var & Sig & Cov & & Bounds & Var & Sig & Cov\\
		\midrule
		\multirow{3}{*}{\shortstack[c]{Indep.\\ $\sim 2\%$ cens.}} & 500 & [0.87, 1.33] & 0.01 & 1.00 & 0.87 && [0.90, 1.34] & 0.02 & 1.00 & 0.81\\
		& 1000 & [0.89, 1.21] & 0.00 & 1.00 & 0.93 && [0.90, 1.21] & 0.01 & 1.00 & 0.87\\
		& 2000 & [0.91, 1.14] & 0.00 & 1.00 & 0.95 && [0.92, 1.15] & 0.00 & 1.00 & 0.92\\
		\midrule
		\multirow{3}{*}{\shortstack[c]{Pos. dep.\\ $\sim 2\%$ cens.}} & 500 & [0.85, 1.36] & 0.01 & 1.00 & 0.92 && [0.87, 1.37] & 0.01 & 1.00 & 0.87\\
		& 1000 & [0.86, 1.23] & 0.00 & 1.00 & 0.96 && [0.89, 1.24] & 0.01 & 1.00 & 0.90\\
		& 2000 & [0.88, 1.16] & 0.00 & 1.00 & 0.97 && [0.90, 1.16] & 0.00 & 1.00 & 0.96\\
		\midrule
		\multirow{3}{*}{\shortstack[c]{Neg. dep.\\ $\sim 2\%$ cens.}} & 500 & [0.87, 1.33] & 0.01 & 1.00 & 0.86 && [0.89, 1.34] & 0.01 & 1.00 & 0.81\\
		& 1000 & [0.89, 1.21] & 0.00 & 1.00 & 0.93 && [0.90, 1.20] & 0.00 & 1.00 & 0.91\\
		& 2000 & [0.91, 1.14] & 0.00 & 1.00 & 0.94 && [0.92, 1.14] & 0.00 & 1.00 & 0.93\\
		\bottomrule
	\end{tabular}
	\caption{Results for the model using the Cox link in a setting with almost no censoring.}
	\label{tab: almost no censoring, Cox}
\end{table}

\begin{table}[ht]
	\centering
	\begin{tabular}{ccccccccccccc}
		\toprule
		& & \multicolumn{3}{c}{Single point} & \phantom{c} & \multicolumn{3}{c}{Intersection} & \phantom{c} & \multicolumn{3}{c}{Majority vote}\\
		\cmidrule{3-5} \cmidrule{7-9} \cmidrule{11-13}
		$t$ & Setting & Bounds & Sig & Cov & & Bounds & Sig & Cov & & Bounds & Sig & Cov\\
		\midrule
		\multirow{6}{*}{\shortstack[c]{1}} & 1-1 & [0.51, 1.37] & 1.00 & 1.00 && [0.65, 1.32] & 1.00 & 1.00 && [0.57, 1.37] & 1.00 & 1.00\\
		& 1-2 & [0.69, 1.51] & 1.00 & 0.99 && [0.74, 1.40] & 1.00 & 0.97 && [0.68, 1.47] & 1.00 & 1.00\\
		& 1-3 & [0.50, 1.43] & 1.00 & 1.00 && [0.66, 1.34] & 1.00 & 1.00 && [0.57, 1.40] & 1.00 & 1.00\\
		& 2-1 & [-0.08, 2.23] & 0.15 & 1.00 && [0.13, 1.67] & 0.94 & 1.00 && [0.00, 1.91] & 0.44 & 1.00\\
		& 2-2 & [0.26, 2.28] & 1.00 & 1.00 && [0.45, 1.73] & 1.00 & 1.00 && [0.35, 2.02] & 1.00 & 1.00\\
		& 2-3 & [-0.31, 5.45] & 0.00 & 1.00 && [0.01, 1.96] & 0.53 & 1.00 && [-0.18, 2.84] & 0.01 & 1.00\\
		\midrule
		\multirow{6}{*}{\shortstack[c]{5}} & 1-1 & [0.32, 2.71] & 1.00 & 1.00 && [0.53, 1.44] & 1.00 & 1.00 && [0.30, 1.88] & 1.00 & 1.00\\
		& 1-2 & [0.53, 3.51] & 1.00 & 0.98 && [0.70, 1.55] & 1.00 & 0.98 && [0.51, 2.23] & 1.00 & 1.00\\
		& 1-3 & [0.21, 5.78] & 0.83 & 1.00 && [0.48, 1.51] & 1.00 & 1.00 && [0.24, 2.46] & 1.00 & 1.00\\
		& 2-1 & [-4.94, 9.79] & 0.00 & 1.00 && [-0.12, 2.40] & 0.05 & 1.00 && [-4.29, 9.90] & 0.01 & 1.00\\
		& 2-2 & [-4.12, 10.00] & 0.03 & 1.00 && [0.26, 2.45] & 1.00 & 1.00 && [0.15, 9.65] & 0.84 & 1.00\\
		& 2-3 & [-5.17, 9.36] & 0.00 & 1.00 && [-0.35, 8.03] & 0.00 & 1.00 && [-5.47, 9.69] & 0.00 & 1.00\\
		\bottomrule
	\end{tabular}
	\caption{Comparison of obtained bounds using the Cox link, testing at a single time point with $\alpha = 0.05$ (\emph{Single point}), or testing at three different time points with Bonferroni corrected level $\alpha  = 0.05/3$ (\emph{Intersection}) or $\alpha = 0.05/2$ (\emph{Majority vote}). When $t = 1$, the tested time points are $0.333$, $0.667$ and $1$. When $t = 5$, the tested time points are $1$, $3$ and $5$.}
	\label{tab: results multiple testing Cox}
\end{table}

\begin{table}[ht]
	\centering
	\begin{tabular}{ccccccccccc}
		\toprule
		& & \multicolumn{4}{c}{$N_{IF} = 12$} & \phantom{c} & \multicolumn{4}{c}{$N_{IF} = 16$}\\
		\cmidrule{3-6} \cmidrule{8-11}
		Setting & $n$ & Bounds & Var & Sig & Cov & & Bounds & Var & Sig & Cov\\
		\midrule
		\multirow{3}{*}{\shortstack[c]{Indep.\\ $\sim 30\%$ cens.}} & 500 & [0.38, 1.67] & 0.03 & 1.00 & 1.00 && [0.39, 1.63] & 0.03 & 1.00 & 1.00\\
		& 1000 & [0.41, 1.45] & 0.01 & 1.00 & 1.00 && [0.44, 1.44] & 0.01 & 1.00 & 1.00\\
		& 2000 & [0.46, 1.35] & 0.01 & 1.00 & 1.00 && [0.48, 1.35] & 0.01 & 1.00 & 1.00\\
		\midrule
		\multirow{3}{*}{\shortstack[c]{Pos. dep.\\ $\sim 30\%$ cens.}} & 500 & [0.52, 1.74] & 0.04 & 1.00 & 1.00 && [0.59, 1.74] & 0.03 & 1.00 & 1.00\\
		& 1000 & [0.57, 1.55] & 0.01 & 1.00 & 1.00 && [0.64, 1.55] & 0.02 & 1.00 & 1.00\\
		& 2000 & [0.62, 1.44] & 0.01 & 1.00 & 1.00 && [0.65, 1.44] & 0.01 & 1.00 & 1.00\\
		\midrule
		\multirow{3}{*}{\shortstack[c]{Neg. dep.\\ $\sim 30\%$ cens.}} & 500 & [0.38, 1.76] & 0.04 & 1.00 & 1.00 && [0.42, 1.69] & 0.04 & 1.00 & 1.00\\
		& 1000 & [0.42, 1.53] & 0.01 & 1.00 & 1.00 && [0.45, 1.51] & 0.01 & 1.00 & 1.00\\
		& 2000 & [0.45, 1.41] & 0.01 & 1.00 & 1.00 && [0.48, 1.38] & 0.01 & 1.00 & 1.00\\
		\midrule
		\multirow{3}{*}{\shortstack[c]{Indep.\\ $\sim 65\%$ cens.}} & 500 & [-0.44,  3.58] &  0.98 &  0.00 &  1.00 && [-0.40,  2.96] &  0.45 &  0.01 &  1.00\\
		& 1000 & [-0.31,  2.67] &  0.25 &  0.00 &  1.00 && [-0.26,  2.36] &  0.07 &  0.01 &  1.00\\
		& 2000 & [-0.24,  2.17] &  0.10 &  0.00 &  1.00 && [-0.19,  2.02] &  0.04 &  0.03 &  1.00\\
		\midrule
		\multirow{3}{*}{\shortstack[c]{Pos. dep.\\ $\sim 65\%$ cens.}} & 500 & [0.00, 2.82] & 0.21 & 0.52 & 1.00 && [0.07, 2.71] & 0.14 & 0.60 & 1.00\\
		& 1000 & [0.09, 2.34] & 0.07 & 0.83 & 1.00 && [0.14, 2.23] & 0.04 & 0.92 & 1.00\\
		& 2000 & [0.15, 2.12] & 0.04 & 0.98 & 1.00 && [0.24, 2.03] & 0.03 & 1.00 & 1.00\\
		\midrule
		\multirow{3}{*}{\shortstack[c]{Neg. dep.\\ $\sim 65\%$ cens.}} & 500 & [-0.77,  9.59] &  1.10 &  0.00 &  1.00 && [-0.63,  8.61] &  3.55 &  0.00 &  1.00\\
		& 1000 & [-0.57,  8.53] &  2.01 &  0.00 &  1.00 && [-0.47,  5.13] &  3.16 &  0.00 &  1.00\\
		& 2000 & [-0.45,  7.01] &  3.14 &  0.00 &  1.00 && [-0.37,  3.53] &  0.60 &  0.00 &  1.00\\
		\bottomrule
	\end{tabular}
	\caption{Results from the simulations regarding dependent covariates, using the Cox link function and including either $N_{IF} = 12$ or $N_{IF} = 16$ instrumental functions \emph{before} possible removal.}
	\label{tab: Dep cov. Cox. 1}
\end{table}

\begin{table}[ht]
	\centering
	\begin{tabular}{ccccccccccc}
		\toprule
		& & \multicolumn{4}{c}{$N_{IF} = 20$} & \phantom{c} & \multicolumn{4}{c}{$N_{IF} = 30$}\\
		\cmidrule{3-6} \cmidrule{8-11}
		Setting & $n$ & Bounds & Var & Sig & Cov & & Bounds & Var & Sig & Cov\\
		\midrule
		\multirow{3}{*}{\shortstack[c]{Indep.\\ $\sim 30\%$ cens.}} & 500 & [0.43, 1.63] & 0.03 & 1.00 & 1.00 && [0.43, 1.67] & 0.05 & 1.00 & 0.99\\
		& 1000 & [0.48, 1.44] & 0.02 & 1.00 & 1.00 && [0.48, 1.44] & 0.02 & 1.00 & 1.00\\
		& 2000 & [0.48, 1.33] & 0.01 & 1.00 & 1.00 && [0.51, 1.32] & 0.01 & 1.00 & 1.00\\
		\midrule
		\multirow{3}{*}{\shortstack[c]{Pos. dep.\\ $\sim 30\%$ cens.}} & 500 & [0.61, 1.72] & 0.04 & 1.00 & 0.98 && [0.65, 1.78] & 0.06 & 1.00 & 0.94\\
		& 1000 & [0.64, 1.54] & 0.02 & 1.00 & 0.99 && [0.69, 1.57] & 0.03 & 1.00 & 0.96\\
		& 2000 & [0.68, 1.44] & 0.01 & 1.00 & 1.00 && [0.71, 1.45] & 0.01 & 1.00 & 0.98\\
		\midrule
		\multirow{3}{*}{\shortstack[c]{Neg. dep.\\ $\sim 30\%$ cens.}} & 500 & [0.44, 1.70] & 0.04 & 1.00 & 1.00 && [0.47, 1.75] & 0.05 & 1.00 & 0.98\\
		& 1000 & [0.46, 1.51] & 0.01 & 1.00 & 1.00 && [0.49, 1.53] & 0.02 & 1.00 & 1.00\\
		& 2000 & [0.48, 1.38] & 0.01 & 1.00 & 1.00 && [0.51, 1.39] & 0.01 & 1.00 & 0.99\\
		\midrule
		\multirow{3}{*}{\shortstack[c]{Indep.\\ $\sim 65\%$ cens.}} & 500 & [-0.37,  2.87] &  0.28 &  0.02 &  1.00 && [-0.39,  2.82] &  0.24 &  0.03 &  1.00\\
		& 1000 & [-0.23,  2.24] &  0.08 &  0.02 &  1.00 && [-0.23,  2.18] &  0.06 &  0.03 &  1.00\\
		& 2000 & [-0.17,  1.94] &  0.03 &  0.04 &  1.00 && [-0.14,  1.87] &  0.03 &  0.06 &  1.00\\
		\midrule
		\multirow{3}{*}{\shortstack[c]{Pos. dep.\\ $\sim 65\%$ cens.}} & 500 & [0.09, 2.58] & 0.10 & 0.72 & 1.00 && [0.12, 2.66] & 0.15 & 0.72 & 1.00\\
		& 1000 & [0.17, 2.20] & 0.04 & 0.92 & 1.00 && [0.21, 2.18] & 0.07 & 0.93 & 1.00\\
		& 2000 & [0.24, 1.97] & 0.02 & 1.00 & 1.00 && [0.30, 1.97] & 0.04 & 1.00 & 1.00\\
		\midrule
		\multirow{3}{*}{\shortstack[c]{Neg. dep.\\ $\sim 65\%$ cens.}} & 500 & [-0.56,  7.88] &  4.98 &  0.00 &  1.00 && [-0.50,  6.52] &  5.10 &  0.00 &  1.00\\
		& 1000 & [-0.43,  4.20] &  1.81 &  0.00 &  1.00 && [-0.42,  3.59] &  0.59 &  0.00 &  1.00\\
		& 2000 & [-0.33,  3.02] &  0.26 &  0.00 &  1.00 && [-0.33,  2.67] &  0.17 &  0.00 &  1.00\\
		\bottomrule
	\end{tabular}
	\caption{Results from the simulations regarding dependent covariates, using the Cox link function and including either $N_{IF} = 20$ or $N_{IF} = 30$ instrumental functions \emph{before} possible removal.}
	\label{tab: Dep cov. Cox. 2}
\end{table}


\begin{table}[ht]
	\centering
	\begin{tabular}{ccccccccccc}
		\toprule
		& & \multicolumn{4}{c}{$N_{IF} = 12$} & \phantom{c} & \multicolumn{4}{c}{$N_{IF} = 16$}\\
		\cmidrule{3-6} \cmidrule{8-11}
		Setting & $n$ & Bounds & Var & Sig & Cov & & Bounds & Var & Sig & Cov\\
		\midrule
		\multirow{3}{*}{\shortstack[c]{Indep.\\ $\sim 30\%$ cens.}} & 500 & [0.57, 1.55] & 0.03 & 1.00 & 0.99 && [0.60, 1.55] & 0.04 & 1.00 & 0.97\\
		& 1000 & [0.62, 1.39] & 0.01 & 1.00 & 0.99 && [0.64, 1.39] & 0.01 & 1.00 & 1.00\\
		& 2000 & [0.66, 1.28] & 0.00 & 1.00 & 1.00 && [0.68, 1.28] & 0.01 & 1.00 & 1.00\\
		\midrule
		\multirow{3}{*}{\shortstack[c]{Pos. dep.\\ $\sim 30\%$ cens.}} & 500 & [0.62, 1.58] & 0.02 & 1.00 & 0.98 && [0.65, 1.57] & 0.03 & 1.00 & 0.94\\
		& 1000 & [0.67, 1.40] & 0.01 & 1.00 & 0.99 && [0.70, 1.40] & 0.02 & 1.00 & 0.98\\
		& 2000 & [0.72, 1.31] & 0.01 & 1.00 & 0.99 && [0.74, 1.30] & 0.01 & 1.00 & 0.98\\
		\midrule
		\multirow{3}{*}{\shortstack[c]{Neg. dep.\\ $\sim 30\%$ cens.}} & 500 & [0.62, 1.60] & 0.02 & 1.00 & 0.98 && [0.65, 1.60] & 0.03 & 1.00 & 0.97\\
		& 1000 & [0.66, 1.43] & 0.01 & 1.00 & 1.00 && [0.68, 1.42] & 0.01 & 1.00 & 0.99\\
		& 2000 & [0.69, 1.31] & 0.00 & 1.00 & 1.00 && [0.71, 1.31] & 0.01 & 1.00 & 1.00\\
		\midrule
		\multirow{3}{*}{\shortstack[c]{Indep.\\ $\sim 65\%$ cens.}} & 500 & [-0.21,  2.57] &  0.15 &  0.07 &  1.00 && [-0.20,  2.47] &  0.11 &  0.10 &  1.00\\
		& 1000 & [-0.11,  2.13] &  0.04 &  0.16 &  1.00 && [-0.08,  2.07] &  0.04 &  0.24 &  1.00\\
		& 2000 & [-0.02,  1.90] &  0.02 &  0.36 &  1.00 && [ 0.01,  1.85] &  0.02 &  0.49 &  1.00\\
		\midrule
		\multirow{3}{*}{\shortstack[c]{Pos. dep.\\ $\sim 65\%$ cens.}} & 500 & [0.13, 2.26] & 0.08 & 0.77 & 1.00 && [0.18, 2.22] & 0.08 & 0.81 & 0.99\\
		& 1000 & [0.24, 1.94] & 0.03 & 0.98 & 1.00 && [0.29, 1.92] & 0.04 & 0.99 & 1.00\\
		& 2000 & [0.31, 1.75] & 0.02 & 1.00 & 1.00 && [0.36, 1.74] & 0.02 & 1.00 & 1.00\\
		\midrule
		\multirow{3}{*}{\shortstack[c]{Neg. dep.\\ $\sim 65\%$ cens.}} & 500 & [-0.72,  7.50] &  4.68 &  0.00 &  1.00 && [-0.67,  6.44] &  3.99 &  0.00 &  1.00\\
		& 1000 & [-0.51,  4.67] &  1.13 &  0.00 &  1.00 && [-0.47,  4.10] &  0.47 &  0.00 &  1.00\\
		& 2000 & [-0.37,  3.70] &  0.29 &  0.00 &  1.00 && [-0.34,  3.33] &  0.13 &  0.00 &  1.00\\
		\bottomrule
	\end{tabular}
	\caption{Results of the main simulation using the AFT link function.}
	\label{tab: main AFT}
\end{table}

\begin{table}[ht]
	\centering
	\begin{tabular}{ccccccccccc}
		\toprule
		& & \multicolumn{4}{c}{$N_{IF} = 20$} & \phantom{c} & \multicolumn{4}{c}{$N_{IF} = 30$}\\
		\cmidrule{3-6} \cmidrule{8-11}
		Setting & $n$ & Bounds & Var & Sig & Cov & & Bounds & Var & Sig & Cov\\
		\midrule
		\multirow{3}{*}{\shortstack[c]{Indep.\\ $\sim 30\%$ cens.}} & 500 & [0.62, 1.57] & 0.04 & 1.00 & 0.97 && [0.62, 1.59] & 0.06 & 1.00 & 0.95\\
		& 1000 & [0.65, 1.38] & 0.02 & 1.00 & 0.99 && [0.67, 1.40] & 0.02 & 1.00 & 0.98\\
		& 2000 & [0.68, 1.28] & 0.01 & 1.00 & 1.00 && [0.70, 1.29] & 0.01 & 1.00 & 0.99\\
		\midrule
		\multirow{3}{*}{\shortstack[c]{Pos. dep.\\ $\sim 30\%$ cens.}} & 500 & [0.65, 1.58] & 0.04 & 1.00 & 0.96 && [0.67, 1.64] & 0.05 & 1.00 & 0.95\\
		& 1000 & [0.71, 1.40] & 0.02 & 1.00 & 0.97 && [0.72, 1.44] & 0.02 & 1.00 & 0.98\\
		& 2000 & [0.74, 1.31] & 0.01 & 1.00 & 0.98 && [0.76, 1.32] & 0.01 & 1.00 & 0.98\\
		\midrule
		\multirow{3}{*}{\shortstack[c]{Neg. dep.\\ $\sim 30\%$ cens.}} & 500 & [0.67, 1.61] & 0.04 & 1.00 & 0.96 && [0.66, 1.65] & 0.05 & 1.00 & 0.96\\
		& 1000 & [0.70, 1.42] & 0.02 & 1.00 & 0.98 && [0.72, 1.46] & 0.03 & 1.00 & 0.95\\
		& 2000 & [0.71, 1.31] & 0.01 & 1.00 & 0.99 && [0.73, 1.33] & 0.01 & 1.00 & 0.99\\
		\midrule
		\multirow{3}{*}{\shortstack[c]{Indep.\\ $\sim 65\%$ cens.}} & 500 & [-0.18,  2.46] &  0.11 &  0.14 &  1.00 && [-0.20,  2.57] &  0.14 &  0.12 &  1.00\\
		& 1000 & [-0.06,  2.06] &  0.04 &  0.27 &  1.00 && [-0.07,  2.10] &  0.05 &  0.24 &  1.00\\
		& 2000 & [ 0.02,  1.84] &  0.02 &  0.51 &  1.00 && [ 0.02,  1.85] &  0.03 &  0.56 &  1.00\\
		\midrule
		\multirow{3}{*}{\shortstack[c]{Pos. dep.\\ $\sim 65\%$ cens.}} & 500 & [0.20, 2.24] & 0.10 & 0.84 & 0.98 && [0.19, 2.32] & 0.13 & 0.78 & 0.97\\
		& 1000 & [0.31, 1.91] & 0.05 & 0.98 & 0.99 && [0.33, 1.95] & 0.09 & 0.98 & 0.96\\
		& 2000 & [0.38, 1.73] & 0.03 & 1.00 & 0.99 && [0.43, 1.75] & 0.05 & 1.00 & 0.96\\
		\midrule
		\multirow{3}{*}{\shortstack[c]{Neg. dep.\\ $\sim 65\%$ cens.}} & 500 & [-0.68,  5.76] &  2.71 &  0.00 &  1.00 && [-0.70,  5.81] &  1.76 &  0.00 &  1.00\\
		& 1000 & [-0.46,  3.94] &  0.34 &  0.01 &  1.00 && [-0.46,  3.92] &  0.27 &  0.00 &  1.00\\
		& 2000 & [-0.33,  3.21] &  0.10 &  0.00 &  1.00 && [-0.32,  3.20] &  0.09 &  0.01 &  1.00\\
		\midrule
		\multirow{3}{*}{\shortstack[c]{Indep.\\ $\sim 2\%$ cens.}} & 500 & [0.86, 1.44] & 0.03 & 1.00 & 0.80 && [0.88, 1.49] & 0.04 & 1.00 & 0.74\\
		& 1000 & [0.87, 1.28] & 0.01 & 1.00 & 0.89 && [0.89, 1.30] & 0.02 & 1.00 & 0.84\\
		& 2000 & [0.89, 1.19] & 0.00 & 1.00 & 0.89 && [0.90, 1.20] & 0.01 & 1.00 & 0.88\\
		\midrule
		\multirow{3}{*}{\shortstack[c]{Pos. dep.\\ $\sim 2\%$ cens.}} & 500 & [0.84, 1.44] & 0.03 & 1.00 & 0.81 && [0.85, 1.47] & 0.04 & 1.00 & 0.81\\
		& 1000 & [0.86, 1.27] & 0.01 & 1.00 & 0.91 && [0.88, 1.30] & 0.02 & 1.00 & 0.85\\
		& 2000 & [0.89, 1.19] & 0.00 & 1.00 & 0.88 && [0.90, 1.21] & 0.01 & 1.00 & 0.88\\
		\midrule
		\multirow{3}{*}{\shortstack[c]{Neg. dep.\\ $\sim 2\%$ cens.}} & 500 & [0.84, 1.42] & 0.03 & 1.00 & 0.85 && [0.88, 1.49] & 0.04 & 1.00 & 0.75\\
		& 1000 & [0.87, 1.28] & 0.01 & 1.00 & 0.88 && [0.89, 1.31] & 0.01 & 1.00 & 0.80\\
		& 2000 & [0.90, 1.18] & 0.00 & 1.00 & 0.89 && [0.91, 1.20] & 0.01 & 1.00 & 0.84\\
		\bottomrule
	\end{tabular}
	\caption{Results for the model using the AFT link under an increased number of instrumental functions.}
	\label{tab: moreIF, AFT}
\end{table}

\begin{table}[ht]
	\centering
	\begin{tabular}{ccccccccccc}
		\toprule
		& & \multicolumn{4}{c}{$N_{IF} = 12$} & \phantom{c} & \multicolumn{4}{c}{$N_{IF} = 16$}\\
		\cmidrule{3-6} \cmidrule{8-11}
		Setting & $n$ & Bounds & Var & Sig & Cov & & Bounds & Var & Sig & Cov\\
		\midrule
		\multirow{3}{*}{\shortstack[c]{Indep.\\ $\sim 2\%$ cens.}} & 500 & [0.81, 1.40] & 0.02 & 1.00 & 0.88 && [0.84, 1.41] & 0.03 & 1.00 & 0.83\\
		& 1000 & [0.86, 1.26] & 0.01 & 1.00 & 0.91 && [0.86, 1.26] & 0.01 & 1.00 & 0.90\\
		& 2000 & [0.88, 1.17] & 0.00 & 1.00 & 0.95 && [0.89, 1.18] & 0.00 & 1.00 & 0.92\\
		\midrule
		\multirow{3}{*}{\shortstack[c]{Pos. dep.\\ $\sim 2\%$ cens.}} & 500 & [0.81, 1.39] & 0.02 & 1.00 & 0.92 && [0.84, 1.42] & 0.02 & 1.00 & 0.83\\
		& 1000 & [0.85, 1.26] & 0.01 & 1.00 & 0.93 && [0.85, 1.26] & 0.01 & 1.00 & 0.93\\
		& 2000 & [0.88, 1.17] & 0.00 & 1.00 & 0.95 && [0.89, 1.17] & 0.00 & 1.00 & 0.94\\
		\midrule
		\multirow{3}{*}{\shortstack[c]{Neg. dep.\\ $\sim 2\%$ cens.}} & 500 & [0.83, 1.41] & 0.02 & 1.00 & 0.88 && [0.83, 1.40] & 0.02 & 1.00 & 0.86\\
		& 1000 & [0.86, 1.25] & 0.01 & 1.00 & 0.92 && [0.88, 1.27] & 0.01 & 1.00 & 0.86\\
		& 2000 & [0.89, 1.17] & 0.00 & 1.00 & 0.92 && [0.89, 1.17] & 0.00 & 1.00 & 0.93\\
		\bottomrule
	\end{tabular}
	\caption{Results for the model using the AFT link in a setting with almost no censoring.}
	\label{tab: almost no censoring, AFT}
\end{table}

\begin{table}[ht]
	\centering
	\begin{tabular}{ccccccccccccc}
		\toprule
		& & \multicolumn{3}{c}{Single point} & \phantom{c} & \multicolumn{3}{c}{Intersection} & \phantom{c} & \multicolumn{3}{c}{Majority vote}\\
		\cmidrule{3-5} \cmidrule{7-9} \cmidrule{11-13}
		$t$ & Setting & Bounds & Sig & Cov & & Bounds & Sig & Cov & & Bounds & Sig & Cov\\
		\midrule
		\multirow{6}{*}{\shortstack[c]{1}} & 1-1 & [0.64, 1.38] & 1.00 & 0.99 && [0.70, 1.36] & 1.00 & 0.99 && [0.65, 1.40] & 1.00 & 1.00\\
		& 1-2 & [0.69, 1.40] & 1.00 & 0.99 && [0.71, 1.38] & 1.00 & 0.98 && [0.66, 1.42] & 1.00 & 1.00\\
		& 1-3 & [0.68, 1.41] & 1.00 & 0.98 && [0.75, 1.37] & 1.00 & 0.96 && [0.68, 1.42] & 1.00 & 1.00\\
		& 2-1 & [-0.07, 2.09] & 0.26 & 1.00 && [0.15, 1.72] & 0.90 & 1.00 && [0.03, 1.91] & 0.59 & 1.00\\
		& 2-2 & [0.30, 1.91] & 1.00 & 0.99 && [0.38, 1.69] & 1.00 & 1.00 && [0.31, 1.81] & 1.00 & 1.00\\
		& 2-3 & [-0.48, 4.01] & 0.00 & 1.00 && [-0.01, 2.15] & 0.39 & 1.00 && [-0.25, 3.03] & 0.02 & 1.00\\
		\midrule
		\multirow{6}{*}{\shortstack[c]{5}} & 1-1 & [0.33, 1.78] & 1.00 & 1.00 && [0.61, 1.43] & 1.00 & 1.00 && [0.44, 1.61] & 1.00 & 1.00\\
		& 1-2 & [0.57, 1.78] & 1.00 & 0.96 && [0.72, 1.47] & 1.00 & 0.94 && [0.61, 1.63] & 1.00 & 1.00\\
		& 1-3 & [0.24, 2.27] & 0.98 & 1.00 && [0.65, 1.48] & 1.00 & 1.00 && [0.42, 1.86] & 1.00 & 1.00\\
		& 2-1 & [-4.16, 9.63] & 0.00 & 1.00 && [-0.13, 2.19] & 0.14 & 1.00 && [-1.03, 5.50] & 0.00 & 1.00\\
		& 2-2 & [-0.87, 6.85] & 0.00 & 1.00 && [0.23, 2.03] & 0.99 & 1.00 && [-0.26, 3.36] & 0.11 & 1.00\\
		& 2-3 & [-5.13, 9.09] & 0.00 & 1.00 && [-0.54, 4.61] & 0.01 & 1.00 && [-5.44, 9.49] & 0.00 & 1.00\\
		\bottomrule
	\end{tabular}
	\caption{Comparison of obtained bounds using the AFT link, testing at a single time point with $\alpha = 0.05$ (\emph{Single point}), or testing at three different time points with Bonferroni corrected level $\alpha  = 0.05/3$ (\emph{Intersection}) or $\alpha = 0.05/2$ (\emph{Majority vote}). When $t = 1$, the tested time points are $0.333$, $0.667$ and $1$. When $t = 5$, the tested time points are $1$, $3$ and $5$.}
	\label{tab: results multiple testing AFT}
\end{table}

\begin{table}[ht]
	\centering
	\begin{tabular}{ccccccccccc}
		\toprule
		& & \multicolumn{4}{c}{$N_{IF} = 12$} & \phantom{c} & \multicolumn{4}{c}{$N_{IF} = 16$}\\
		\cmidrule{3-6} \cmidrule{8-11}
		Setting & $n$ & Bounds & Var & Sig & Cov & & Bounds & Var & Sig & Cov\\
		\midrule
		\multirow{3}{*}{\shortstack[c]{Indep.\\ $\sim 30\%$ cens.}} & 500 & [0.40, 1.70] & 0.03 & 1.00 & 1.00 && [0.43, 1.71] & 0.03 & 0.99 & 1.00\\
		& 1000 & [0.47, 1.50] & 0.01 & 1.00 & 1.00 && [0.52, 1.49] & 0.02 & 1.00 & 1.00\\
		& 2000 & [0.53, 1.37] & 0.01 & 1.00 & 1.00 && [0.56, 1.37] & 0.01 & 1.00 & 1.00\\
		\midrule
		\multirow{3}{*}{\shortstack[c]{Pos. dep.\\ $\sim 30\%$ cens.}} & 500 & [0.47, 1.70] & 0.03 & 1.00 & 0.99 && [0.54, 1.71] & 0.04 & 1.00 & 0.97\\
		& 1000 & [0.56, 1.50] & 0.01 & 1.00 & 1.00 && [0.61, 1.50] & 0.02 & 1.00 & 1.00\\
		& 2000 & [0.62, 1.38] & 0.01 & 1.00 & 1.00 && [0.64, 1.37] & 0.01 & 1.00 & 0.99\\
		\midrule
		\multirow{3}{*}{\shortstack[c]{Neg. dep.\\ $\sim 30\%$ cens.}} & 500 & [0.46, 1.77] & 0.03 & 1.00 & 1.00 && [0.48, 1.74] & 0.04 & 1.00 & 0.98\\
		& 1000 & [0.52, 1.57] & 0.01 & 1.00 & 1.00 && [0.55, 1.54] & 0.01 & 1.00 & 1.00\\
		& 2000 & [0.59, 1.45] & 0.01 & 1.00 & 1.00 && [0.61, 1.42] & 0.01 & 1.00 & 1.00\\
		\midrule
		\multirow{3}{*}{\shortstack[c]{Indep.\\ $\sim 65\%$ cens.}} & 500 & [-0.77,  3.37] &  0.28 &  0.00 &  1.00 && [-0.67,  3.09] &  0.25 &  0.00 &  1.00\\
		& 1000 & [-0.54,  2.70] &  0.11 &  0.00 &  1.00 && [-0.47,  2.53] &  0.08 &  0.00 &  1.00\\
		& 2000 & [-0.42,  2.27] &  0.06 &  0.00 &  1.00 && [-0.34,  2.22] &  0.03 &  0.01 &  1.00\\
		\midrule
		\multirow{3}{*}{\shortstack[c]{Pos. dep.\\ $\sim 65\%$ cens.}} & 500 & [-0.13,  2.46] &  0.11 &  0.25 &  1.00 && [-0.06,  2.37] &  0.06 &  0.32 &  1.00\\
		& 1000 & [ 0.03,  2.10] &  0.04 &  0.57 &  1.00 && [ 0.08,  2.04] &  0.04 &  0.68 &  1.00\\
		& 2000 & [ 0.11,  1.85] &  0.02 &  0.86 &  1.00 && [ 0.16,  1.84] &  0.02 &  0.92 &  1.00\\
		\midrule
		\multirow{3}{*}{\shortstack[c]{Neg. dep.\\ $\sim 65\%$ cens.}} & 500 & [-1.41,  8.62] &  2.58 &  0.00 &  1.00 && [-1.25,  7.17] &  3.41 &  0.00 &  1.00\\
		& 1000 & [-1.04,  6.69] &  2.28 &  0.00 &  1.00 && [-0.92,  4.91] &  0.66 &  0.00 &  1.00\\
		& 2000 & [-0.83,  5.30] &  1.54 &  0.00 &  1.00 && [-0.70,  3.98] &  0.29 &  0.00 &  1.00\\
		\bottomrule
	\end{tabular}
	\caption{Results from the simulations regarding dependent covariates, using the AFT link function and including either $N_{IF} = 12$ or $N_{IF} = 16$ instrumental functions \emph{before} possible removal.}
	\label{tab: Dep cov. AFT. 1}
\end{table}

\begin{table}[ht]
	\centering
	\begin{tabular}{ccccccccccc}
		\toprule
		& & \multicolumn{4}{c}{$N_{IF} = 20$} & \phantom{c} & \multicolumn{4}{c}{$N_{IF} = 30$}\\
		\cmidrule{3-6} \cmidrule{8-11}
		Setting & $n$ & Bounds & Var & Sig & Cov & & Bounds & Var & Sig & Cov\\
		\midrule
		\multirow{3}{*}{\shortstack[c]{Indep.\\ $\sim 30\%$ cens.}} & 500 & [0.45, 1.69] & 0.05 & 1.00 & 0.98 && [0.42, 1.71] & 0.05 & 0.98 & 0.98\\
		& 1000 & [0.51, 1.49] & 0.02 & 1.00 & 1.00 && [0.54, 1.51] & 0.03 & 1.00 & 0.99\\
		& 2000 & [0.58, 1.37] & 0.01 & 1.00 & 1.00 && [0.58, 1.37] & 0.01 & 1.00 & 1.00\\
		\midrule
		\multirow{3}{*}{\shortstack[c]{Pos. dep.\\ $\sim 30\%$ cens.}} & 500 & [0.52, 1.69] & 0.04 & 1.00 & 0.99 && [0.55, 1.74] & 0.08 & 1.00 & 0.96\\
		& 1000 & [0.62, 1.52] & 0.03 & 1.00 & 0.99 && [0.64, 1.52] & 0.03 & 1.00 & 0.99\\
		& 2000 & [0.67, 1.38] & 0.01 & 1.00 & 0.99 && [0.68, 1.40] & 0.01 & 1.00 & 0.98\\
		\midrule
		\multirow{3}{*}{\shortstack[c]{Neg. dep.\\ $\sim 30\%$ cens.}} & 500 & [0.50, 1.80] & 0.05 & 1.00 & 1.00 && [0.50, 1.78] & 0.07 & 1.00 & 0.97\\
		& 1000 & [0.57, 1.55] & 0.02 & 1.00 & 1.00 && [0.56, 1.59] & 0.03 & 1.00 & 0.99\\
		& 2000 & [0.60, 1.42] & 0.01 & 1.00 & 1.00 && [0.61, 1.43] & 0.01 & 1.00 & 1.00\\
		\midrule
		\multirow{3}{*}{\shortstack[c]{Indep.\\ $\sim 65\%$ cens.}} & 500 & [-0.63,  2.97] &  0.19 &  0.00 &  1.00 && [-0.68,  3.03] &  0.19 &  0.00 &  1.00\\
		& 1000 & [-0.44,  2.48] &  0.07 &  0.01 &  1.00 && [-0.45,  2.43] &  0.08 &  0.01 &  1.00\\
		& 2000 & [-0.32,  2.15] &  0.04 &  0.00 &  1.00 && [-0.32,  2.14] &  0.03 &  0.00 &  1.00\\
		\midrule
		\multirow{3}{*}{\shortstack[c]{Pos. dep.\\ $\sim 65\%$ cens.}} & 500 & [-0.04,  2.34] &  0.07 &  0.36 &  1.00 && [-0.07,  2.39] &  0.10 &  0.30 &  1.00\\
		& 1000 & [ 0.10,  2.02] &  0.03 &  0.68 &  1.00 && [ 0.10,  2.04] &  0.06 &  0.73 &  0.99\\
		& 2000 & [ 0.20,  1.81] &  0.02 &  0.98 &  1.00 && [ 0.22,  1.81] &  0.03 &  0.96 &  1.00\\
		\midrule
		\multirow{3}{*}{\shortstack[c]{Neg. dep.\\ $\sim 65\%$ cens.}} & 500 & [-1.19,  6.50] &  3.28 &  0.00 &  1.00 && [-1.16,  6.18] &  2.26 &  0.00 &  1.00\\
		& 1000 & [-0.86,  4.43] &  0.44 &  0.00 &  1.00 && [-0.87,  4.42] &  0.39 &  0.00 &  1.00\\
		& 2000 & [-0.65,  3.65] &  0.19 &  0.00 &  1.00 && [-0.62,  3.58] &  0.16 &  0.00 &  1.00\\
		\bottomrule
	\end{tabular}
	\caption{Results from the simulations regarding dependent covariates, using the AFT link function and including either $N_{IF} = 20$ or $N_{IF} = 30$ instrumental functions \emph{before} possible removal.}
	\label{tab: Dep cov. AFT. 2}
\end{table}

\section{Pancreas data application} \label{supp: Pancreas data application}
\subsection{Analysis using AFT link}
\begin{figure}[ht]
	\centering
	\includegraphics[width = \linewidth]{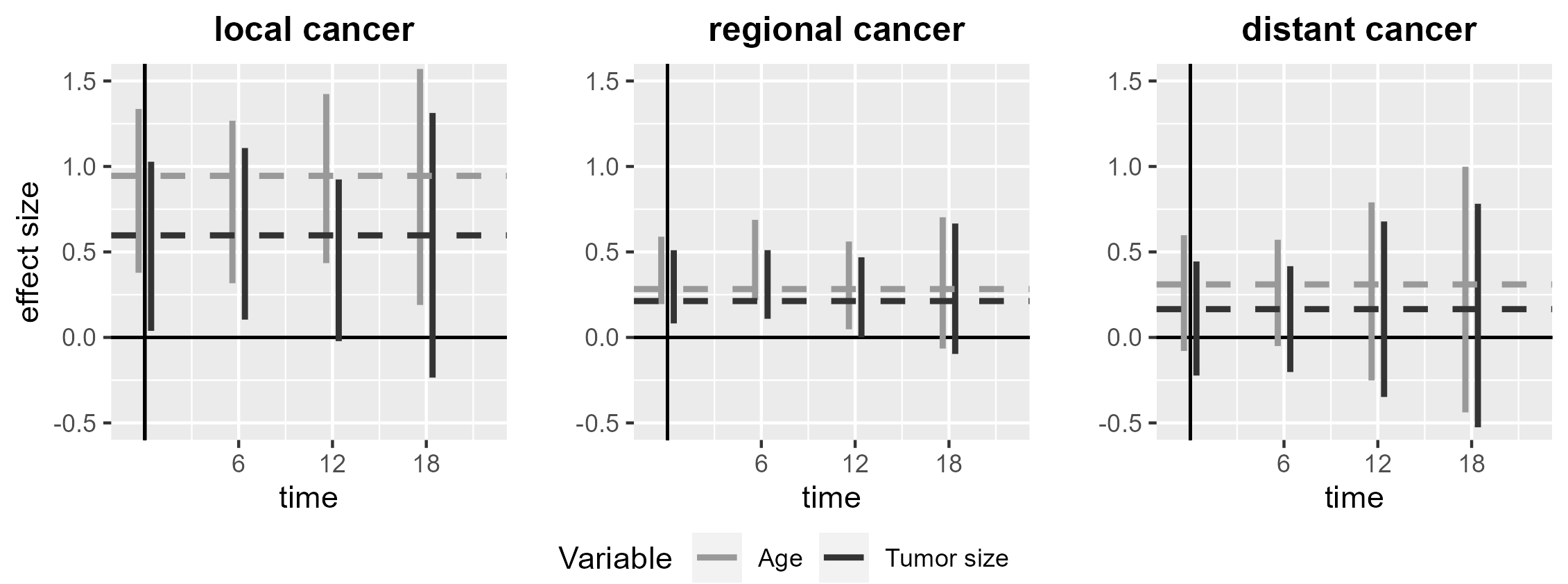}
	\caption{The results for the SEER pancreatic cancer data set using the AFT link function. Interpretation is analogous to Figure \ref{fig: Results Pancreas Cox}.}
	\label{fig: Results Pancreas AFT}
\end{figure}

We perform a similar analysis on the SEER Pancreas data set as in Section \ref{sec: Data applications}, this time using the AFT link function instead of the Cox link function. Plots of the results are shown in Figure \ref{fig: Results Pancreas AFT}.

Many of the same conclusions as were drawn in Section \ref{sec: Data applications} can be made here. We again see that both age and the size of the tumor can be determined to have a positive and significant effect on the probability of dying before six months from local pancreatic cancer. That is, black males of older age or having larger tumors tend to die sooner. The same conclusion can be made for age when looking at the probability of dying before 12 or 18 months, though no significant statements can be made for the effect of tumor size. If one is willing to make the additional assumption that the covariate effects are time independent, both identified intervals become significant.

The plots for the case of regional and distant cancer can be interpreted in a similar fashion. Notably, in the case of distant cancer, the covariate effects could not be concluded to be significant at any of the selected time points, nor when assuming that they are time-independent. For reference, the results of a classical AFT model with log-logistic baseline survival (i.e. proportional odds model) are overlaid.

\subsection{Comparison of combination methods} \label{supp: Comparison of combination methods}
In Section \ref{sec: Data applications}, we estimate the identified intervals under the additional assumption of time-independent effects of covariates by considering the model at three different time points, $(6, 12, 18)$, using the adjusted level $\alpha = 0.05 / 3$, and then combining the separate identified intervals by means of intersection. In doing so, however, two choices have been made: (i) we selected the time points at which to consider the model, and (ii) we selected the combination method. In this section, we investigate the results under different choices.

In terms of the combination method, we will consider three options. The first and third options are the intersection and majority vote method as discussed in Section \ref{sec: Time-independent effects of covariates}. The second option is similar to the first, but, motivated by the observation that earlier time points often lead to narrower bounds, only differs in the way in which the corrected levels are chosen: while the first option gives equal confidence level $\alpha = 0.05 / A$ to all time points, the second will allow earlier time points to be estimated at a lower level by assigning time point $t_a$ the level $\alpha_a = (0.05/a)(1 + \dots + A^{-1})^{-1}$, for $a \in \{1, \dots, A\}$. In terms of the grid of time points, we will consider $\tau_1 = (6, 12, 18)$, $\tau_2 = (2, 6, 10, 14, 18)$ and $\tau_3 = (2, 4, 6, 8, 10, 12, 14, 16, 18)$.

\begin{figure}[ht]
	\centering
	\includegraphics[width = \linewidth]{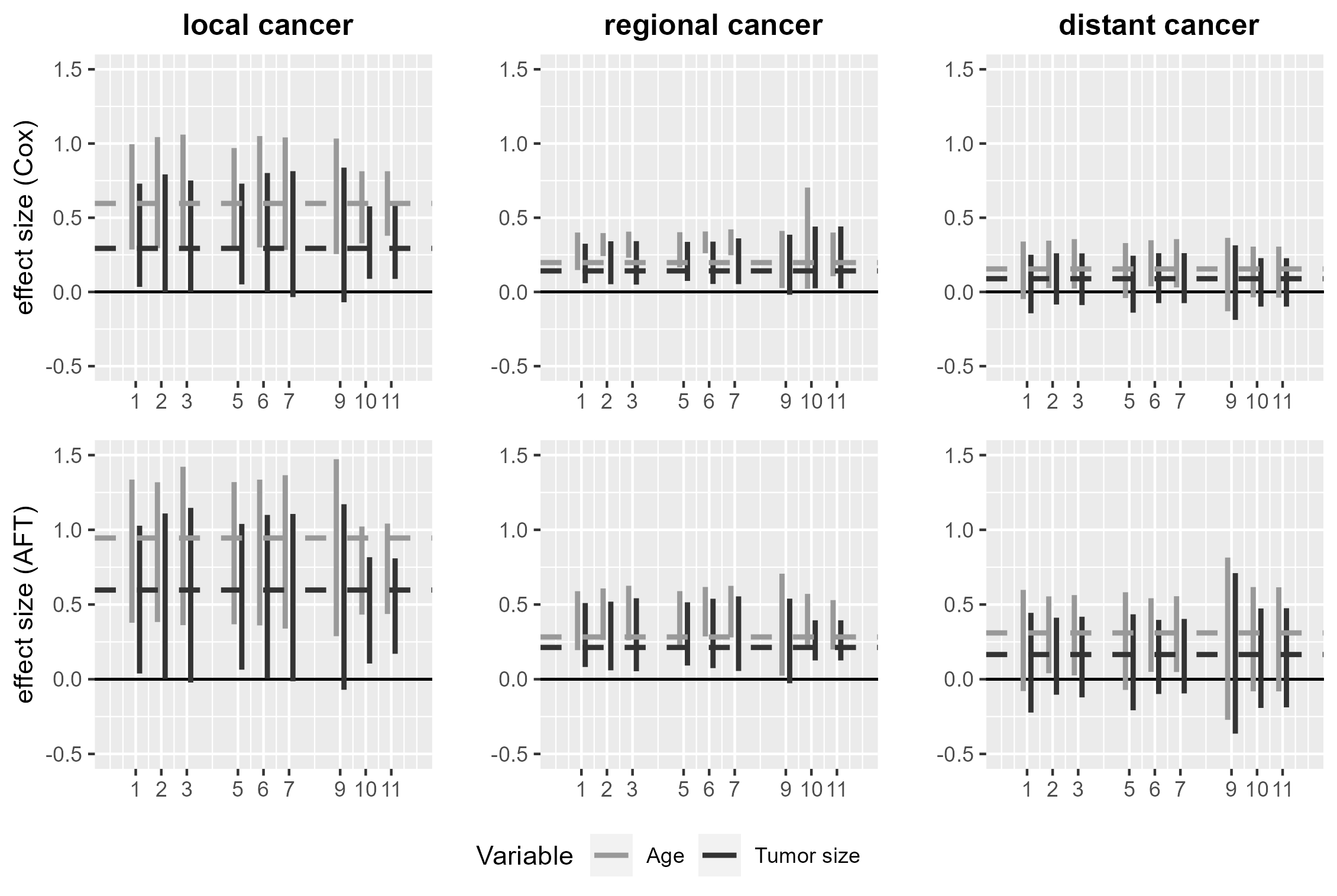}
	\caption{Comparison of different combination rules and considered sets of time points for the model with Cox link (top row) and AFT link (bottom row). Labels $(1, 2, 3)$, $(5, 6, 7)$ and $(9, 10, 11)$ correspond to the first, second and third combination method, respectively. Within these three groups, the smallest label corresponds to using $\tau_1$, the second smallest to using $\tau_2$ and the largest to using $\tau_3$.}
	\label{fig: Comparison}
\end{figure}

The results of such an analysis under both the AFT and Cox link are plotted in Figure \ref{fig: Comparison}. From these plots, no definitive \emph{best} strategy can be discerned. We do note that the majority vote combination rule, combined with considering many time points often leads to relatively narrow bounds. It can also be remarked that the estimates of the identified AFT or Cox model (assuming independence) sometimes fall outside of the obtained identified intervals when considering the subjects with regional cancer. This is an indication that the model might be misspecified for this stratum in the data.

\section{Alternative approach to time-independent coefficients}\label{supp: Alternative approach to time-independent coefficients}
Section \ref{sec: Time-independent effects of covariates} describes an approach to include the additional assumption of time-independent effects of covariates into the proposed methodology. As already noted in that section, it can \emph{partially} do this, as it only imposes that the parameter of interest, $\beta_k$, is independent of time. All other coefficients could still be time-dependent. While this might be a desirable property in some cases, in many other cases the approach might be overly flexible.

Motivated by this observation, we investigated an alternative approach to impose time-independent effects of all covariates, which start by augmenting the unconditional moment restrictions at a given time point $t$ to immediately consider the moment restrictions at all time points in the grid $(t_1, \dots, t_A)$, as follows:
\begin{equation} \label{eq: unconditional moment restrictions, alternative}
	\forall g \in \mathcal{G}:
	\begin{cases}
		\mathbb{E}[(\mathbbm{1}(Y \leq t_1) - \Lambda(\beta_0^{t_1} + \tilde{X}^\top\tilde{\beta}))g(X)] \geq 0\\
		\mathbb{E}[( \Lambda(\beta_0^{t_1} + \tilde{X}^\top\tilde{\beta}) - \mathbbm{1}(Y \leq t_1, \Delta = 1))g(X)] \geq 0\\
		\qquad \vdots\\
		\mathbb{E}[(\mathbbm{1}(Y \leq t_A) -  \Lambda(\beta_0^{t_A} + \tilde{X}^\top\tilde{\beta}))g(X)] \geq 0\\
		\mathbb{E}[(\Lambda(\beta_0^{t_A} + \tilde{X}^\top\tilde{\beta}) - \mathbbm{1}(Y \leq t_A, \Delta = 1))g(X)] \geq 0,
	\end{cases}
\end{equation}
with parameter vector $(\beta_0^{t_1}, \dots, \beta_0^{t_A}, \beta_1, \dots, \beta_d)$, denoting $\tilde{\beta} = (\beta_1, \dots, \beta_d)$, and with the restriction that $\beta_0^{t_1} < \dots < \beta_0^{t_A}$. Estimation again proceeds by test inversion, using the developments of \cite{Bei2024}.

It can be seen that this approach indeed imposes time-independence of all coefficients, as the condition to be tested requires that there exists a single vector $\beta$ such that the moment conditions at each time point are satisfied. In the combination approach, it would be possible that there exist different $\beta_1, \dots, \beta_A$, all with $k$-th element equal to $r$, such that the conditions at $t_1$ hold for $\beta_1$, the conditions at $t_2$ hold for $\beta_2$, etc.

Unfortunately, preliminary simulations showed that the approach proposed here is too computationally demanding. Moreover, due to the increased complexity of the condition to be tested, the variance of the test increases, leading to little or no improvements with respect to the computationally simpler combination approach of Section \ref{sec: Time-independent effects of covariates}. For this reason, the approach as presented here was not investigated further.


\bibliography{bibliographySuppArXiv}